\documentclass[manuscript2]{aastex701}

\newcommand{\gmin}{\gamma_{\rm min}}
\newcommand{\gmax}{\gamma_{\rm max}}

\usepackage{tikz, caption}
\newcommand{\labeledimage}[3]{%
  \begin{tikzpicture}
    \node[anchor=south west, inner sep=0] (image) at (0,0) {\includegraphics[width=#2\linewidth]{#3}};
    \begin{scope}[x={(image.south east)}, y={(image.north west)}]
      \node[anchor=north west, font=\bfseries\normalsize, fill=white, inner sep=2pt, rounded corners=2pt] at (-.08, 0.95) {#1};
    \end{scope}
  \end{tikzpicture}%
}

\usepackage{physics, caption, subcaption, siunitx, enumitem, bm, gensymb, graphicx}

\begin{document}

\title{Broad-band Spectral Modeling of Large-Scale X-ray Jets in High-Redshift Quasars: \\ An MHD-Informed Approach}

\author[orcid=0009-0008-3575-3965,sname='Liniewicz']{Patryk Liniewicz}
\affiliation{Astronomical Observatory of the Jagiellonian University, ul. Orla 171, 30-244 Krak\'ow, Poland}
\affiliation{Jagiellonian University, Doctoral School of Exact and Natural Sciences, ul. Prof. St. Łojasiewicza 11, 30-348 Kraków, Poland}
\email[show]{patryk.liniewicz@doctoral.uj.edu.pl}  

\author[orcid=0000-0002-7263-7540,sname='Stawarz']{\L ukasz Stawarz}
\affiliation{Astronomical Observatory of the Jagiellonian University, ul. Orla 171, 30-244 Krak\'ow, Poland}
\email[]{lukasz.1.stawarz@uj.edu.pl}  

\author[orcid=0000-0002-4377-0174,sname='Cheung']{C. C. Cheung}
\affiliation{Space Science Division, Naval Research Laboratory, Washington, DC 20375, USA}
\email[]{chi.c.cheung2.civ@us.navy.mil}

\author[orcid=0000-0003-0216-8053, sname='Migliori']{Giulia Migliori}
\affiliation{INAF—Istituto di Radioastronomia, Via Gobetti 101, I-40129, Bologna, Italy}
\email[]{giulia.migliori@inaf.it}  

\author[orcid=0000-0002-0905-7375, sname='Siemiginowska']{Aneta Siemiginowska}
\affiliation{Center for Astrophysics, Harvard \& Smithsonian, Cambridge MA 02138, USA}
\email[]{asiemiginowska@cfa.harvard.edu}  

\begin{abstract}
We present a systematic spectral analysis of kiloparsec-scale jets in high-redshift quasars, modeling their radio-to-X-ray emission as synchrotron radiation and inverse-comptonization of CMB by relativistic electrons. In contrast to the homogeneous one-zone approximation commonly adopted in the literature, we describe the jet as a current-carrying, axially symmetric outflow with a purely toroidal magnetic field in magnetohydrostatic equilibrium and with radial velocity shear. In this framework, the pressure, magnetic-field, and bulk-velocity profiles are linked self-consistently, capturing the radial stratification of the emitting region without introducing additional free parameters. For any individual source, the model effectively retains only a small number of free parameters, including the total jet power, $L_{\rm j}$, and the on-axis bulk Lorentz factor, $\Gamma_0$. We consider two prescriptions for the radial distribution of the radiating electrons -- proportional either to the gas pressure or to the rest-frame magnetic energy density -- and two toroidal-field profiles, yielding four model variants. Applying the model to a sample of ten quasar jets at $z \geq 2.5$ with X-ray features resolved by \textit{Chandra}, we perform Bayesian parameter inference and model comparison. The Bayesian evidence systematically favors electron distributions that follow the gas pressure rather than the magnetic energy density, while the data discriminate only weakly between the assumed field profiles. The inferred jet powers, reaching $L_{\rm j} \sim 10^{49}\,\mathrm{erg\,s^{-1}}$, are systematically larger than those obtained from one-zone models, and the corresponding global jet magnetization parameters are low. None of the derived quantities, including $\Gamma_0 \sim \mathcal{O}(10)$, shows a significant monotonic trend with redshift.
\end{abstract}


\section{Introduction} 

Understanding of the global structure of relativistic outflows in astrophysical systems requires the framework of magnetohydrodynamics (MHD), in which collisionless, highly conductive jet plasma is approximated as a neutral, magnetized fluid. Due to the non-linear character of the MHD equations, numerical simulations are the primary tool for obtaining maps of fluid density~$\rho$, bulk velocity~$\bm{u} \equiv \bm{\beta} c$, pressure~$P$, and magnetic field~$\bm{B}$, for a given set of initial and boundary conditions; throughout, vector quantities are typeset in boldface. The presence of a dynamically important large-scale magnetic field furthermore requires fully three-dimensional simulations, since axisymmetric 2D setups artificially suppress intrinsically non-axisymmetric MHD instabilities and modes of jet evolution \citep[see, e.g.,][]{1999MNRAS.308.1069K,Nakamura04, 10.1111/j.1365-2966.2009.15642.x}.
 
Even given reliable MHD solutions, simulating non-thermal emission images introduces further difficulties. The observed non-thermal radiation is produced by a population of ultra-relativistic electrons and positrons~$n_e(\gamma_e)$, where $\gamma_e = E / m_e c^2\gg 1$ is the electron Lorentz factor. However, (i)~this population may be spatially distinct from the background jet plasma treated within the MHD approximation (which can be of either purely leptonic or a mixed electron–positron–proton composition) and from the jet magnetic field, and (ii)~its energy distribution cannot be determined self-consistently within the MHD framework. Based on the simulation output $(\rho, \bm{u}, P, \bm{B})$, additional assumptions are therefore required to constrain the energy spectra of the radiating electrons.
 
The standard approach to the latter problem~(ii) is to assume a power-law electron distribution $n_e(\gamma_e) = n_0 \gamma_e^{-s}$ with energy index $s \simeq 2$ as expected from Fermi-type particle acceleration processes, where the minimum and maximum electron energies are free parameters. Regarding the former issue~(i), the most common choice is to assume that the integrated energy density of radiating electrons, $U_e = \int \gamma_e m_e c^2 \, n_e(\gamma_e)\, d\gamma_e$, is proportional to the gas pressure, $U_e \propto P$, or equivalently the gas density~$\rho$ via the equation of state. This would imply that $U_e$ constitutes a constant fraction of the jet fluid energy density, while the MHD output maps could be post-processed to produce synchrotron pseudo-emissivity $j_\nu \propto P\, B^\alpha$ instead of $j_\nu \propto n_0\, B^{(s+1)/2}$, where $\alpha$ is the spectral index set to a particular value \citep[see, e.g.,][and references therein]{fuentes2018total}.
 
Alternative prescriptions have also been explored, in which the radiating electron distribution instead follows the magnetic field energy density, $U_e \propto B^2$ \citep{porth2011synchrotron}, or is coupled to bulk velocity compression ($\bm{\nabla} \cdot \bm{u}$), shear ($\bm{\nabla}_{\perp} u$), or electric current density ($\bm{\nabla} \times \bm{B}$) as proxies for the local dissipation rate \citep{galaxies6010031}. A complementary approach is to approximately account for electron transport by solving the corresponding diffusion--advection equation based on the bulk velocity and magnetic field from MHD simulations \citep{Jones_1999, Tregillis_2001, TREGILLIS2002387}.
 
Once pseudo-emissivity maps have been produced, observed intensity maps can be calculated by further post-processing with the radiative transfer equation, incorporating Doppler effects relevant for relativistic outflows with radial velocity stratification. These procedures are active areas of research, but they are computationally expensive and hardly enable systematic exploration of the parameter space, particularly when broad-band spectra resulting from multiple emission processes---such as synchrotron and inverse-Compton (IC) scattering---are concerned.
 
In this context, our approach is especially relevant. We explore, in a systematic manner, the broad-band non-thermal emission of kiloparsec-scale relativistic jets, assuming the simplest ideal MHD model for a current-carrying, axially symmetric jet, far from the launching site where it can be considered collimated, and in magneto-hydrostatic equilibrium.

Cylindrical equilibrium models of magnetized jets have been widely employed for studies of jet structure and stability, including both pressure-supported MHD equilibria \citep[see, e.g.,][]{1992A&A...256..354A,Nalewajko12,ONeill12,Das19,2022ApJ...929..181K}, as well as force-free configurations \citep[see, e.g.,][]{Lyubarskii99,Narayan09,Sobacchi18}. This framework is therefore well-motivated and physically rigorous; it is best applicable to quasar sources where, at large distances, rotational motion of the flow and the poloidal field component can be neglected. Within this framework, we parametrize the relativistic electron energy density distribution as a power law, following either the gas pressure or the rest-frame magnetic pressure, and calculate broad-band spectra due to synchrotron emission and IC scattering of Cosmic Microwave Background (CMB) photons, noting that the CMB constitutes the dominant target photon field at large distances from the host galaxy center. We focus on radio (GHz) and X-ray (keV) frequencies, as these two bands offer arcsecond-resolution imaging of kpc-scale jets with instruments such as the Very Large Array (VLA) and Chandra X-ray Observatory (Chandra) \citep{2000SPIE.4012....2W}, respectively. Our modeling can then be applied to direct fitting of observational data, enabling estimation of main physical parameters---such as total kinetic jet power and the jet global magnetization---from basic observables including radio and X-ray fluxes of resolved jet segments.
  
Two distinct scenarios can account for X-ray emission in extragalactic large-scale jets \citep[see, e.g.,][]{harris2006}. For low-luminosity FR-I sources, synchrotron radiation is the preferred explanation for the radio-to-X-ray emission, given observational constraints from variability, morphology, the overall shape of the broad-band continuum, and the steep X-ray spectral slopes \citep{Worrall09}. This interpretation requires electrons with $\gamma_e > 10^7$, posing questions about the acceleration mechanism at large distances from the core. In contrast, the IC/CMB scenario is often favored for high-luminosity quasar and FR-II jets \citep{10.1046/j.1365-8711.2001.04160.x}, where the observed X-ray fluxes typically exceed the extrapolation of the synchrotron continuum from lower frequencies. In this picture, IC/CMB emission from low-energy electrons can in principle account for the enhanced X-ray fluxes, provided that the CMB energy density, $u'_{\rm CMB}$, is significantly boosted in the emitting plasma rest frame owing to relativistic bulk motion of the jet, corresponding to bulk Lorentz factors $\Gamma \equiv (1 - \beta^2)^{-1/2} \gg 1$, as well as its redshift dependence:
\begin{equation}
    u'_{\rm CMB} = (1+z)^4 \, \Gamma^2\, u_{\rm CMB},
\end{equation}
where primes denote quantities measured in the jet comoving frame.

We note that the IC/CMB model has been challenged by several observational constraints, including radio studies of jet bulk velocity profiles and distributions in quasar and FR-II jets \citep[e.g.,][]{1997MNRAS.286..425W, Mullin09}, deep optical polarimetry available for a few quasar jets \citep{Cara13}, X-ray morphological studies \citep[e.g.,][]{2007ApJ...657..145S,Marchenko17,2023ApJS..265....8R}, and, finally, \textit{Fermi}-LAT upper limits on the predicted GeV emission from several nearby quasar sources \citep[e.g.,][]{2015ApJ...805..154M, 2023MNRAS.518.3222B}. Nevertheless, the model remains a viable framework, particularly for high-redshift jets, and is therefore adopted here. We compute two-component SEDs consisting of synchrotron radio emission and IC/CMB scattering by a single electron population. Synchrotron self-Compton emission is neglected, as it is subdominant at kpc scales \citep{10.1046/j.1365-8711.2001.04160.x,2004ApJ...608...95S}.

Most IC/CMB modeling of large-scale quasar jets presented in the literature relies on a major simplifying assumption: large portions of the outflow, corresponding to bright knots with kpc-scale sizes or larger, are approximated as spherical emission regions moving with a uniform bulk velocity and containing isotropically distributed particles embedded in a tangled magnetic field. Both the particle and magnetic field distributions are typically assumed to be homogeneous throughout the entire emitting region \citep{2000ApJ...544L..23T,10.1046/j.1365-8711.2001.04160.x}.

Extensions of this scenario usually involve introducing an additional emission zone, commonly identified with the jet boundary layer, characterized by a lower bulk velocity or even an explicit radial velocity shear. Since such boundary layers are expected to be efficient sites of particle acceleration, a separate population of radiating electrons is often introduced in these regions, effectively increasing the number of free model parameters compared to the simplest homogeneous one-zone models \citep[see, e.g.,][]{Stawarz02,Aloy08,Wang21,tavecchio21}.

The main objective of this work is to propose an alternative approach in which the assumption of a homogeneous emission region is relaxed while avoiding the introduction of an arbitrary set of model parameters. Instead, quantities such as the jet pressure, magnetic field, and bulk velocity profiles are linked self-consistently within a well-defined MHD framework. Although this framework still represents a simplified description of realistic relativistic jets (see the model description in the following section), it nevertheless constitutes a significant improvement over previous modeling approaches.

All model spectra are computed in \textrm{Python 3} using \textrm{JAX} \citep{jax2018github} for automatic differentiation and hardware-accelerated array computation. Throughout the paper we assume a modern, flat $\Lambda$CDM cosmology with $H_0 = 67.97$~km\,s$^{-1}$\,Mpc$^{-1}$, $\Omega_m = 0.307$ and $\Omega_{\Lambda}=0.693$ \citep{Adame_2025}.

\section{Methods} \label{sec:methods}
 
\subsection{Jet MHD Structure} \label{sec:jet-structure}
 
We follow the simplest formulation of MHD models for collimated current-carrying outflows in magneto-hydrostatic equilibrium, as most recently analyzed by \citet{2022ApJ...929..181K}. In particular, we assume an axially symmetric cylindrical jet with a radially stratified magnetic field, $\bm{B}=\bm{B}(r)$, and bulk Lorentz factor, $\Gamma=\Gamma(r)$. The jet is taken to be uniform along the $z$-direction, i.e.\ no velocity or magnetic-field gradients are present along the outflow. The jet is assumed to satisfy radial magneto-hydrostatic equilibrium and pressure balance with the ambient medium at its boundary $r=R_{\rm j}$, where $\Gamma (R_{\rm j}) \equiv 1$. The jet fluid is assumed to obey an ultra-relativistic equation of state, such that the specific fluid enthalpy is given by $w \simeq 4P$. Most importantly, we consider a purely toroidal magnetic-field configuration, $\bm{B}(r)=B_\phi\!(r)\,\hat{\phi}$, anticipating negligible poloidal field components at large distances from the jet origin. Both transverse velocity shear and toroidal-field dominance are supported by radio polarimetric observations and basic theoretical considerations regarding large-scale quasar jets \citep[see, e.g.,][]{1984ARA&A..22..319B,BBR84}. We emphasize that the cylindrical, $z$-uniform, axisymmetric equilibrium adopted here is a deliberate \emph{per-segment} idealization: each modeled jet feature, whether a knot or a more extended segment, is treated as a locally cylindrical outflow, such that the knotty large-scale structure of powerful jets is represented as a sequence of independent regions rather than as a single global flow.
 
The equilibrium condition 
\begin{equation}\label{eq:equilibrium}
\frac{1}{c} \, \bm{J} \times \bm{B} = \bm{\nabla} P ,
\end{equation}
where $\bm{J} = (c/4\pi) \, \bm{\nabla}\times \bm{B}$ is the electric current density, allows for a self-consistent determination of the internal jet pressure profile, $P(r)$, once the profiles of $\Gamma(r)$ and $B_\phi(r)$ are specified. Introducing the normalized jet radius $x \equiv r/R_{\rm j}$ and dimensionless functions $b(x) \equiv B_\phi\!(x)/B_\phi\!(R_{\rm j})$ and $p(x) \equiv P\!(x)/P\!(R_{\rm j})$, one obtains in particular:
\begin{equation}\label{eq:pressure}
    p(x) = 1 +  \beta_{\rm pl}^{-1} -  \beta_{\rm pl}^{-1}\,f(x) + 2 \beta_{\rm pl}^{-1} \int_x^1 \frac{f(s)}{s}\,ds,
\end{equation}
where $f(x) \equiv \left[b(x)/\Gamma\!(x)\right]^2$ is the normalized rest-frame magnetic pressure, since $B'_\phi = B_\phi/\Gamma$, and 
\begin{equation}\label{eq:beta}
\beta_{\rm pl} \equiv \frac{P(1)}{P_B(1)}
\end{equation}
is the plasma-beta parameter at the jet boundary for the corresponding fluid and magnetic pressures $P(1)$ and $P_B(1) = B^2_\phi(1)/8\pi$, respectively. Throughout the following analysis we assume the fiducial value $\beta_{\rm pl} =1$, meaning the exact pressure equipartition at the jet boundary.
 
We parametrize the radial bulk velocity and magnetic profiles as:
\begin{equation}\label{eq:gamma_profile}
    \Gamma(x) = 1 + (\Gamma_0 - 1)(1 - x^{k_0}), \quad \textrm{and}
\end{equation}
\begin{equation}\label{eq:b_profile}
    b(x) = \frac{(1+a)\,x^{k_1}}{1 + a\,x^{k_2}},
\end{equation}
which have simple closed forms yet are flexible enough to span a broad range of physically relevant configurations. We note that vastly different choices of $k_0$, $a$, $k_1$, and $k_2$ yield nearly identical results provided they correspond to similar values of the global jet magnetization parameter~$\sigma$ (see the next section for the definition) for a fixed $\Gamma_0$.

In this work, we fix $k_0 = 2$ in all calculations and adopt two specific forms of the magnetic-field profile given by Eq.~\eqref{eq:b_profile}: {\bf model (1)} with $(a,k_1,k_2)=(0,1,-)$, corresponding to a linear increase of the toroidal field from the jet axis to the boundary, and {\bf model (2)} with $(a,k_1,k_2)=(100,1,2)$, characterized by a magnetic-field maximum at $x_{\rm peak}=0.1$ followed by a gradual decline toward the jet boundary; note that for such, the maximum value of the magnetic field is $B_\phi\!(x_{\rm peak}) \simeq 5 \times B_\phi\!(R_{\rm j})$.

\subsection{Radiating Electrons} \label{sec:electrons}
 
As specified above, in this work we assume that the jet plasma is hot, meaning $w \simeq 4P$. This plasma component, however, does not necessarily correspond exactly to the radial distribution of particles responsible for the bulk of the observed non-thermal jet emission (synchrotron and IC radiation). The latter component is identified here with a population of ultra-relativistic electrons, for which we assume a simple power-law energy distribution:
\begin{equation}\label{eq:edist}
    n'_e(\gamma_e') = K'_e(x) \times \zeta(\gamma_e') \quad \textrm{where} \quad \zeta(\gamma_e') = {\gamma_e'}^{-p} \, e^{-\gmin/\gamma_e'} \, e^{-\gamma_e'/\gmax}, 
\end{equation}
and $K'_e(x)$ is the normalization constant, while $\gmin$ and $\gmax$ denote the minimum and maximum electron Lorentz factors, respectively. Here and throughout the paper, primes denote quantities measured in the plasma rest frame at a given jet radius $x \in [0,1]$. The normalization constant is determined by assuming that the total energy density of radiating electrons, $U'_e(x) = m_e c^2 \int \!d\gamma_e' \, \gamma_e'\, n'_e(\gamma_e')$, constitutes a constant fraction $\eta_e$ of either the rest-frame magnetic energy density $U'_B(x) = {B'_\phi}^2(x)/8\pi$, or the fluid pressure $P(x)$, yielding
\begin{equation}\label{eq:modelsAB}
K'_e(x)= \dfrac{\eta_e \, P_B\!(1) \, }
{m_e c^2 \int \!d\gamma_e'\,\gamma_e'\,\zeta\!(\gamma_e')} \times 
\begin{cases}
f\!(x)
& \text{{\bf model (A)}}, \\
\beta_{\rm pl} \, p\!(x)
& \text{{\bf model (B)}},
\end{cases}
\end{equation}
respectively. Throughout the following analysis we assume fiducial values $\eta_e = 0.1$ and $\gmin=30$, although we also discuss the sensitivity of the model spectra to these parameters (see Section~\ref{sec:elasticity} below). As for the maximum electron Lorentz factors, for the majority of the analyzed sources we assume a fiducial value of $\gmax = 10^5$, except for four cases in which the broader radio-continuum coverage allows $\gmax$ to be treated as a free parameter.

Combining the two above specified normalization prescriptions (A) and (B) with the two magnetic field profiles (1) and (2) from the previous Section~\ref{sec:jet-structure}, we obtain four model variants summarized in Table~\ref{tab:models}. We adopt this labeling convention throughout.

\begin{deluxetable}{lll}
\tablecaption{Definition of the four model variants. The label digit
(1 or 2) refers to the magnetic field profile, while the letter
(A or B) denotes the electron normalization
prescription.\label{tab:models}}
\tablehead{
  \colhead{Model} &
  \colhead{Normalization} &
  \colhead{Magnetic field profile}
}
\startdata
1A & $U'_e \propto U'_B$ & $B_{\phi}(r)$ linearly increasing with $r\in[0,R_{\rm j}]$ \\
1B & $U'_e \propto P$    & $B_{\phi}(r)$ linearly increasing with $r \in [0,R_{\rm j}]$ \\
2A & $U'_e \propto U'_B$ & $B_{\phi}(r)$ reaches maximum at $r=0.1 \, R_{\rm j}$ \\
2B & $U'_e \propto P$    & $B_{\phi}(r)$ reaches maximum at $r=0.1 \, R_{\rm j}$  \\
\enddata
\end{deluxetable}

\subsection{Jet Power and Magnetization} \label{sec:jetpower}

By integrating over the jet cross-section, we obtain the powers carried by particles and by the jet magnetic field (i.e., the Poynting flux), respectively:
\begin{equation}\label{eq:Lp}
    L_p = 8\pi c \int_0^{R_{\rm j}} \! dr \, r\, \beta(r)\, \Gamma^2(r)\, P(r),
\end{equation}
and
\begin{equation}\label{eq:LB}
    L_B = \frac{1}{2} c \int_0^{R_{\rm j}} \! dr \, r\, \beta(r)\, B^2_{\phi}(r).
\end{equation}
The total jet power is then given by
\begin{equation}\label{eq:Lj}
    L_{\rm j} = L_p + L_B \equiv L_p \, (1+ \sigma) \simeq L_p ,
\end{equation}
where $\sigma \equiv L_B/L_p$ is the jet magnetization parameter. Here we have anticipated $\sigma < 1$ as expected for the configurations considered here according to the analysis of \citet{2022ApJ...929..181K}. This yields the following expression for the magnetic pressure at the jet boundary:
\begin{equation}\label{eq:P1}
    P_B(1) \simeq \frac{\beta_{\rm pl}^{-1} \, L_{\rm j}}{8\pi c\, R_{\rm j}^2 \displaystyle\int_0^1 \! dx \, x\, \beta(x)\, \Gamma^2(x)\, p(x)}.
\end{equation}
Thus, the jet parameters $L_{\rm j}$, $R_{\rm j}$, and $\beta_{\rm pl}$, together with the dimensionless profiles $b(x)$ and $\Gamma(x)$, fully determine the magnetic and dynamical properties of the model, including the toroidal magnetic-field structure and the pressure profile. The radiative properties are then specified through the additional parameter $\eta_e$ and the electron energy distribution function $\zeta(\gamma_e')$.
  
\subsection{Synchrotron and Inverse-Compton Emissivities} \label{sec:emission}
 
Assuming an isotropic distribution of radiating electrons in the plasma rest frame, the comoving synchrotron emissivity is given by
\begin{equation}\label{eq:jsyn}
    [\nu' j'_{\nu'}]^{\rm syn} = \frac{\sqrt{3}\,e^3 \, B' }{4\pi \, m_e c^2} \, \nu' \, \int\! d\gamma_e'\;  n'_e(\gamma_e') \,\, \mathcal{R}\!\left(\frac{\nu'}{\nu_c}\right),
\end{equation}
where $\nu_c = 3eB' {\gamma_e'}^2 / 4\pi m_e c$ is the critical frequency and $\mathcal{R}(x)$ is given in terms of modified Bessel functions $K_{4/3}$ and $K_{1/3}$ \citep{1986A&A...164L..16C}. Since these special functions are not available in \textrm{JAX}, in practice we evaluate the kernel $\mathcal{R}(x)$ using the analytical approximation of \citet{2010PhRvD..82d3002A}.

If the target photon field is also approximately isotropic in the plasma rest frame, the corresponding IC emissivity is given by
\begin{eqnarray}\label{eq:jnu-ic}
    [\nu' j'_{\nu'}]^\text{IC} & = & \frac{3\,m_e c^3\,\sigma_T}{16\pi} \\
    && \times {\epsilon'}^2 \, \int d\epsilon'_0 \int d\gamma_e'\, {\gamma_e'}^{-2} {\epsilon'_0}^{-1} \, n'_\text{ph}(\epsilon'_0)\, n'_e(\gamma_e') \, \mathcal{F}_{\rm IC}(\epsilon', \epsilon'_0, \gamma_e'), \nonumber
\end{eqnarray}
where $\epsilon' = h\nu' / m_e c^2$, $\epsilon'_0$ is the target photon energy expressed in the same $m_e c^2$ units, $n'_\text{ph}(\epsilon'_0)$ is the plasma rest-frame distribution of seed photons, and $\mathcal{F}_{\rm IC}$ denotes the standard Jones--Blumenthal \& Gould kernel, valid in both the Thomson and Klein--Nishina regimes of inverse Compton scattering \citep{1970RvMP...42..237B}. 

We further approximate the CMB as a monochromatic photon field and neglect the anisotropy induced in the jet rest frame by relativistic bulk motion\footnote{See in this context \citet{1995ApJ...446L..63D} who discussed the exact beaming pattern emerging from the additional anisotropy of seed photons in IC scattering.}. With such, we write
\begin{equation}\label{eq:cmb-delta}
    n'_{\rm ph}(\epsilon'_0) \simeq \frac{u_{\rm CMB} \, \Gamma^2 (1+z)^4}{m_e c^2\,\epsilon'_0}\times \delta\left[\epsilon'_0 - \Gamma (1+z) \epsilon_{\rm CMB}\right],
\end{equation}
where $u_{\rm CMB} \simeq 4.2 \times 10^{-13}$~erg~cm$^{-3}$ is the CMB energy density and $\epsilon_{\rm CMB} \simeq 1.2 \times 10^{-9}$ corresponds to the mean CMB photon energy in $m_e c^2$ units, $\langle h\nu_0 \rangle \simeq 6.3 \times 10^{-4}$~eV, both evaluated at the present epoch $z=0$. This approximation reduces the dimensionality of the integration in Eq.~\eqref{eq:jnu-ic} by eliminating one integral.
 
\subsection{Observed Luminosities} \label{sec:luminosity}

The observed synchrotron and IC/CMB luminosities are obtained by integrating the corresponding emissivities over a given jet volume, $\int dV'$, with the appropriate Doppler boosting, namely
\begin{equation}\label{eq:lum}
[\nu L_\nu] = 4 \pi \int \ dV' \, \mathcal{D}^4(x) \, [\nu' j'_{\nu'}]_{\nu' = (1+z) \nu / \mathcal{D}(x) },
\end{equation}
where the radius-dependent jet Doppler factor is
\begin{equation}\label{eq:doppler}
    \mathcal{D}(x) = \frac{1}{\Gamma(x)\left[1 - \beta(x)\cos\vartheta\right]},
\end{equation}
for a jet inclination angle $\vartheta$. Since the jet is uniform along $z$ and axially symmetric, the volume integral reduces to an integral over the normalized radial coordinate~$x$ multiplied by the length of a given segment of the outflow~$\ell$, namely $dV' = 2 \pi R_{\rm j}^2 \, \ell' \, x \, dx$. Note also that the emissivities $[\nu' j'_{\nu'}](x)$ carry an implicit radial dependence through $b(x)$ and $\Gamma(x)$. Finally, we note that for a moving jet $\ell' = \ell / \mathcal{D}$ \citep{Lind85}.
 
The full expressions, incorporating Doppler boosting effects, are:
\begin{eqnarray}\label{eq:final-synch}
    && [\nu L_\nu]_{\rm obs}^{\rm syn} = \frac{2\pi\sqrt{3}\,e^3 R_{\rm j}^2\,\ell\,(1+z) }{m_e c^2} \\
    && \times \nu \, \int_0^1\! dx \int\! d\gamma_e'\; x\, B'(x)\, \mathcal{D}^2(x)\, K'_e(x)\, \zeta(\gamma_e')\, \mathcal{R}\!\left(\frac{4\pi m_e c\,(1+z) \nu}{3e\,\mathcal{D}(x)\,B'(x)\,{\gamma_e'}^2}\right), \nonumber
\end{eqnarray}
and:
\begin{eqnarray}\label{eq:final-ic}
    && [\nu L_\nu]_{\rm obs}^{\rm IC} = \frac{3\pi\,c\,\sigma_T\,R_{\rm j}^2\,\ell\,u_{\rm CMB} \, (1+z)^4 }{2 \, \epsilon_{\rm CMB}^2}\, \\
    && \times \epsilon^2 \, \int_0^1 \!dx \int \! d\gamma_e'\; x\, \mathcal{D}(x)\, K'_e(x)\, \zeta(\gamma_e') \, {\gamma_e'}^{-2}\, \mathcal{F}_{\rm IC}\!\left(\frac{(1+z) \epsilon}{\mathcal{D}(x)},\,(1+z) \Gamma(x)\,\epsilon_{\rm CMB},\,\gamma_e'\right), \nonumber
\end{eqnarray}
where $B'(x) = B(x)/\Gamma(x) \equiv \sqrt{8 \pi P_B\!(1) \, f\!(x)}$. Observed fluxes are obtained via $[\nu F_\nu] = [\nu L_\nu] / 4\pi d_{\rm L}^2$, where $d_L$ is the source luminosity distance. Note that the volume integration and frame transformations introduced additional model parameters $z$, $\ell$, and $\vartheta$. Because Eqs.~\eqref{eq:final-synch}--\eqref{eq:final-ic} map the rest-frame emissivities directly to observed-frame $\nu F_\nu$ through the explicit $(1+z)$ frequency factors and the luminosity distance $d_{\rm L}$, the cosmological K-correction is built into the forward model for both the synchrotron and IC/CMB components, and no separate correction factor is needed.
 
\section{Broad-band Spectra} \label{sec:synthetic}

Altogether, the model consists of 15 parameters, summarized in Table~\ref{tab:params}. Some of these are constrained directly by the data for individual sources, and therefore are input parameters, including the source redshift $z$, and the characteristic sizes of the emitting region, $R_{\rm j}$ and $\ell$. The electron energy index is also determined by the photon index of the X-ray continuum, $\Gamma_{\rm X}$, obtained from the X-ray data analysis, according to the standard relation $p = 2\,\Gamma_{\rm X}-1$. The other parameters are treated here as fiducial, meaning that they are fixed to reference values throughout the fitting procedure for all analyzed sources; these include the boundary plasma-beta parameter $\beta_{\rm pl}$, the electron normalization parameter $\eta_e$, the minimum and maximum electron Lorentz factors $\gmin$ and $\gmax$ (unless otherwise stated), the parameters $(a,k_1,k_2)$ describing the magnetic field profile $b(x)$, and the shear index $k_0$ specifying the bulk Lorentz factor profile $\Gamma(x)$. The remaining three parameters --- including the jet inclination angle $\vartheta$, the total jet power $L_{\rm j}$, and the central bulk Lorentz factor $\Gamma_0$ --- are treated as free parameters in the fitting procedure in all cases. Once all those parameters are specified, the remaining jet properties can be determined self-consistently, including the jet magnetization parameter $\sigma$, magnetic field intensity $B_{\phi}(R_{\rm j})$ at the jet boundary, and the external medium pressure needed for the jet confinement, $P_{\rm ext} = (1 + \beta_{\rm pl})\, P_B\!(1) = 2 \, \, P_B\!(1)$.

\begin{deluxetable}{lllc}
\scriptsize
\tablecaption{Parameters of the model\label{tab:params}}
\tablehead{
  \colhead{Symbol} &
  \colhead{Parameter} &
    \colhead{Type} &
  \colhead{Data/Value/Prior}
}
\startdata
\multicolumn{4}{c}{{\it Jet global parameters}} \\
$z$ ($d_{\rm L}$)  & redshift (luminosity distance)  & input & optical spectroscopy\\
$R_{\rm j}$  & jet radius    & input & radio/X-ray imaging\\
$\ell$ & jet segment length & input & radio/X-ray imaging\\
$\vartheta$ & jet inclination & free & $\sin\vartheta$ (isotropic) over $\vartheta \in [0\fdg1,\,25\degr]$\\
\hline
\multicolumn{4}{c}{{\it MHD model}} \\
$L_{\rm j}$ & total jet power & free & log-uniform on $L_{\rm j}$ over $\log L_{\rm j}\in[46,50]$\\
$\beta_{\rm pl}$  & boundary plasma-beta & fiducial & 1\\
$\Gamma_0$  & central bulk Lorentz factor & free & log-uniform on $(\Gamma_0-1)$ over $\Gamma_0\in[2,\,60]$  \\
$k_0$   & shear index in $\Gamma(x)$ profile   & fiducial & 2 \\
$(a,k_1,k_2)$ & magnetic field profile $b(x)$   & fiducial  & {\it model 1:} $(0,1,-)$; {\it model 2:} $(100,1,2)$\\
\hline
\multicolumn{4}{c}{{\it Radiating electrons}} \\
$\eta_e$ & electron normalization & fiducial & 0.1\\
$p$   &   electron energy index   & input & X-ray spectroscopy\\
$\gmin$   &  minimum electron Lorentz factor   & fiducial & 30 \\
$\gmax$   &  maximum electron Lorentz factor   & fiducial/free & $10^5$/log-uniform over $\log \gmax \in [3,6]$ \\
\enddata
\end{deluxetable}

\begin{figure}
    \centering
    \includegraphics[width=1.0\linewidth]{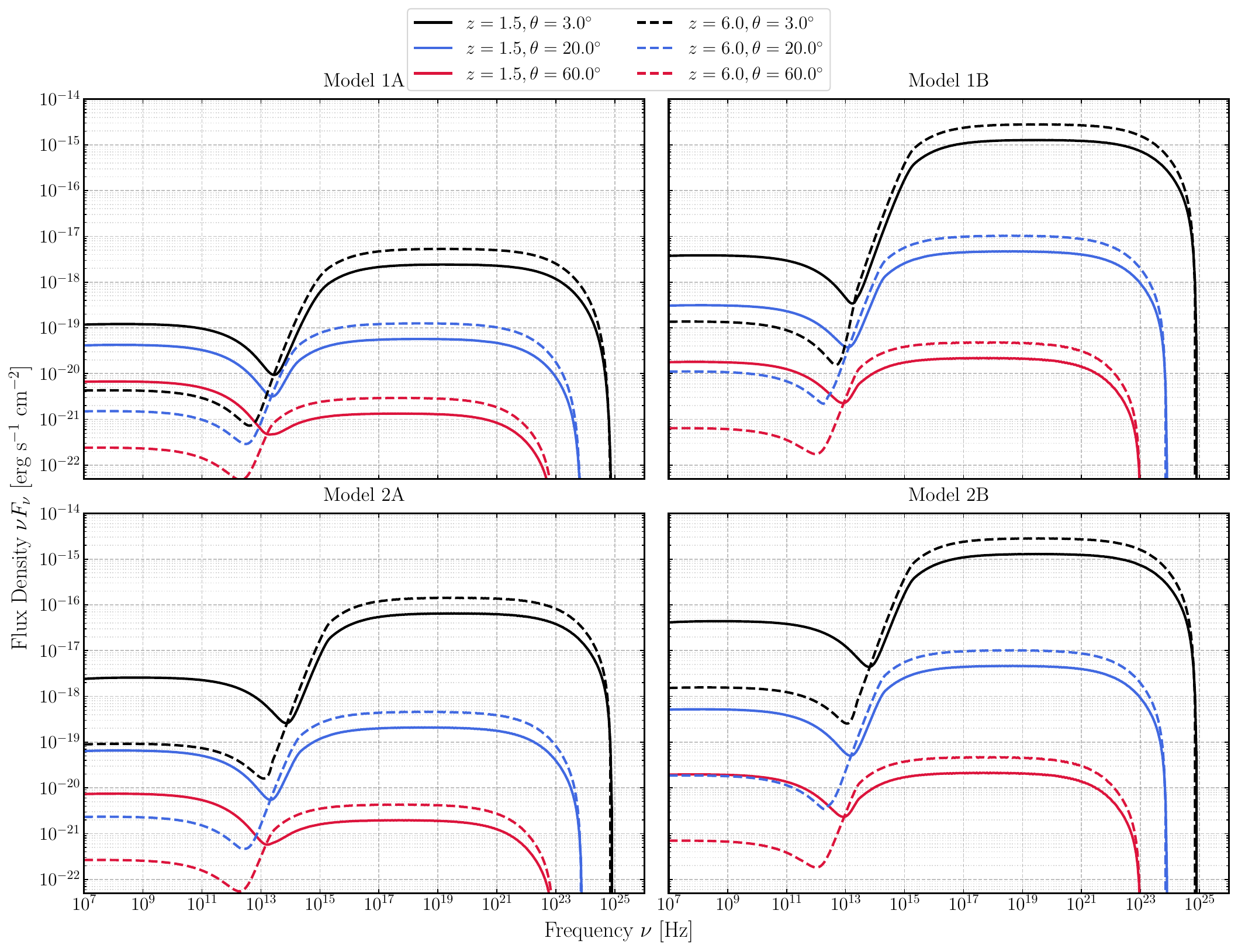}
    \caption{Illustrative broad-band Spectral Energy Distributions (SEDs) computed for a representative set of free and input model parameters, namely $R_{\rm j}=1$~kpc, $\ell=10$~kpc, $\Gamma_0=10$, $p=3$, and $L_{\rm j}=10^{47}$~erg~s$^{-1}$, together with the fiducial parameters listed in Table~\ref{tab:params}. The spectra are shown for all four model variants (1A, 1B, 2A, and 2B). Solid and dashed lines correspond to redshifts $z=1.5$ and $z=6.0$, respectively, while black, blue, and red curves correspond to viewing angles $\vartheta=3\degree$, $20\degree$, and $60\degree$. On each panel, the lower bump in the radio to infra-red range is due to the synchrotron emission, while the upper bump in the ultraviolet to $\gamma$-ray range is due to the IC/CMB process.  High-energy $\gamma$-ray fluxes are not corrected for the EBL absorption.}
    \label{fig:synthetic-spectra}
\end{figure}

As a first test, we verify that the model produces spectra with reasonable shapes and flux levels, broadly comparable to those observed in quasar jets and obtained in other modeling approaches. Figure~\ref{fig:synthetic-spectra} presents the results for an illustrative set of free and input model parameters --- namely $R_{\rm j}=1$~kpc, $\ell=10$~kpc, $p=3$, $\Gamma_0=10$, and $L_{\rm j}=10^{47}$~erg~s$^{-1}$ --- together with the fiducial parameters listed in Table~\ref{tab:params}; the spectra are computed for all four model variants (1A, 1B, 2A, and 2B), and for various redshifts $z$ and jet inclination angles $\vartheta$. We adopt $p=3$ as a representative of the X-ray photon indices in our sample ($\Gamma_X\approx2$, with $p=2\Gamma_X -1$, see Table~\ref{tab:sample}), which are typically steeper than the canonical $p\approx2$. Since the present work does not focus on the high-energy $\gamma$-ray emission from jets, the calculated spectra are not corrected for the cosmological attenuation of $\gamma$-ray fluxes due to interactions with the Extragalactic Background Light (EBL). We also note that the observed IC/CMB flux scales as $F_{\rm IC}\propto u_{\rm CMB}\,(1+z)^{4}/d_{\rm L}^2 \propto 1/d_{\rm A}^2$, where $d_{\rm A} = d_{\rm L}/(1+z)^2$ is the angular diameter distance, which is roughly constant for $z>0.5$. In other words, the $(1+z)^4$ enhancement of the seed CMB energy density is almost exactly cancelled by cosmological surface-brightness dimming \citep{Schwartz2002}. The IC/CMB component is therefore nearly independent of redshift over $0.5\lesssim z\lesssim6$, with only a mild ($\sim$factor-of-two) dip near $z\simeq1.5$. In Figure~\ref{fig:synthetic-spectra} this is apparent as the near-overlap of the IC portions of the solid and dashed curves.

Several qualitative differences between the model variants are apparent. Model~1A exhibits the weakest dependence on viewing angle, followed by model~2A; models~1B and~2B show a stronger angular sensitivity. Family~B models generally produce higher flux levels than their family~A counterparts. The IC/CMB components of models~1B and~2B are nearly indistinguishable, while their synchrotron components differ, offering a potential diagnostic to discriminate between the two normalization prescriptions using radio data.

\begin{figure}[ht]
    \centering
    \includegraphics[width=1.0\linewidth]{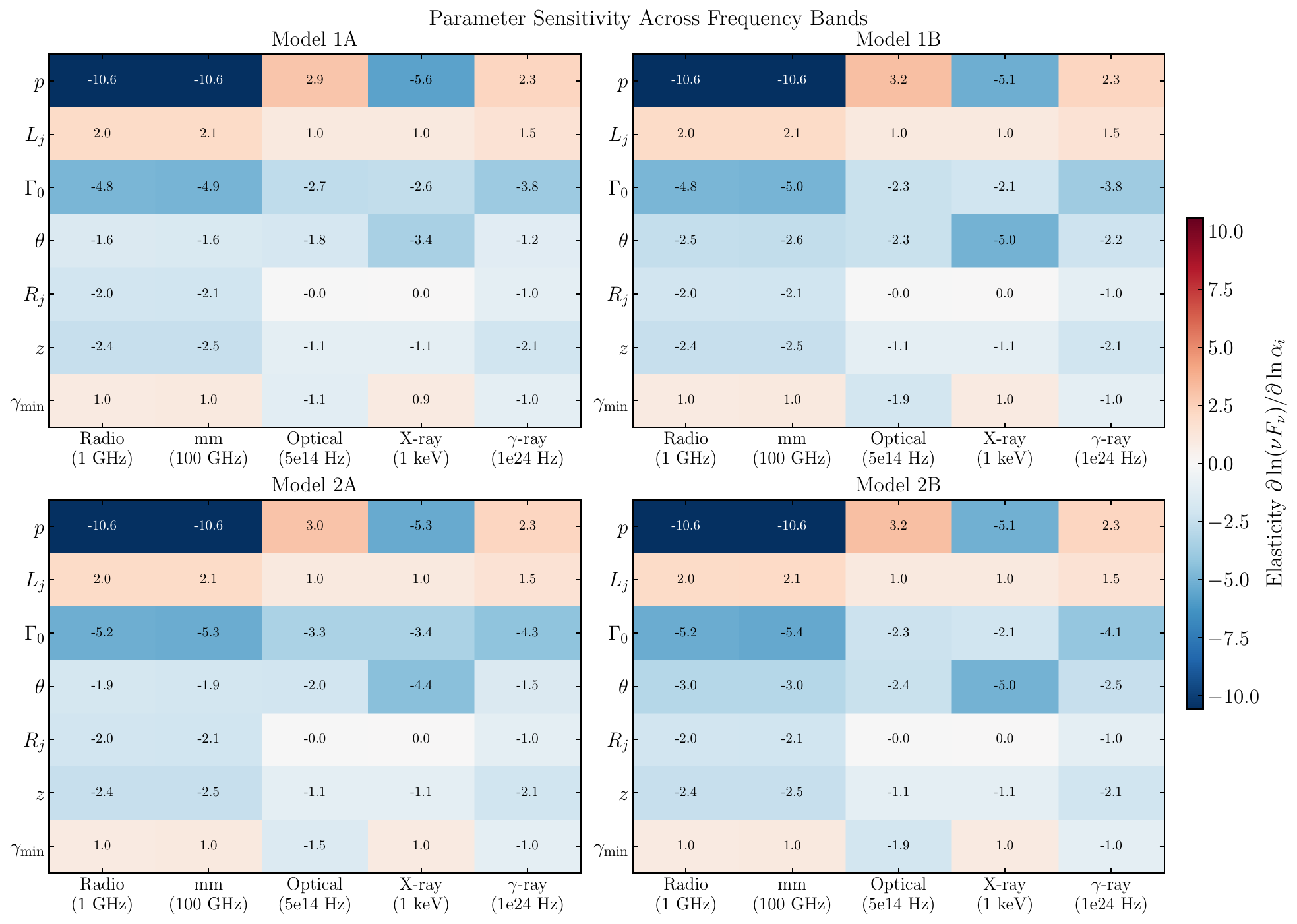}
    \caption{Elasticity of the four model variants under a 5\% change in each parameter, evaluated at representative observing frequencies.}
    \label{fig:sens-heatmap}
\end{figure}

\subsection{Elasticity Analysis} \label{sec:elasticity}

To quantify the sensitivity of the model SEDs to individual parameters beyond visual inspection, we employ the concept of \textit{elasticity}, widely used in economics \citep[e.g.,][]{varian2014intermediate}. For a positive function $\mathcal{F}(x; \bm{\kappa})$ of a positive argument $x$ and parameters $\bm{\kappa}$, the elasticity with respect to parameter $\kappa_i$ is defined as:
\begin{equation}\label{eq:elasticity}
    \mathcal{E}_{\kappa_i}[\mathcal{F}] = \left.\pdv{\ln \mathcal{F}}{\ln \kappa_i}\right|_{x = x_0},
\end{equation}
and represents the fractional change in $\mathcal{F}$ in response to a fractional change in $\kappa_i$, evaluated at a fixed argument $x_0$. In our case, $\mathcal{F} = [\nu L_\nu]$ and $x_0$ corresponds to a set of representative observing frequencies spanning from radio through $\gamma$-ray bands. We evaluate the elasticity numerically under a 5\% perturbation of each free model parameter, and additionally include the fiducial and input parameters $z$, $R_{\rm j}$, $p$ and $\gmin$ as a consistency check (with $z$ serving here as a proxy for the uncertainty in the luminosity distance $d_L$, since quasar redshifts themselves are known to $\lesssim1\%$), allowing them to vary in this part of the analysis. The results are presented as a heatmap in Figure~\ref{fig:sens-heatmap}.

The elasticity analysis confirms and refines the qualitative trends noted above. Model~1A is indeed the least sensitive to changes in viewing angle~$\vartheta$, followed by model~2A. Family~A models---and model~2A in particular---show greater sensitivity to the on-axis Lorentz factor~$\Gamma_0$ than their family~B counterparts. 

A noteworthy feature is the frequency-dependent sensitivity to the electron power-law index~$p$: model~1A shows the weakest response in the optical band but the strongest in the X-ray band. This behavior likely reflects differences in the spectral break position between models rather than a genuine insensitivity, since the elasticity is a local quantity and the spectral curvature near a break can amplify or suppress the response depending on the evaluation frequency. The radio and mm bands are by far the most sensitive to the electron index $p$, as expected since $p$ directly sets the synchrotron slope; this sensitivity is identical across all four variants. The vanishing elasticity with respect to $\gmax$ (not shown in Figure~\ref{fig:sens-heatmap} for compactness, equal to zero everywhere) across all bands indicates that the maximum electron energy lies well above the range probed by the observing frequencies considered here; changes in $\gmax$ affect the SED only at frequencies beyond the 10~GeV range for the fiducial parameters adopted. It is worth noting that this near-vanishing $\gmax$ elasticity holds only for \emph{broad} power-law distributions with $\gmin\ll\gmax$. For \emph{narrow} spectra, with $\log \gmax/\gmin$ of order a few, the high-frequency cutoff shifts into the observed bands, and the SED, and hence the model discrimination, becomes strongly sensitive to $\gmax$. This is precisely what motivates treating $\gmax$ as a free parameter for the sources with the best-sampled radio continua (Section~\ref{sec:results}). 

Finally, we do not test here the sensitivity of the model spectra to the parameters $\eta_e$ and $\ell$, as variations in these quantities would only result in a vertical scaling of the broad-band SEDs, without changing their spectral shapes.

\subsection{Spectral Fitting Procedure} \label{sec:fitting}

For parameter inference and model comparison we use nested sampling, introduced by Skilling \citep{skilling2004nested, 10.1214/06-BA127} as a method to recover the Bayesian evidence $\mathcal{Z}$ and, as a byproduct, the posterior distribution. It works by iteratively removing a point (or a number of points) with the lowest likelihood value from a set of randomly scattered ``live points'' within the prior volume. This procedure samples the entire volume and naturally allows for the computation of the evidence as a simple sum (e.g., using the trapezoidal rule). It is a natural choice for tasks such as probing (potentially) multi-modal posterior distributions and performing model comparison, both of which are relevant for the cylindrical jet model considered here. In particular, posterior distributions and model evidence within the Bayesian framework were computed using the MLFriends \citep{2016S&C....26..383B, 2019PASP..131j8005B} algorithm implemented in the UltraNest~\footnote{\url{https://johannesbuchner.github.io/UltraNest/}} package \citep{2021JOSS....6.3001B}.

\begin{deluxetable*}{lllccccc}[h]
  \tabletypesize{\scriptsize}
  \tablewidth{\textwidth}
  \tablecaption{%
      High-redshift quasars with resolved X-ray jets%
      \label{tab:sample}%
  }
  \tablehead{
      \colhead{Name}          &
      \colhead{$z$} &
      \colhead{Feature}&
      \colhead{$R_{\rm j} \times \ell$}           &
      \colhead{$\Gamma_{\rm X}$}    &
      \colhead{$F_{\rm X}$ @ 1\,keV} &
            \colhead{$F_{\rm opt}$ @ $10^{14}$\,Hz} &
          \colhead{$F_{\rm R}$ @ GHz} \\
                \colhead{}          &
      \colhead{} &
      \colhead{}&
      \colhead{[arcsec]}           &
      \colhead{}    &
      \colhead{[nJy]} &
            \colhead{[$\mu$Jy]} &
          \colhead{[mJy]} 
  }
  \startdata
PSO J030947.49
      & $6.10$
      & Jet
      & $2.0 \times 2.5$
      & $1.79^{+0.74}_{-0.69}$
      & $0.42^{+0.07}_{-0.04}$
      & ---
      & $0.63\pm 0.14$ @ 3.0 \\
GB 1428$+$4217
      & $4.72$
      & Jet
      & $0.5 \times 3.6 $
      & $1.7^f$
      & $1.75^{+0.7}_{-0.54}$
      & $<0.13$ @ $3.7$
      & $1.4\pm0.14$ @ 1.4 \\
            & & & & & & 
      & $0.41\pm 0.04$ @ 4.9\\
GB 1508$+$5714
      & $4.30$
      & Jet
      & $0.4 \times 1.6 $
      & $1.9 \pm 0.36$
      & $1.1\pm0.2$
      & $<0.36$ @ 3.75
      & $20.5 \pm 2.6$ @ 0.14\\
      & & & & & &  $<1.1$ @ 6.2 
      & $1.2\pm 0.12$ @ 1.4\\
      & & & & &  &
      & $0.2 \pm 0.04$ @ 4.9\\
      & & & & & & 
      & $<0.3$ @ 8.4   \\
1745$+$624
      & $3.89$
      & K1.4+K1.8
      & $0.2 \times 1.1 $
      & $1.62^{+0.16}_{-0.17}$
      & $7.8\pm 0.8$
      & $<0.06$ @ $5.2$
      & $49.7\pm 7.5$ @ 1.5 \\
                  & & & & & & 
      & $17.6\pm 1.8$ @ 4.9 \\
                  & & & & & & 
      & $9.5\pm 1.0$ @ 8.5\\
                  & & & & & & 
      & $6.4\pm 1.0$ @ 14.9\\
 PKS J1421$-$0643
      & $3.69$
      & A+B+C+D
      & $0.55 \times 3.0$
      & $1.65 \pm 0.19$
      & $4.2 \pm 0.6$
      & $<0.67$ @ $3.73$
      & $2.77 \pm 0.04$ @ 5.1 \\
       & & & & &
      & $<0.37$ @ $5.65$ 
      & $1.58\pm 0.02$ @ 7.0\\
          & & & & & & 
      & $0.78\pm 0.04$ @ 9.0\\
          & & & & & & 
      & $0.41\pm 0.04$ @ 10.7\\
PMN J0909$+$0354
      & $3.29$
      & NNW
      & $0.4 \times 0.8$
      & $1.42\pm 0.44$
      & $0.45\pm 0.14$
      & ---
      & $20.0 \pm 4.4$ @ 1.5 \\
         & & & & & & 
      & $5.3 \pm 1.2$ @ 6.2\\
         & & & & & & 
      & $2.8 \pm 1.0$ @ 8.5\\
J1405$+$0415
      & $3.22$
      & Jet
      & $0.75 \times 2.0$
      & $2.37^{+0.34}_{-0.32}$
      & $1.39^{+0.23}_{-0.33}$
      & ---
      & $<0.39$ @ 6.0\\
0805$+$046
      & $2.88$
      & Region 2
      & $0.35 \times 2.3$
      & $2.04^{+0.77}_{-0.71}$
      & $0.34^{+0.18}_{-0.13}$
      & ---
      & $12.4\pm 1.2$ @ 4.9 \\
       & & & & & & 
      & $6.8 \pm 0.7$ @ 8.4\\
0730$+$257
      & $2.69$
      & Region 2
      & $0.45 \times 1.8$
      & $1.67^{+0.49}_{-0.47}$
      & $0.64^{+0.25}_{-0.20}$
      & ---
      &  $20.0 \pm 2.0$ @ 4.9 \\
      & & & & & & 
      & $13.6\pm 1.4$ @ 8.7\\
      & 
      & Region 3
      & $0.6 \times 4.0$
      & $1.93^{+0.51}_{-0.48}$
      & $0.71^{+0.26}_{-0.21}$
      & ---
      &  $96.8 \pm 9.7$ @ 4.9 \\
                       & & & & & & 
      & $55.3\pm 5.5$ @ 8.7\\
B3 0727$+$409
      & $2.50$
      & Knot1.4
      & $0.4 \times 0.4$
      & $1.74^f$
      & $0.44\pm 0.10$
      & ---
      & $4.5 \pm 0.9$ @ 1.4 \\
                       & & & & & & 
      & $1.5 \pm 0.3$ @ 4.9\\
      & 
      & Extended
      & $1.8 \times 10$
      & $1.74^{+0.34}_{-0.32}$
      & $2.7 \pm 0.7$
      & ---
      & $<2.2$ @ 1.4 \\
                            & & & & & & 
      & $<1.8$ @ 4.7
 \enddata
  \tablecomments{%
 {\bf PSO J030947.49$+$271757.31:} $R_{\rm j}$ taken as half of the angular extension of the jet feature, $4^{\prime\prime}$; the energy flux density at 1\,keV estimated from the provided 0.5--7.0\,keV energy flux; \citet{Ighina_2022}. {\bf GB 1508$+$5714:} \citet{Siem03,Cheung_2004,Cheung05,Kappes22}. {\bf GB 1428$+$4217:} the energy flux density at 1\,keV estimated from the provided 2--10\,keV energy flux; \citet{Cheung2012}. {\bf 1745$+$624:} jet radius taken as the radius of the apertures used for extraction of the {\it HST} upper limits, while the jet length is taken as the length of the {\it Chandra} source extraction region for the total jet (K1.4+K1.8); \citet{Cheung2006}. {\bf PKS J1421$-$0643:} $\ell$ taken as the projected extent over which net X-ray emission from the entire jet (knots A+B+C+D) is detected above the background and outside the quasar core PSF; $R_{\rm j}$ taken as the radius of a cylinder equivalent to the median transverse width (Gaussian $1\sigma$) measured along the jet; radio fluxes are taken as the sum of the fluxes measured for the individual knots, while optical upper limits are taken as the highest values reported for the individual knots; \citet{Worrall_2020}. {\bf PMN J0909$+$0354:} the energy flux density at 1\,keV estimated from the provided 0.5--7.0\,keV energy flux for the region defined in a $0\farcs4$ circle centered on the observed $2\farcs3$-distant radio knot, with a 30\% uncertainty; \citet{Perger21}. {\bf J1405$+$0415:} \citet{Snios21,Maithil_2025}. {\bf 0805$+$046:} \citet{MK16}, this work. {\bf 0730$+$257:} \citet{MK16}, this work. {\bf B3 0727$+$409:} linear sizes of the K1.4 feature, taken as the knot's radio extent corresponding to the VLA 4.86\,GHz beam size. \citet{Simionescu_2016}.}
\end{deluxetable*}

We adopt independent priors for all sampled free parameters, summarized in Table~\ref{tab:params}. The viewing angle $\vartheta$ is assigned an isotropic prior, $p(\vartheta)\propto\sin\vartheta$, implemented via the inverse-CDF transform of $\cos\vartheta$ over the interval $[0\fdg1,\,25\degr]$. The upper bound excludes configurations geometrically inconsistent with the observed jet-to-counterjet brightness ratio in quasar sources, while the lower bound avoids the coordinate singularity at $\vartheta=0\degr$. The jet kinetic luminosities $L_{\rm j}$ are assigned log-uniform (Jeffreys) priors over $\log_{10}(L_{\rm j}/\mathrm{erg\,s^{-1}})\in[46,50]$, reflecting prior ignorance of the luminosity scale while remaining bounded by physically plausible values for quasar jets. The bulk Lorentz factors $\Gamma_0$ are assigned log-uniform priors on $\Gamma_0-1$ over $\Gamma_0\in[2,60]$, spanning the range of Lorentz factors inferred from superluminal motions in parsec-scale quasar jets \citep[e.g.,][]{Homan21}. We verified that the posteriors of both preferred models remain well within the prior support, confirming that the upper bound $\Gamma_0=60$ is non-informative. Finally, the prior for $\gmax$, whenever treated as a free parameter, is taken as log-uniform over $\log \gmax \in [3,6]$.

Sampled parameters were inferred via nested sampling with 800 live points. All best-fit values quoted in the next section are posterior medians with asymmetric $1\sigma$ credible intervals (16th and 84th percentiles); credible intervals are propagated directly through the posterior samples without Gaussian approximation. Boundary quantities are derived from 1,000 posterior draws for each component separately.

\section{Application to High-redshift Quasar Jets} \label{sec:application}

The framework presented in this paper --- namely X-ray emission from large-scale quasar jets produced through IC/CMB scattering --- is naturally best suited for application to high-redshift sources, for reasons discussed in the previous sections. Accordingly, we have compiled a sample of quasar jets resolved in X-rays, with and without radio counterparts. Throughout this work, we define ``high-$z$'' sources as those with $z\geq2.5$. This threshold lies around the peak of the quasar luminosity function \citep[e.g.,][]{Richards16,Shen20} and therefore probes quasar populations at earlier cosmic epochs, where in addition IC/CMB effects are expected to become increasingly important.

\begin{deluxetable}{llc cccc}[h]
  \tabletypesize{\footnotesize}
  \tablewidth{0pt}
  \tablecaption{%
      Bayesian model comparison via the Jeffreys scale.
      $\Delta\ln\mathcal{Z}$ is quoted relative to the preferred model
      (reference set to zero).%
      \label{tab:jeffreys}%
  }
  \tablehead{
      \colhead{Source}    &
      \colhead{$z$}       &
      \colhead{Best}      &
      \multicolumn{4}{c}{$\Delta\ln\mathcal{Z}$ / interpretation}  \\
      \colhead{}          &
      \colhead{}          &
      \colhead{model}     &
      \colhead{1A}        &
      \colhead{1B}        &
      \colhead{2A}        &
      \colhead{2B}
  }
  \startdata
  PSO J030947.49$+$271757.31         & $6.10$ & 1B
      & $-1.03$ / weak
      & $+0.00$ / \textbf{pref.}
      & $-0.11$ / indist.
      & $-0.23$ / indist. \\
  PSO J030947.49$+$271757.31$^a$ & $6.10$ & 1B
      & $-2.74$ / mod.
      & $+0.00$ / \textbf{pref.}
      & $-1.72$ / weak
      & $-0.01$ / indist. \\
  PSO J030947.49$+$271757.31$^b$  & $6.10$ & 2B
      & $-3.23$ / mod.
      & $-0.04$ / indist.
      & $-1.96$ / weak
      & $+0.00$ / \textbf{pref.} \\
  GB 1428$+$4217                     & $4.72$ & 1B
      & $-4.40$ / mod.
      & $+0.00$ / \textbf{pref.}
      & $-2.60$ / mod.
      & $-0.06$ / indist. \\
  GB 1508$+$5714                     & $4.30$ & 1B
      & $-3.14$ / mod.
      & $+0.00$ / \textbf{pref.}
      & $-2.05$ / weak
      & $-0.21$ / indist. \\
  1745$+$624                         & $3.89$ & 1B
      & $-45.01$ / decis.
      & $+0.00$ / \textbf{pref.}
      & $-3.47$ / mod.
      & $-0.32$ / indist. \\
  PKS J1421$-$0643                   & $3.69$ & 1B
      & $-3.93$ / mod.
      & $+0.00$ / \textbf{pref.}
      & $-2.55$ / mod.
      & $-0.80$ / indist. \\
  PMN J0909$+$0354                   & $3.29$ & 1B
      & $-4.06$ / mod.
      & $+0.00$ / \textbf{pref.}
      & $-2.32$ / weak
      & $-0.22$ / indist. \\
  J1405$+$0415                       & $3.22$ & 1B
      & $-3.44$ / mod.
      & $+0.00$ / \textbf{pref.}
      & $-1.80$ / weak
      & $-0.25$ / indist. \\
  0805$+$046                         & $2.88$ & 1B
      & $-1.67$ / weak
      & $+0.00$ / \textbf{pref.}
      & $-0.44$ / indist.
      & $-0.35$ / indist. \\
  0730$+$257                         & $2.69$ & 1B
      & $-3.23$ / mod.
      & $+0.00$ / \textbf{pref.}
      & $-1.90$ / weak
      & $-0.18$ / indist. \\
  B3 0727$+$409                      & $2.50$ & 1B
      & $-5.29$ / decis.
      & $+0.00$ / \textbf{pref.}
      & $-2.55$ / mod.
      & $-0.27$ / indist. \\
  \enddata
  \tablecomments{%
      Interpretation follows the Jeffreys scale: $|\Delta\ln\mathcal{Z}| < 1$ indistinguishable, $1$--$2.5$ weak evidence against, $2.5$--$5$ moderate evidence against, $>5$ decisive evidence against. Superscripts $^{a,b}$ indicate the PSO J030947.49$+$271757.31 models with $R_{\rm j}=0\farcs5$ and $0\farcs2$, respectively, instead of the nominal value of $2\farcs0$.
  }
\end{deluxetable}

\subsection{High-$z$ Jet Sample} \label{sec:sample}

A systematic search of the literature yielded a sample of 10 quasars at redshifts $z\geq2.5$ for which \textit{Chandra} observations revealed extended jet features. The selected sources are listed in Table~\ref{tab:sample}, together with the main input quantities used in the analysis: (i) source name, (ii) source redshift $z$, (iii) identification of the X-ray jet feature, (iv) projected linear sizes $R_{\rm j}\times\ell$, (v) X-ray photon index $\Gamma_{\rm X}$ (superscript $^f$ indicates frozen $\Gamma_{\rm X}$ in the {\it Chandra} data analysis), (vi) X-ray flux at 1\,keV, (vii) optical upper limits, and (viii) radio flux(es)/upper limit(s). Sources in the table are ordered by decreasing redshift. The corresponding references for all sources are provided in the table caption. In the cases of 0805$+$046 and 0730$+$257, we used an additional VLA image at 8.4~GHz (0\farcs4 beam; program AH480 from 1992-12-06) and a 4.9~GHz image (0\farcs35 beam; program AL093 from 1985-02-27), respectively, that one of us (C.C.C.) had processed as part of the work reported in \citet{MK16} to provide additional measurements of the flux densities of the jet regions of interest.

The sample is not homogeneous in terms of source selection and observational properties. Some X-ray detections resulted from targeted \textit{Chandra} observations of jet-candidate sources selected based on the presence of large-scale radio jets resolved on arcsecond scales, whereas others correspond to serendipitous discoveries \citep[e.g., B3 0727+409][]{Simionescu_2016}. In addition, in some cases the X-ray fluxes were extracted from regions defined by radio imaging, while other prominent X-ray features have no identified radio counterparts. Consequently, not all detected X-ray features have measured radio fluxes, although radio upper limits are available in all such cases.

It should be emphasized that our analysis excludes terminal hotspots and counter-hotspots, focusing only on extended structures that can be classified as jets or jet knots. Specifically, we exclude the hotspot in 1745$+$624, Region~1 in 0730$+$257, Regions~1 and 3 in 0805$+$046, and the hotspot in B3~0727$+$409. We also exclude candidate X-ray jet sources for which the available \textit{Chandra} observations yielded only upper limits. This category comprises 1239$+$376 at $z=3.82$ and 1754$+$676 at $z=3.6$ \citep{MK16}, J1610$+$1811 at $z=3.12$ \citep{Snios21, Maithil_2025}, PSO J352.4034$-$15.3373 at $z=5.83$ \citep{Connor21}, and the peculiar source J1016$+$2037 at $z=3.11$ analyzed by \citet{Snios21}, which exhibits non-coincident X-ray and radio features misaligned by approximately $180\degr$. Finally, we note that, within our sample, 0805$+$046 is characterized by the lowest-significance (marginal) detection \citep{MK16}.

\begin{figure}[th]
    \centering \includegraphics[width=0.38\linewidth]{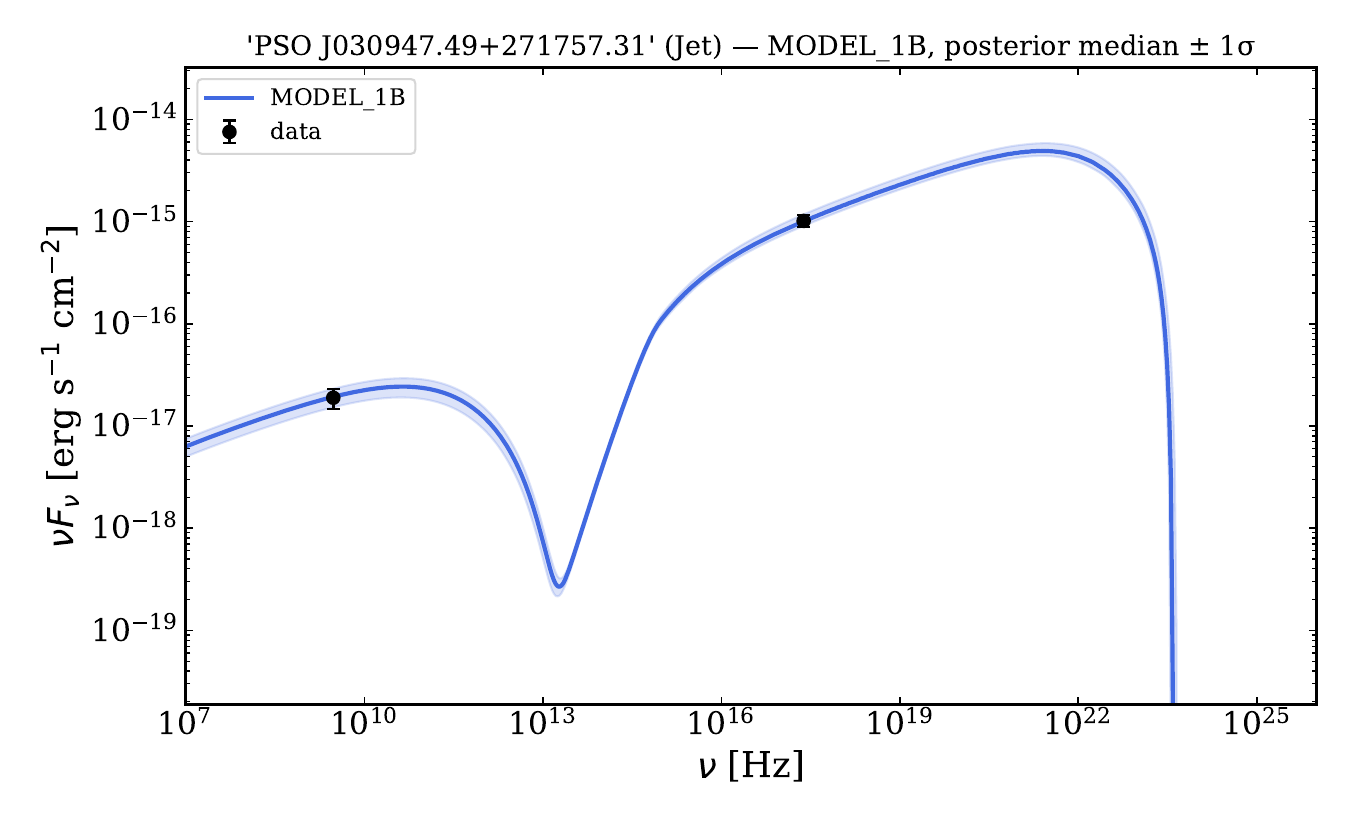}
     \centering \includegraphics[width=0.38\linewidth]{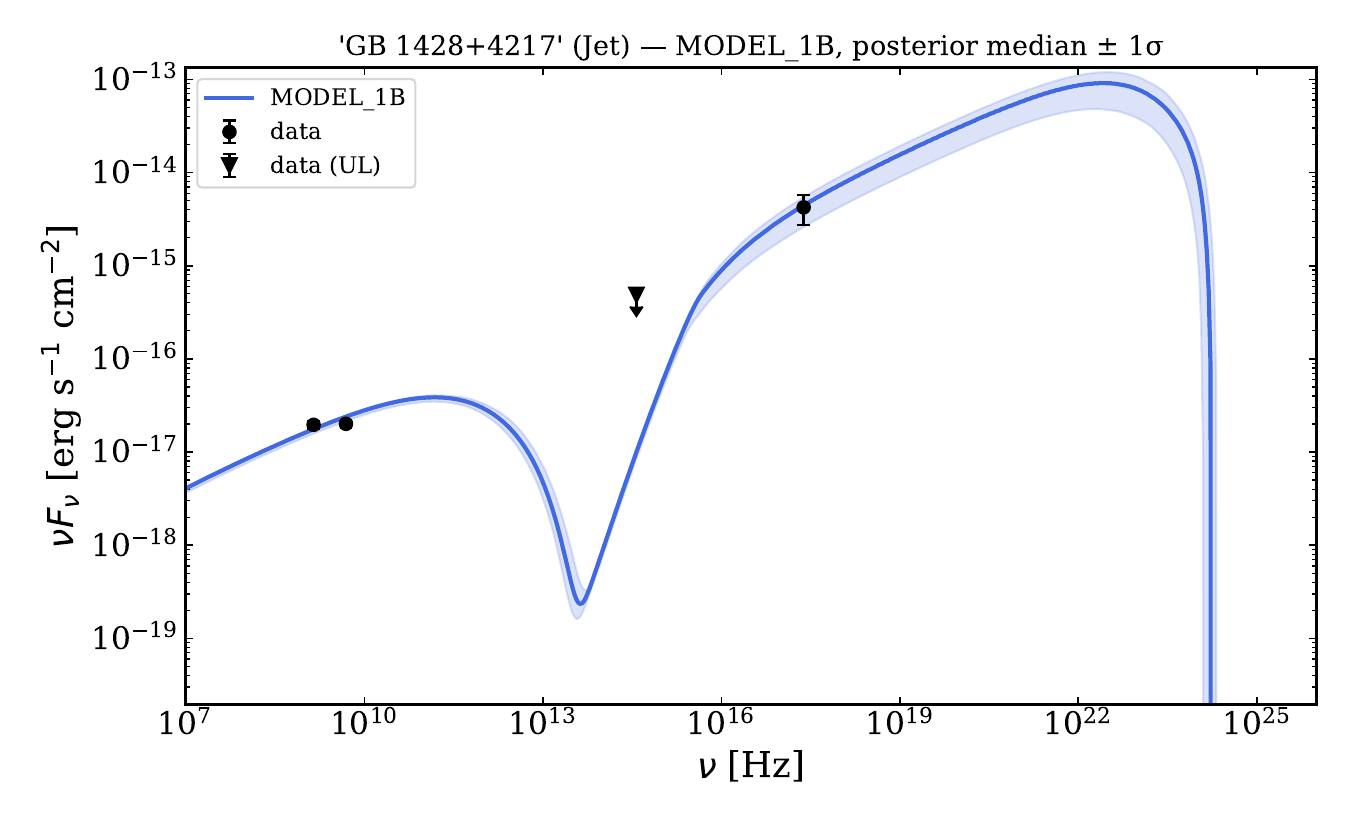}
     \centering \includegraphics[width=0.38\linewidth]{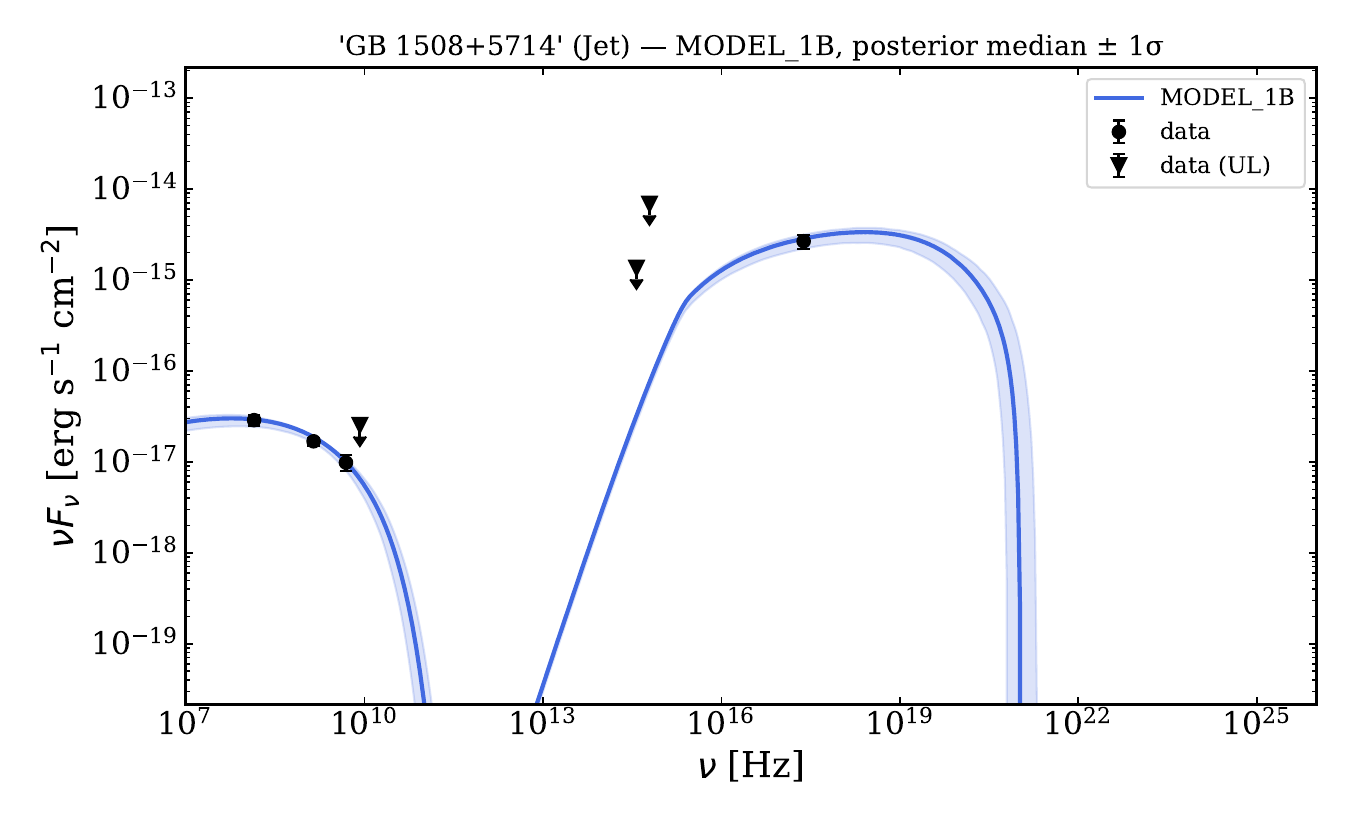}
     \centering \includegraphics[width=0.38\linewidth]{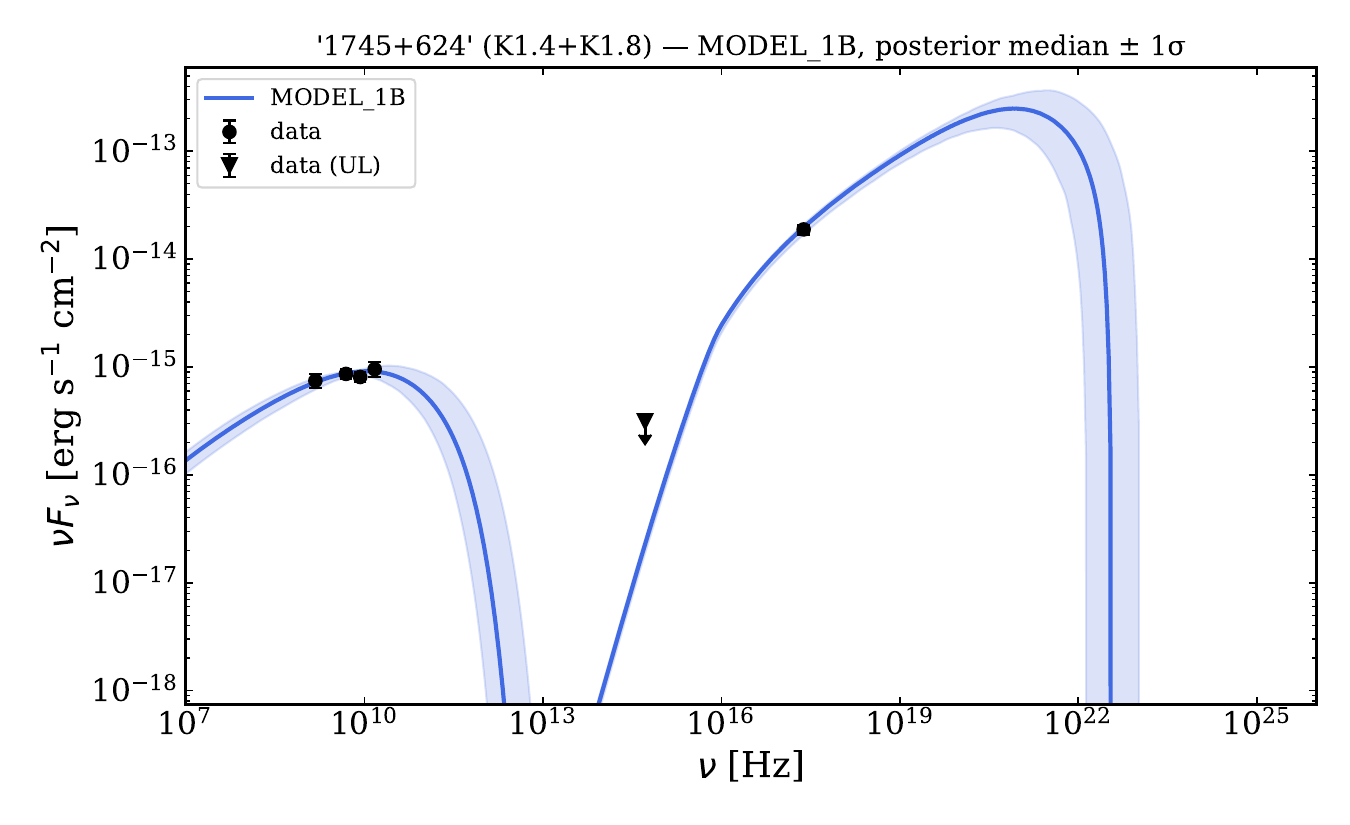}
     \centering \includegraphics[width=0.38\linewidth]{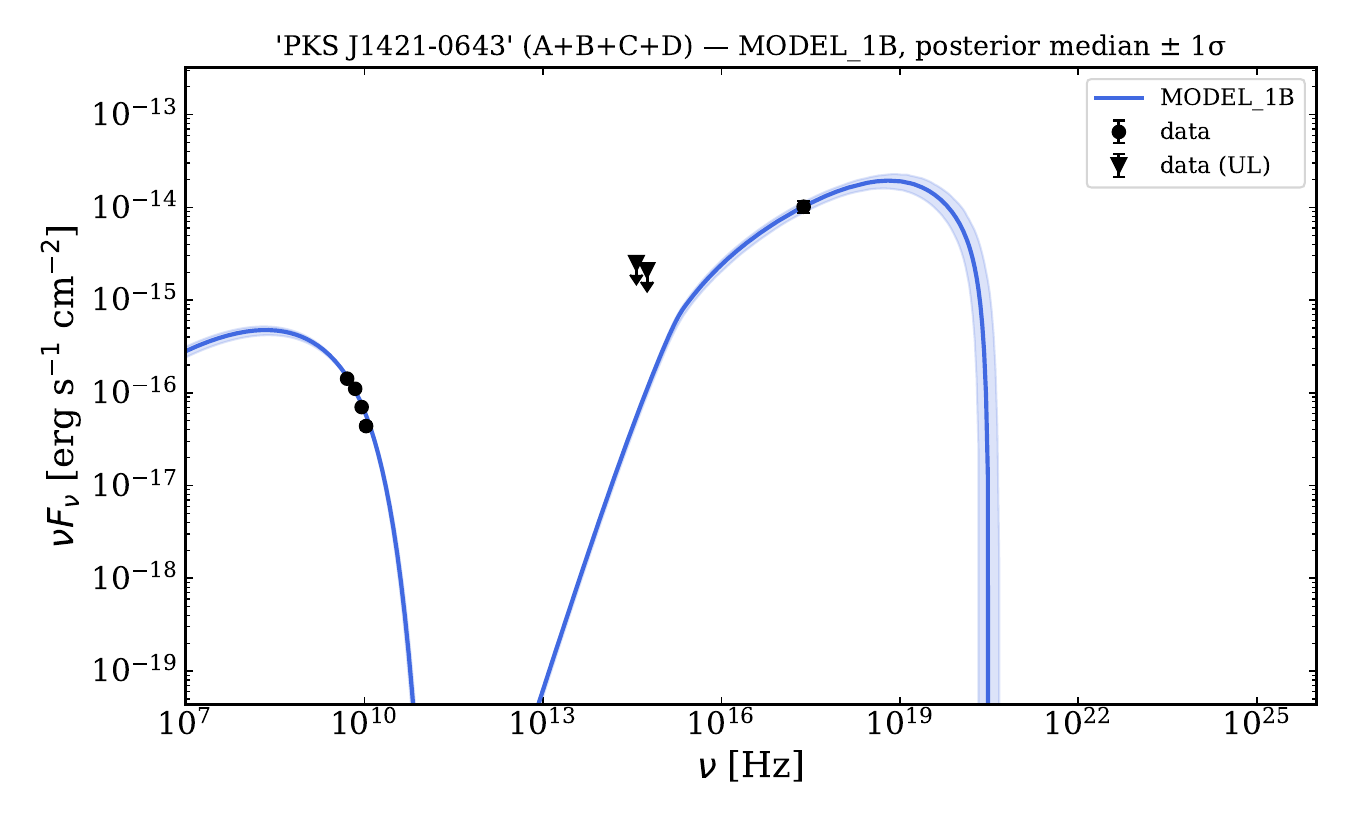}
       \centering \includegraphics[width=0.38\linewidth]{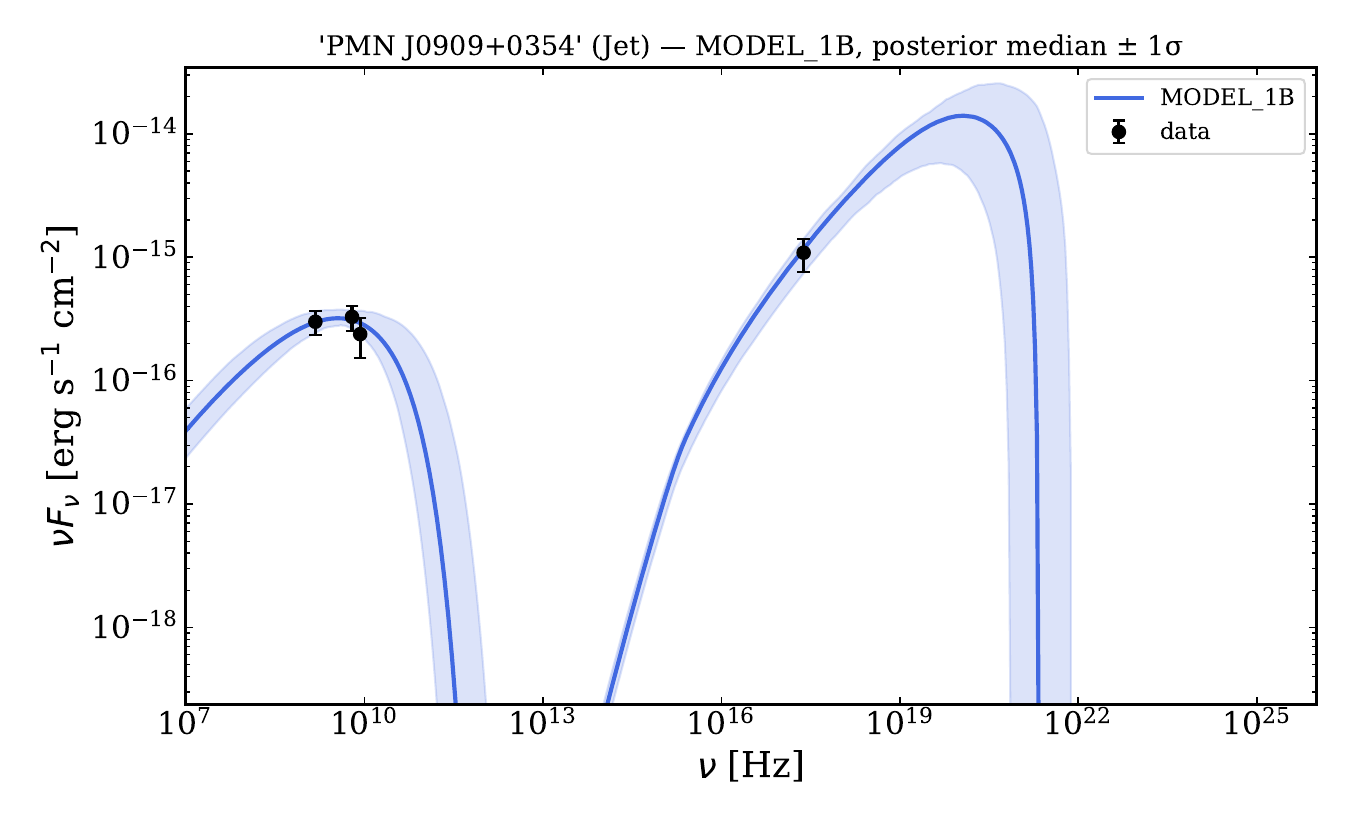}
      \centering \includegraphics[width=0.38\linewidth]{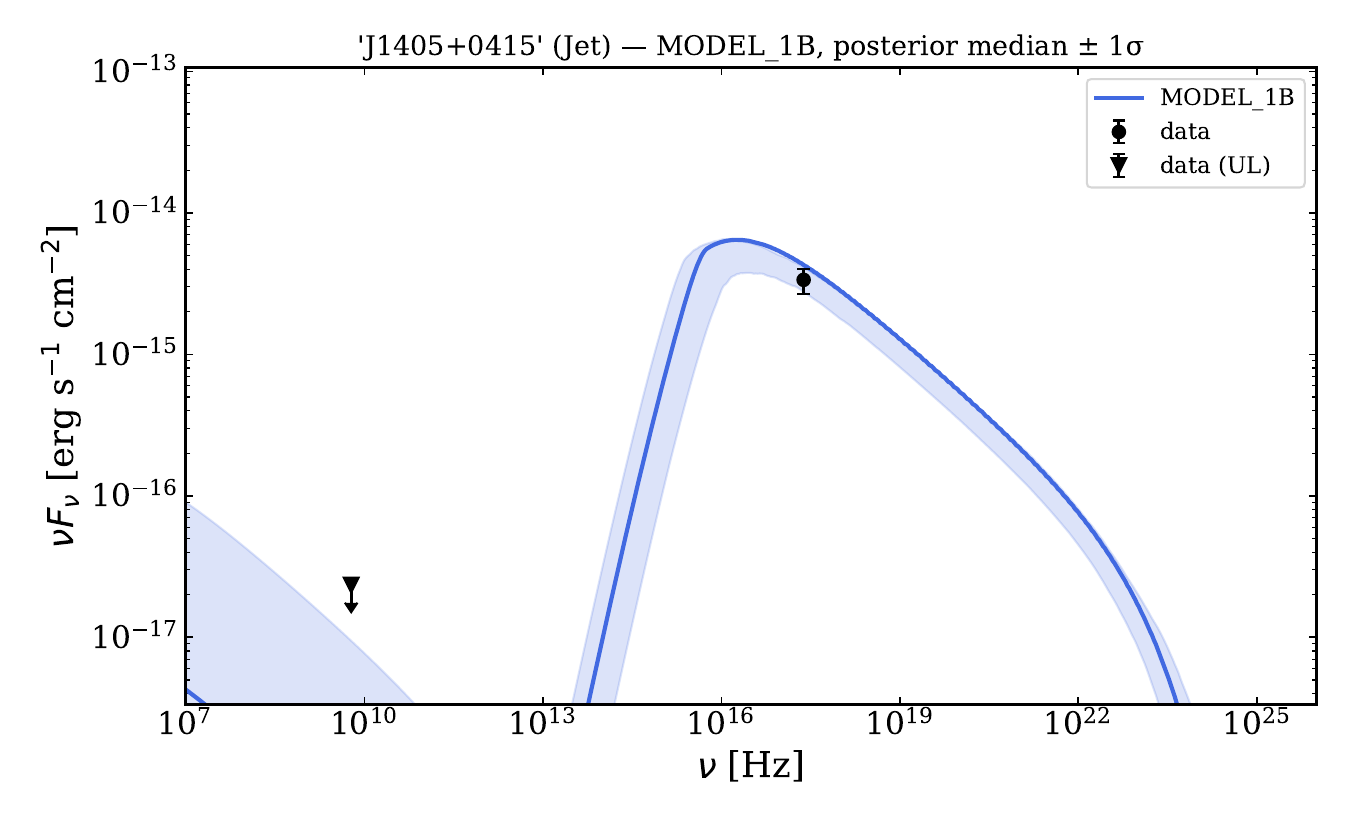}
        \centering \includegraphics[width=0.38\linewidth]{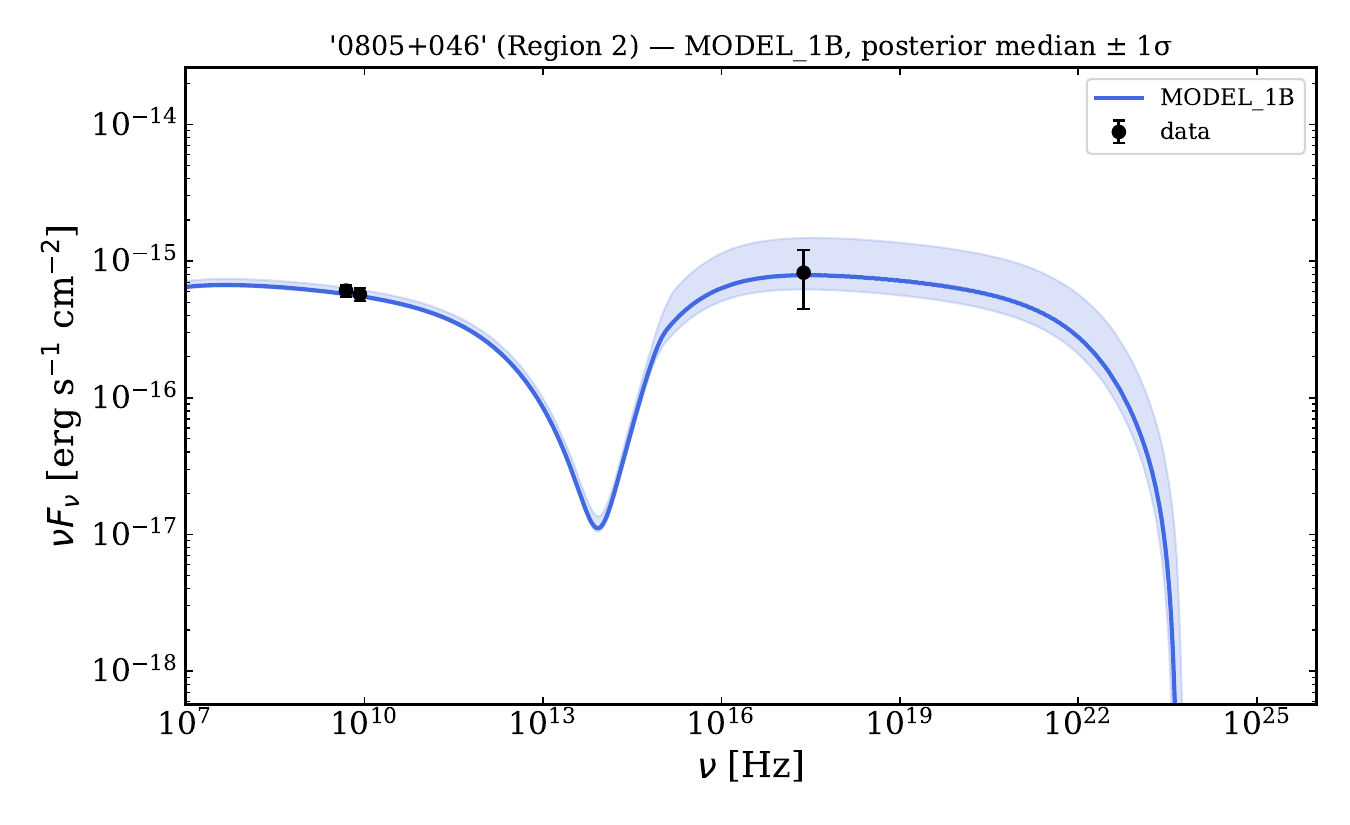}
        \centering \includegraphics[width=0.38\linewidth]{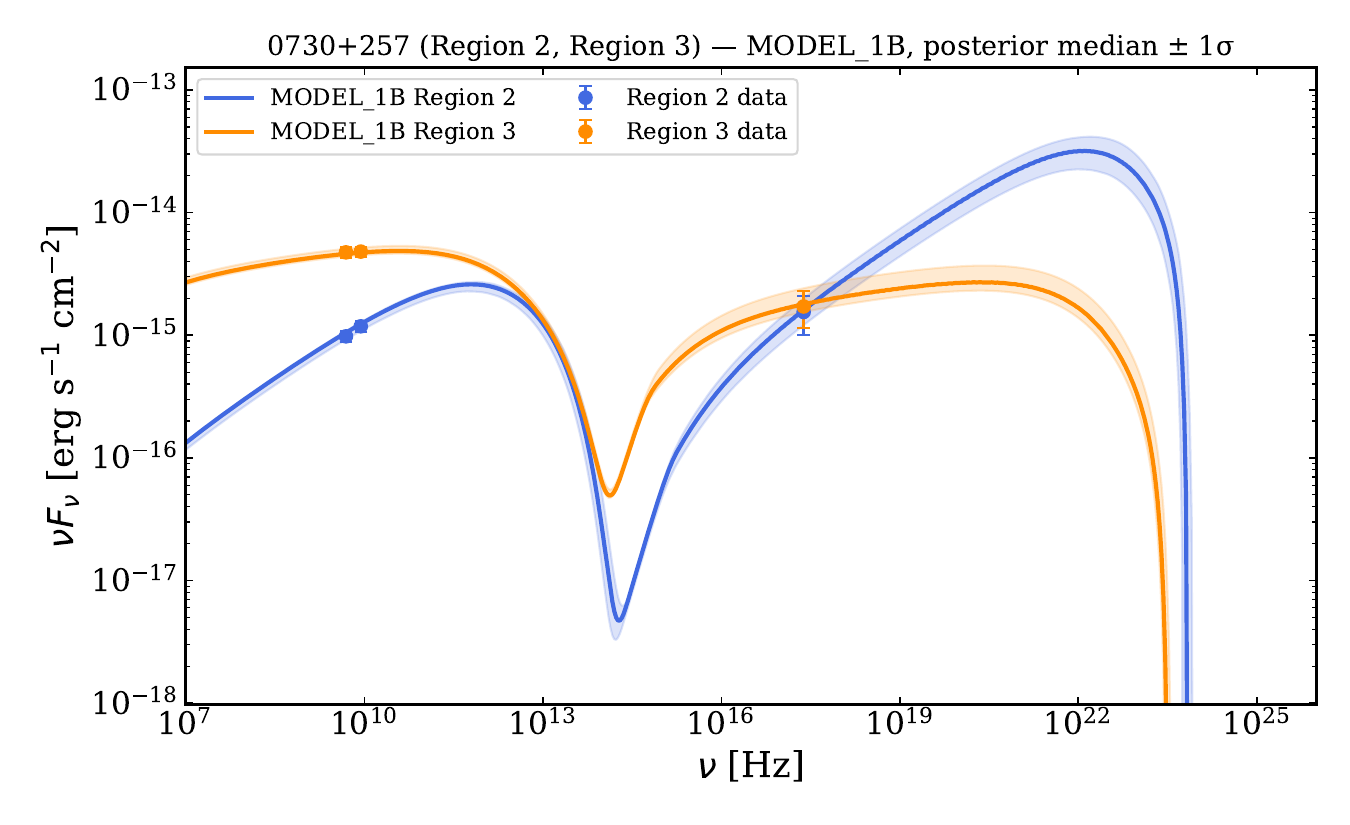}
         \centering \includegraphics[width=0.38\linewidth]{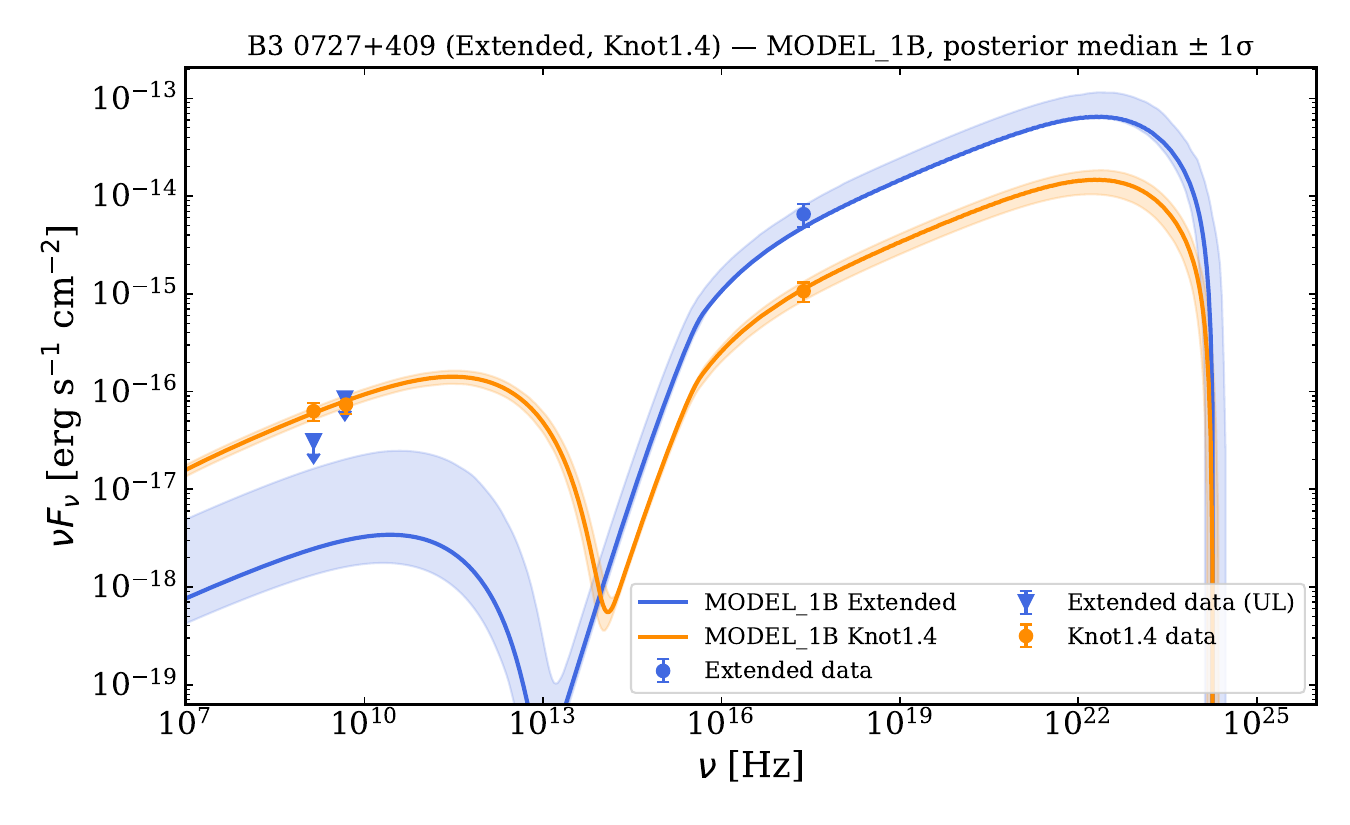}
    \caption{SEDs of the analyzed high-$z$ quasar jets fitted with the statistically preferred model (1B) and plotted with $1\sigma$ error bands. For the four sources with multi-frequency radio coverage (GB\,1508$+$5714, 1745$+$624, PKS\,J1421$-$0643, and PMN\,J0909$+$0354) the maximum electron Lorentz factor $\gmax$ is treated as a free parameter (Table~\ref{tab:results}); for the remaining sources it is held at the fiducial value of $\gmax=10^5$.}
    \label{fig:b3fit}
\end{figure}

\subsection{Results of the Model Fitting} \label{sec:results}

In general, as discussed below in more detail, the Bayesian evidence consistently favors models in which the spatial distribution of radiating electrons is linked to the thermal plasma properties rather than to the magnetic-energy density, with the possible exception of one source. By contrast, the data provide only weak discrimination between alternative parametrizations of the jet magnetic-field structure. Within the IC/CMB framework coupled to the adopted MHD description of a current-carrying jet in magneto-hydrostatic equilibrium, these trends emerge systematically across the source sample.

Before turning to the global trends, we address the maximum electron Lorentz factor $\gmax$. For the six sources whose radio coverage is sparse, $\gmax$ is held at the fiducial value of $10^5$. For the four sources with multi-frequency radio data (GB\,1508$+$5714, 1745$+$624, PKS\,J1421$-$0643, and PMN\,J0909$+$0354), whose synchrotron continua show a steep high-frequency decline that is incompatible with $\gmax=10^5$, we instead treated $\gmax$ as a free parameter rather than fixing it by hand, obtaining the best-fit values $\gmax \lesssim 10^4$ (Table~\ref{tab:results}).

The detailed quantitative outcome of the model comparison for all sources is summarized in Table~\ref{tab:jeffreys}, which lists the log-evidence differences $\Delta\ln\mathcal{Z}$ for each model variant relative to the best-fitting model for a given source. Model~1B attains the highest evidence for every source, the only nominal exception being the smallest-radius ($R_{\rm j}=0\farcs2$) refit of PSO\,J030947.49$+$271757.31, where models 1B and 2B are statistically tied (2B preferred by only $\Delta\ln\mathcal{Z}=0.04$). At its nominal radius $R_{\rm j}=2\farcs0$ the same source has all variants but 1A mutually indistinguishable ($|\Delta\ln\mathcal{Z}|<0.25$), with 1B formally best; such tiny margins lie far below the threshold of statistical significance \citep[following the prescription of][]{Trotta01032008}.

More tellingly, model~2B is statistically indistinguishable from 1B across the \emph{entire} sample, with $|\Delta\ln\mathcal{Z}|\leq0.80$ throughout, so that the two family-B variants, which differ only in the toroidal-field profile $b(x)$, cannot be separated by the present data. The preference for the gas-pressure normalization (family~B) over the magnetic-energy-density normalization (family~A) is, by contrast, systematic. The evidence against family~A ranges from weak ($\Delta\ln\mathcal{Z}\simeq-1.7$ for 0805$+$046) through moderate for the bulk of the sample to decisive for the best-constrained jets ($\Delta\ln\mathcal{Z}\simeq-5.3$ for B3\,0727$+$409 and $\simeq-45$ for 1745$+$624, in both cases relative to 1B). This A-versus-B discrimination is driven predominantly by model~1A: model~2A is consistently favored over 1A, typically by $\Delta\ln\mathcal{Z}\sim1$--$2$ and far more strongly for 1745$+$624, although it remains disfavored relative to family~B. The most decisive discrimination is obtained for the sources with multi-frequency radio coverage (1745$+$624, B3\,0727$+$409), whereas for the X-ray-only, highest-redshift source PSO\,J030947.49$+$271757.31 the four variants remain mutually indistinguishable at the nominal radius; the reduced-radius refits of the same source ($R_{\rm j}=0\farcs5$ and $0\farcs2$ in Table~\ref{tab:jeffreys}) do sharpen this into a clear family-B preference.

\begin{deluxetable}{lllllllll}[h]
  \tabletypesize{\scriptsize}
  \tablewidth{0pt}
  \tablecaption{%
      Best-fit values of the model free parameters $(\vartheta, \, \Gamma_0, \, \log L_{\rm j}, \, \gamma_{\rm max})$, and the derived parameters $(\sigma, \, B(R_{\rm j}))$.%
      \label{tab:results}%
  }
  \tablehead{
      \colhead{Source}          &
      \colhead{Feature}         &
      \colhead{Model}           &
      \colhead{$\vartheta$}     &
      \colhead{$\Gamma_0$}      &
      \colhead{$\log L_{\rm j}$} &
      \colhead{$\gamma_{\rm max}$} &
      \colhead{$\sigma$}        &
      \colhead{$B(R_{\rm j})$}  \\
      \colhead{}                &
      \colhead{}                &
      \colhead{}                &
      \colhead{[deg]}           &
      \colhead{}                &
      \colhead{[erg\,s$^{-1}$]} &
      \colhead{$\times 10^{3}$} &
      \colhead{$\times 10^{-2}$} &
      \colhead{[$100\,\mu$G]}
  }
  \startdata
  PSO J030947.49$+$271757.31     & Jet          & 1B
      & $19.3_{-7.0}^{+4.2}$
      & $2.5_{-0.3}^{+0.4}$
      & $47.70_{-0.16}^{+0.16}$
      & $-$
      & $2.85_{-0.55}^{+0.58}$
      & $0.67_{-0.18}^{+0.24}$ \\
                                 &              & 2B
      & $19.1_{-8.2}^{+4.3}$
      & $3.7_{-0.6}^{+0.7}$
      & $47.68_{-0.40}^{+0.31}$
      & $-$
      & $12.92_{-3.20}^{+3.96}$
      & $0.43_{-0.20}^{+0.28}$ \\
  PSO J030947.49$+$271757.31$^a$ & Jet          & 1B
      & $22.2_{-6.2}^{+1.9}$
      & $9.0_{-4.9}^{+14.7}$
      & $48.35_{-0.87}^{+1.04}$
      & $-$
      & $0.36_{-0.30}^{+0.97}$
      & $1.73_{-1.48}^{+10.06}$ \\
                                 &              & 2B
      & $21.7_{-6.2}^{+2.0}$
      & $12.0_{-5.9}^{+15.0}$
      & $48.55_{-1.02}^{+0.94}$
      & $-$
      & $1.67_{-1.31}^{+3.97}$
      & $1.66_{-1.42}^{+7.45}$ \\
  PSO J030947.49$+$271757.31$^b$  & Jet          & 1B
      & $17.8_{-4.5}^{+1.4}$
      & $13.4_{-7.7}^{+21.0}$
      & $48.27_{-0.96}^{+1.03}$
      & $-$
      & $0.17_{-0.14}^{+0.61}$
      & $2.71_{-2.35}^{+17.14}$ \\
                                 &              & 2B
      & $17.5_{-4.9}^{+1.5}$
      & $18.1_{-10.1}^{+22.3}$
      & $48.51_{-1.16}^{+0.92}$
      & $-$
      & $0.77_{-0.61}^{+2.71}$
      & $2.67_{-2.34}^{+13.88}$ \\
  GB 1428$+$4217                 & Jet          & 1B
      & $13.0_{-3.6}^{+2.0}$
      & $15.7_{-8.4}^{+21.3}$
      & $48.35_{-0.85}^{+0.95}$
      & $-$
      & $0.13_{-0.10}^{+0.39}$
      & $0.89_{-0.75}^{+4.64}$ \\
                                 &              & 2B
      & $12.9_{-3.1}^{+1.9}$
      & $20.6_{-9.3}^{+20.4}$
      & $48.54_{-0.85}^{+0.82}$
      & $-$
      & $0.60_{-0.44}^{+1.26}$
      & $0.86_{-0.69}^{+3.01}$ \\
  GB 1508$+$5714                 & Jet          & 1B
      & $16.6_{-3.8}^{+1.5}$
      & $13.2_{-7.0}^{+13.5}$
      & $48.87_{-0.84}^{+0.77}$
      & $2.94_{-0.53}^{+0.81}$
      & $0.18_{-0.13}^{+0.51}$
      & $2.31_{-1.87}^{+9.19}$ \\
                                 &              & 2B
      & $15.8_{-5.4}^{+1.6}$
      & $15.1_{-7.7}^{+13.7}$
      & $48.91_{-1.12}^{+0.75}$
      & $2.89_{-0.53}^{+0.77}$
      & $1.08_{-0.77}^{+2.94}$
      & $2.12_{-1.80}^{+7.65}$ \\
  1745$+$624                     & K1.4+K1.8    & 1B
      & $8.7_{-2.7}^{+1.1}$
      & $18.8_{-8.8}^{+15.7}$
      & $49.00_{-0.73}^{+0.67}$
      & $9.36_{-3.15}^{+6.22}$
      & $0.09_{-0.06}^{+0.20}$
      & $3.65_{-2.77}^{+10.84}$ \\
                                 &              & 2B
      & $8.4_{-2.2}^{+1.0}$
      & $25.5_{-10.1}^{+14.9}$
      & $49.20_{-0.79}^{+0.57}$
      & $9.30_{-2.96}^{+6.09}$
      & $0.40_{-0.23}^{+0.64}$
      & $3.41_{-2.53}^{+7.21}$ \\
  PKS J1421$-$0643               & A+B+C+D      & 1B
      & $18.2_{-6.1}^{+2.0}$
      & $9.5_{-4.8}^{+9.5}$
      & $48.93_{-0.81}^{+0.77}$
      & $1.80_{-0.21}^{+0.22}$
      & $0.33_{-0.23}^{+0.75}$
      & $2.30_{-1.84}^{+8.45}$ \\
                                 &              & 2B
      & $17.2_{-5.0}^{+2.3}$
      & $11.8_{-4.5}^{+8.3}$
      & $49.01_{-0.81}^{+0.70}$
      & $1.81_{-0.22}^{+0.22}$
      & $1.72_{-1.09}^{+2.40}$
      & $2.04_{-1.56}^{+5.06}$ \\
  PMN J0909$+$0354               & NNW          & 1B
      & $17.8_{-5.3}^{+3.8}$
      & $10.5_{-5.5}^{+13.1}$
      & $48.68_{-0.80}^{+0.90}$
      & $4.56_{-1.66}^{+3.83}$
      & $0.27_{-0.21}^{+0.69}$
      & $2.08_{-1.70}^{+9.47}$ \\
                                 &              & 2B
      & $17.2_{-5.1}^{+3.8}$
      & $14.8_{-6.9}^{+11.6}$
      & $48.96_{-0.95}^{+0.71}$
      & $4.56_{-1.69}^{+4.12}$
      & $1.12_{-0.75}^{+2.45}$
      & $2.06_{-1.66}^{+6.29}$ \\
  J1405$+$0415                   & Jet          & 1B
      & $11.4_{-5.1}^{+4.5}$
      & $13.3_{-6.5}^{+16.0}$
      & $48.78_{-0.88}^{+0.79}$
      & $-$
      & $0.18_{-0.14}^{+0.41}$
      & $0.99_{-0.82}^{+3.63}$ \\
                                 &              & 2B
      & $11.1_{-4.8}^{+3.8}$
      & $16.8_{-7.8}^{+16.8}$
      & $48.80_{-1.00}^{+0.80}$
      & $-$
      & $0.89_{-0.65}^{+1.96}$
      & $0.80_{-0.67}^{+2.80}$ \\
  0805$+$046                     & Region 2     & 1B
      & $20.0_{-7.0}^{+3.7}$
      & $3.1_{-0.7}^{+1.7}$
      & $48.07_{-0.25}^{+0.44}$
      & $-$
      & $2.02_{-0.99}^{+0.85}$
      & $3.41_{-1.68}^{+3.66}$ \\
                                 &              & 2B
      & $20.2_{-7.9}^{+3.7}$
      & $4.8_{-1.2}^{+2.8}$
      & $48.13_{-0.57}^{+0.68}$
      & $-$
      & $8.62_{-4.80}^{+5.56}$
      & $2.47_{-1.63}^{+4.59}$ \\
  0730$+$257                     & Region 2     & 1B
      & $21.6_{-3.6}^{+2.3}$
      & $9.9_{-4.8}^{+10.4}$
      & $48.82_{-0.73}^{+0.78}$
      & $-$
      & $0.30_{-0.22}^{+0.66}$
      & $2.15_{-1.68}^{+7.68}$ \\
                                 &              & 2B
      & $20.4_{-5.9}^{+2.9}$
      & $11.4_{-5.2}^{+10.9}$
      & $48.85_{-0.94}^{+0.79}$
      & $-$
      & $1.82_{-1.31}^{+3.66}$
      & $1.93_{-1.59}^{+6.51}$ \\
                                 & Region 3     & 1B
      & $21.6_{-3.6}^{+2.3}$
      & $2.3_{-0.2}^{+0.3}$
      & $48.16_{-0.08}^{+0.12}$
      & $-$
      & $3.25_{-0.55}^{+0.42}$
      & $2.96_{-0.58}^{+0.75}$ \\
                                 &              & 2B
      & $20.4_{-5.9}^{+2.9}$
      & $3.2_{-0.4}^{+0.6}$
      & $48.02_{-0.28}^{+0.24}$
      & $-$
      & $16.61_{-3.75}^{+4.43}$
      & $1.75_{-0.65}^{+0.88}$ \\
  B3 0727$+$409                  & Extended     & 1B
      & $12.8_{-3.9}^{+1.6}$
      & $19.8_{-11.4}^{+22.3}$
      & $48.19_{-0.79}^{+0.89}$
      & $-$
      & $0.08_{-0.06}^{+0.32}$
      & $0.13_{-0.11}^{+0.71}$ \\
                                 &              & 2B
      & $12.5_{-3.6}^{+1.6}$
      & $22.9_{-11.3}^{+21.2}$
      & $48.32_{-0.76}^{+0.75}$
      & $-$
      & $0.49_{-0.35}^{+1.26}$
      & $0.13_{-0.10}^{+0.47}$ \\
                                 & Knot1.4      & 1B
      & $12.8_{-3.9}^{+1.6}$
      & $13.2_{-6.6}^{+16.8}$
      & $48.64_{-0.77}^{+0.89}$
      & $-$
      & $0.18_{-0.14}^{+0.43}$
      & $1.48_{-1.20}^{+6.39}$ \\
                                 &              & 2B
      & $12.5_{-3.6}^{+1.6}$
      & $17.4_{-7.4}^{+15.7}$
      & $48.80_{-0.86}^{+0.80}$
      & $-$
      & $0.83_{-0.59}^{+1.48}$
      & $1.36_{-1.09}^{+4.40}$ \\
  \enddata
  \tablecomments{%
      Values quoted as $x_{-\sigma_{\rm lo}}^{+\sigma_{\rm hi}}$ where
      $\sigma_{\rm lo}$ and $\sigma_{\rm hi}$ are the 16th--84th percentile
      half-widths of the posterior. Superscripts $^{a,b}$ indicate the PSO J030947.49$+$271757.31 models with $R_{\rm j}=0\farcs5$ and $0\farcs2$, respectively, instead of the nominal value of $2\farcs0$.
  }
\end{deluxetable}

Given the above, a summary of the best-fit parameters $(\vartheta,\,\Gamma_0,\,\log L_{\rm j})$, together with the derived quantities $(\sigma,\,B(R_{\rm j}),\,P_{\rm ext})$, is provided in Table~\ref{tab:results} for models 1B and 2B for all sources. For the four sources with multi-frequency radio coverage the fitted maximum electron Lorentz factor $\gmax$ is reported as an additional column; for the remaining sources $\gmax$ is fixed at the fiducial value of $10^5$. The corresponding SEDs of the analyzed jets fitted with the statistically preferred models are presented in Figure~\ref{fig:b3fit}. The resulting radial profiles $\Gamma(x)$, $B'(x)$, $p(x)$, and $\mathcal{D}(x)$ as functions of the normalized jet radius, $x\equiv r/R_{\rm j}$, are presented for each source in Appendix~\ref{sec:AppA}. A comparison of the best-fit model parameters for families A and B is provided in Appendix~\ref{sec:AppB} in the form of corner plots.

One issue that should be noted is that the modeled jets, while typically resolved along their longitudinal direction, are often unresolved or only marginally resolved in the transverse direction. As a result, the jet radius $R_{\rm j}$ used as an input parameter is frequently set by the effective \textit{Chandra} angular resolution, $\sim 0\farcs5$, and should therefore be regarded as an upper limit. 

To assess the impact of this limitation, we analyze in more detail the case of PSO\,J030947.49$+$271757.31, for which the nominal radius $R_{\rm j}=2\farcs0$ corresponds to half of the projected extent (perpendicular to the jet position angle) of the source extraction region applied by \citet{Ighina_2022} to the {\it Chandra} maps. We re-fitted the source with the radius reduced to $0\farcs5$ and $0\farcs2$ (superscripts $a$ and $b$ in Tables~\ref{tab:jeffreys} and~\ref{tab:results}, respectively). With these changes, we find that the best-fit value of the jet viewing angle $\vartheta \approx 20\degr$ is essentially unaffected ($\vartheta = 19\fdg3,\ 22\fdg2,\ 17\fdg8$ for $R_{\rm j} = 2\farcs0,\ 0\farcs5,\ 0\farcs2$, respectively). The three values agree well within their $1\sigma$ credible intervals, confirming that $\vartheta$ is largely insensitive to the assumed transverse size. The remaining parameters respond systematically. The posterior medians of both $\Gamma_0$ and $\log L_{\rm j}$ increase as $R_{\rm j}$ is reduced --- from $\Gamma_0\simeq 2.5$, $\log L_{\rm j}/{\rm erg\,s^{-1}}\simeq 47.7$ at $2\farcs0$ to $\Gamma_0\simeq 9-14$, $\log L_{\rm j}/{\rm erg\,s^{-1}}\simeq 48.3 - 48.4$ at $0\farcs5 - 0\farcs2$ --- while their credible intervals widen by roughly an order of magnitude. The increase in $\log L_{\rm j}$ remains formally consistent with the fiducial-radius result once these enlarged uncertainties are taken into account, whereas the shift in $\Gamma_0$ exceeds the (tight) fiducial-radius interval and therefore reflects a genuine trend rather than a sampling artifact; the inferred values also move toward the range $\Gamma_0\sim 10$, which is more typical of quasar jets. This is expected in the sense that a smaller emitting volume ($\propto R_{\rm j}^2\,\ell$) must be compensated by stronger Doppler boosting and/or a larger total energy budget to reproduce the same observed IC/CMB flux.

\section{Discussion} \label{sec:discussion}

The IC/CMB framework, embedded in the simple MHD model of a current-carrying jet in magneto-hydrostatic equilibrium, reproduces the broad-band SEDs of the high-redshift quasar jet sample surprisingly well. We emphasize that, for any individual source, the model effectively has only three free parameters: the jet inclination $\vartheta$, the total jet power $L_{\rm j}$, and the central bulk Lorentz factor $\Gamma_0$. Moreover, six of the ten quasar jets at $z\geq2.5$ are fitted with a single common set of fiducial parameters ($\beta_{\rm pl}$, $\eta_{\rm e}$, $\gamma_{\rm min}$, $\gamma_{\rm max}$, $a$, $k_1$, $k_2$, $k_0$), supplemented only by source-specific input quantities ($z$, $R_{\rm j}$, $\ell$, $p$) derived directly from the data. For the four sources with the best-sampled radio continua we additionally fit $\gmax$, which is driven to $\lesssim 10^4$ rather than the fiducial value of $10^5$, indicating that their radiating electrons reach systematically lower maximum energies; even so, the gas-pressure (family-B) preference is unchanged. Given the diversity of the sample, this economy of parameters is itself noteworthy.

Within this framework the data robustly favor a radial distribution of radiating electrons that follows the gas pressure, $U_{\rm e}'\propto P(r)$ (family~B), rather than the rest-frame magnetic energy density, $U_{\rm e}'\propto B_\phi'^2(r)$ (family~A), while remaining relatively insensitive to the parametrization of the magnetic-field profile $B_\phi(r)$ (models~1 versus~2). Physically, $U_{\rm e}'\propto P(r)$ implies that the non-thermal electron pressure is a roughly constant fraction of the thermal pressure across the jet cross-section, as expected if the radiating particles remain tied to the thermal pool through shock acceleration processes. 

A direct consequence of this preference is that the synchrotron and IC/CMB emissivities have different radial distributions. For family~B the frequency-integrated synchrotron emissivity scales as $\propto K_{\rm e}'(x)\,{B'}^2(x)\propto P(x)\,U_B'(x)$ (with the monochromatic emissivity at fixed frequency scaling as $\propto P(x)\,{B'}^{(p+1)/2}(x)$), whereas the IC/CMB emissivity scales as $\propto K_{\rm e}'(x)\propto P(x)$. Because magneto-hydrostatic equilibrium imposes distinct radial profiles for the gas pressure and the magnetic field, the two components are subject to different effective Doppler boosting $\mathcal{D}(x)$ for a given inclination $\vartheta$ and shear profile $\Gamma(x)$. The resulting synchrotron-to-IC flux and maximum energy ratios can therefore depart substantially from the value predicted by homogeneous one-zone models.

\begin{figure}
    \centering
    \includegraphics[width=0.49\textwidth]{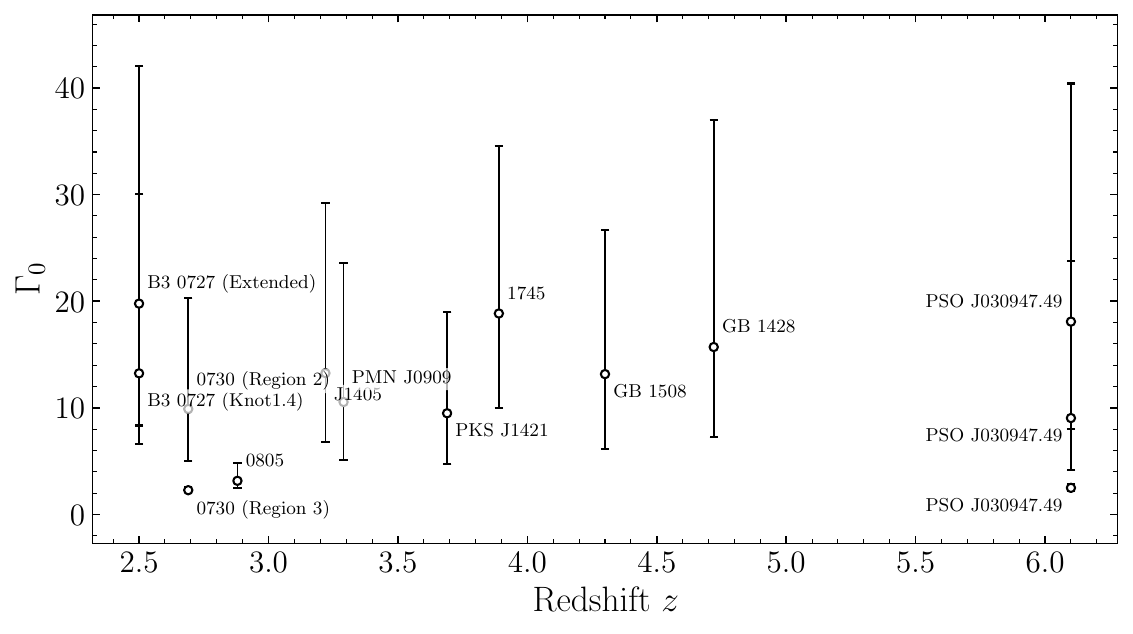}
    \includegraphics[width=0.49\textwidth]{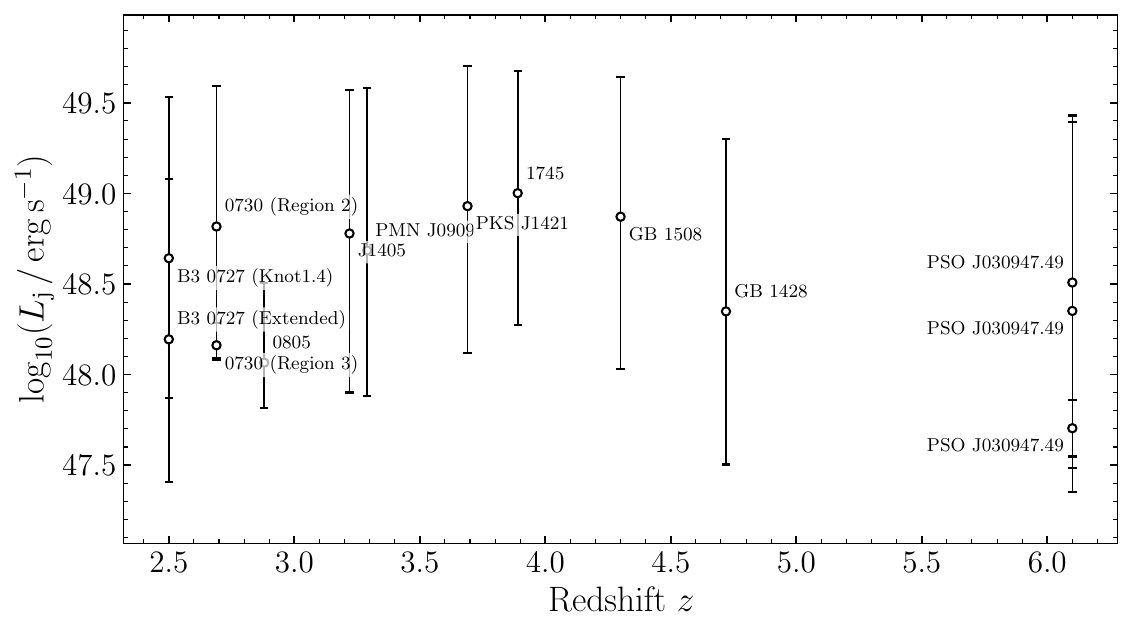}
    \includegraphics[width=0.49\textwidth]{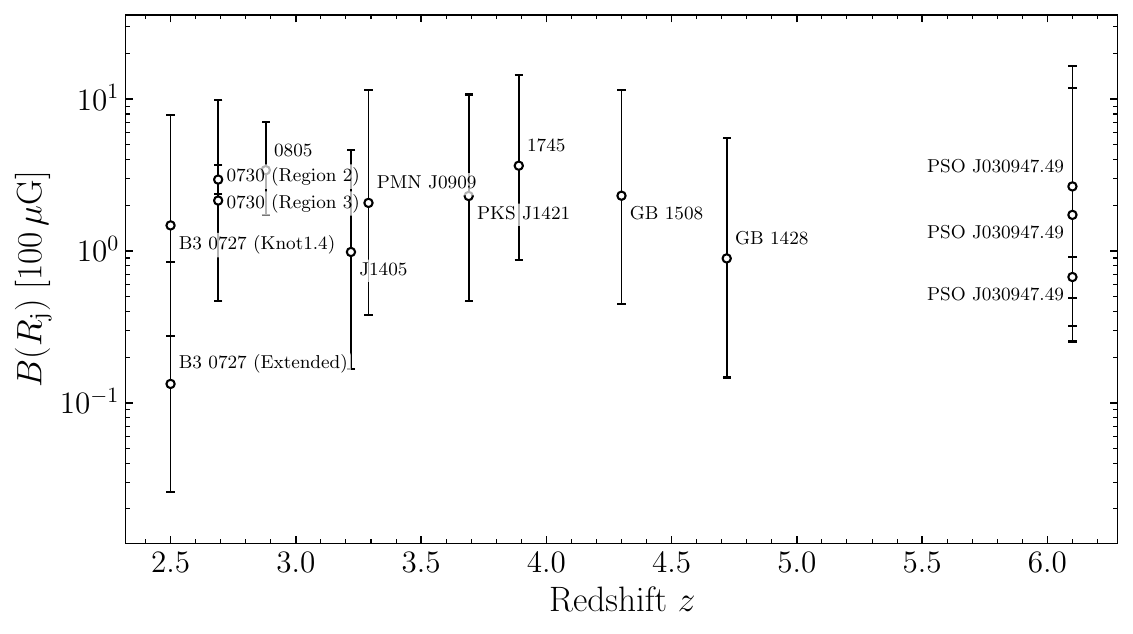}
    \includegraphics[width=0.49\textwidth]{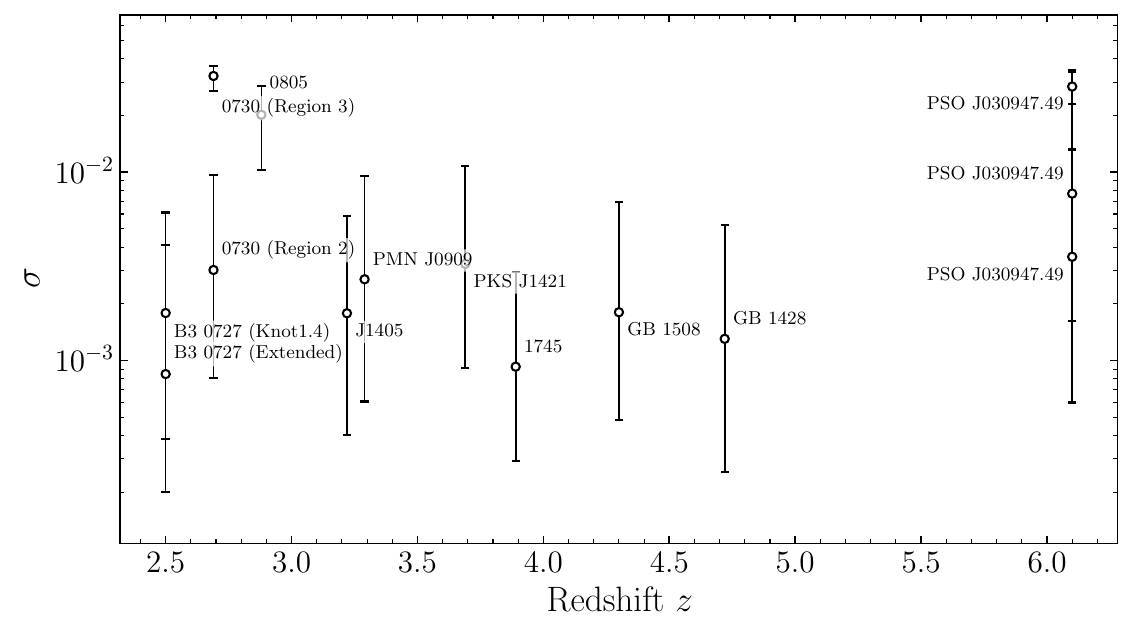}
    \caption{The inferred $\Gamma_0$, $L_{\rm j}$, $B(R_{\rm j})$, and $\sigma$ as functions of redshift for the ten fitted sources. Values for the best-fit model are presented.}
    \label{fig:redshifts}
\end{figure}

Figure~\ref{fig:redshifts} presents the inferred $\Gamma_0$, $L_{\rm j}$, $B(R_{\rm j})$, and $\sigma$ as functions of redshift for the ten fitted sources. The main conclusion is that none of these quantities exhibits a systematic monotonic trend with redshift. This is encouraging, as otherwise one might suspect an artificial dependence of the inferred model parameters on distance, driven by the increase of the CMB energy density with redshift. Although the uncertainties are relatively large, two clear outliers can be identified in terms of their $\Gamma_0$ and $\sigma$ values: 0805$+$046 and Region~3 in 0730$+$257. In particular, these sources are characterized by the smallest inferred central bulk Lorentz factors, $\Gamma_0<5$, and at the same time by the largest magnetization parameters, $\sigma>0.01$. Interestingly, Region 3 in 0730$+$257 could plausibly be interpreted to contain a recessed hotspot as seen in some lobe-dominated quasars \citep[see, e.g.,][]{1994AJ....108..766B}, consistent with the low value of $\Gamma_0$ returned by our analysis. The case of 0805$+$046 is more problematic. However, as noted in the previous section, this jet constitutes the lowest-significance {\it Chandra} detection among all the sources analyzed here. Still, the fact that our modeling systematically favors the largest jet inclination angles, of order $20\degr$, and only moderate bulk Lorentz factors for 0805$+$046 and 0730$+$257, is consistent with their classification as lobe-dominated quasars.

A robust feature of the fits is that the inferred total jet powers are systematically larger than those obtained in homogeneous one-zone models, reaching $\log(L_{\rm j}/{\rm erg\,s^{-1}})\lesssim 49$ in the most powerful sources. Under the order-of-magnitude scaling $M_{\rm BH}\sim (\eta_{\rm acc}/0.1)\,(L_{\rm j}/10^{49}\,{\rm erg\,s^{-1}})\,10^{10}\,M_\odot$, which follows from assuming both a maximal jet-production efficiency, $L_{\rm j}\sim \dot{M}_{\rm acc}c^2$ for the mass accretion rate $\dot{M}_{\rm acc}$, and Eddington-limited accretion, $\eta_{\rm acc}\dot{M}_{\rm acc}c^2\sim L_{\rm Edd}$ for the radiative efficiency of the accretion disk $\eta_{\rm acc}$, such powers would imply central black-hole masses of $M_{\rm BH}\lesssim 10^{10}\,M_\odot$, close to the upper envelope of the SMBH mass function at these epochs \citep{2026ApJ...998...48P}. We stress, however, that this is only an order-of-magnitude consistency argument rather than a measurement, and that it depends sensitively on $\eta_{\rm e}$ (the normalization of the radiating electron population): increasing $\eta_{\rm e}$ from $0.1$ to unity lowers the inferred jet powers, and hence the implied black-hole masses, by about an order of magnitude, to $M_{\rm BH}\lesssim 10^{9}\,M_\odot$.

To test this implication, we compiled independent black-hole mass estimates for the entire sample, as summarized in Table~\ref{tab:mbh}. For five sources, these estimates are based on accretion disk modeling or dedicated rest-frame-UV broad-line virial analyses that are independent of the jet energetics: PSO\,J030947 \citep{Belladitta2022}, GB\,1428$+$4217 \citep{Ghisellini2009}, GB\,1508$+$5714 \citep[accreting at only $\lambda_{\rm Edd}\simeq0.02$;][]{Alzati2025}, 1745$+$624 \citep{Cheung2006}, and B3\,0727$+$409 \citep[with consistent C\,{\sc iv} and Mg\,{\sc ii} estimates;][]{Simionescu_2016}. For four additional sources, we adopt single-epoch C\,{\sc iv} virial masses from the SDSS\,DR16Q spectral-property catalogue \citep{Rakshit2020}: PMN\,J0909$+$0354, 0805$+$046, 0730$+$257, and J1405$+$0415. Finally, for PKS\,J1421$-$0643, the quasar bolometric luminosity ($L_{\rm bol}\simeq4\times10^{47}\,{\rm erg\,s^{-1}}$; \citealt{Worrall_2020}) implies an Eddington-limit lower bound of $M_{\rm BH}\gtrsim3\times10^{9}\,M_\odot$. The independently estimated black-hole masses therefore all fall within $\sim0.5$\,dex of $10^{9}\,M_\odot$. We caution, however, that the single-epoch C\,{\sc iv} masses carry an additional systematic uncertainty of $\sim0.3$--$0.5$\,dex beyond their formal statistical errors, owing to the wind-contaminated and partly non-virial character of the C\,{\sc iv} line. More generally, single-epoch virial masses tend to be systematically overestimated for the most luminous, highly accreting quasars at high redshift \citep{2022MNRAS.515..491M}. Nevertheless, we conclude that the independently estimated black-hole masses are broadly consistent with those implied by the inferred jet powers, provided that the normalization of the radiating-electron population is as high as $\eta_{\rm e}\sim1$, rather than the fiducial value $\eta_{\rm e}\sim0.1$, even without the need to invoke super-Eddington accretion rates. 

If, on the other hand, the large jet powers returned by our modeling do genuinely require exceptionally massive central black holes, they may contribute to the apparent scarcity of {\it Chandra}-detected jets among high-$z$ quasars: the currently detected systems could represent only the high-mass tail of the underlying SMBH population at these redshifts.

\begin{deluxetable}{lcccl}
  \tabletypesize{\scriptsize}
  \tablewidth{0pt}
  \tablecaption{Independent central black-hole masses for the sample.\label{tab:mbh}}
  \tablehead{
      \colhead{Source} &
      \colhead{$z$} &
      \colhead{$\log\!\left(M_{\rm BH}/M_\odot\right)$} &
      \colhead{Method} &
      \colhead{Reference}
  }
  \startdata
  PSO J030947.49$+$271757.31 & $6.10$ & $9.16_{-0.38}^{+0.36}$ & C\,{\sc iv} virial              & \citet{Belladitta2022} \\
  GB 1428$+$4217             & $4.72$ & $9.2$           & accretion disk SED            & \citet{Ghisellini2009} \\
  GB 1508$+$5714             & $4.30$ & $8.65\pm0.19$          & accretion disk modeling & \citet{Alzati2025} \\
  1745$+$624                 & $3.89$ & $\gtrsim9.0$           & C\,{\sc iv} virial              & \citet{Cheung2006} \\
  PKS J1421$-$0643           & $3.69$ & $\gtrsim9.5$           & Eddington floor & \citet{Worrall_2020} \\
  PMN J0909$+$0354           & $3.29$ & $9.55\pm0.05$          & C\,{\sc iv} (SDSS)             & \citet{Rakshit2020} \\
  J1405$+$0415               & $3.22$ & $8.86\pm0.04$          & C\,{\sc iv} (SDSS)             & \citet{Rakshit2020} \\
  0805$+$046                 & $2.88$ & $9.25\pm0.04$          & C\,{\sc iv} (SDSS)             & \citet{Rakshit2020} \\
  0730$+$257                 & $2.69$ & $8.86\pm0.03$          & C\,{\sc iv} (SDSS)             & \citet{Rakshit2020} \\
  B3 0727$+$409              & $2.50$ & $8.4$           & C\,{\sc iv}, Mg\,{\sc ii} virial & \citet{Simionescu_2016} \\
\enddata
  \tablecomments{``C\,{\sc iv} (SDSS)'' denotes single-epoch
      virial masses from the SDSS\,DR16Q spectral-property catalogue, \citet{Rakshit2020}, with
      \citet{2006ApJ...641..689V} calibration; the errors are statistical only, with a further
      $\sim0.3$--$0.5$\,dex systematic intrinsic to the wind-affected C\,{\sc iv} line. For sources
      with a dedicated spectral analysis we adopt the published values instead: the two $z>4$ quasars
      from accretion disk modeling (since at those redshifts C\,{\sc iv} falls at the unreliable red
      edge of the SDSS range), and B3 0727$+$409 from its discovery paper
      (\citealt{Simionescu_2016}, consistent C\,{\sc iv} and Mg\,{\sc ii} estimates,
      $2.3$ and $3.3\times10^{8}\,M_\odot$). For PKS\,J1421$-$0643, we assumed Eddington floor given by the quasar bolometric luminosity ($L_{\rm bol}\simeq4\times10^{47}\,{\rm erg\,s^{-1}}$; \citealt{Worrall_2020}).}
 \end{deluxetable}

The inferred jet magnetization parameters are, by contrast, small: $\sigma\ll1$ throughout family~1B, with family~2B reaching somewhat higher values ($\sigma\sim0.1$ for the lowest-$\Gamma_0$ sources) but still remaining below unity. This appears to be a natural consequence of the IC/CMB framework: a large power carried by radiating electrons is required to reproduce the observed X-ray emission through inverse-Compton scattering of the CMB, while the same single electron population must simultaneously account for the observed synchrotron radio emission. This feature cannot easily be removed by an \emph{ad hoc} modification of the magnetic-field structure. Introducing an additional uniform vertical component $B_z(r)=\mathrm{const}$, for instance, leaves the magneto-hydrostatic equilibrium condition unchanged and, because a uniform axial field does not contribute to the Poynting flux along the jet axis, also does not alter $\sigma$ as defined here \citep[see][]{2022ApJ...929..181K}; it would instead primarily enhance the synchrotron radio output, making the elevated X-ray fluxes harder to explain within a single universal electron distribution. Larger $\sigma$ could be achieved only by allowing a radial stratification of the electron energy spectrum, which would weaken the original motivation for the IC/CMB interpretation relative to the competing synchrotron scenario.

Finally, the inferred total pressure at the jet boundary is relatively high, ranging from $ \gtrsim 10^{-11}\,\mathrm{erg\,cm^{-3}}$ for the extended jet in B3\,0727$+$409 up to $\lesssim 10^{-8}\,\mathrm{erg\,cm^{-3}}$ for 1745$+$624 and 0805$+$046. This places direct constraints on the density and pressure of the ambient medium surrounding these high-$z$ jets within the adopted equilibrium MHD framework. However, we note the uncertainties in the $P_{\rm ext}$ determination, not reported in Table~\ref{tab:results}, are large, with the upper bound in many cases basically unconstrained.

\section{Summary and conclusions} \label{sec:conclusions}

In this work we have developed a self-consistent framework for modeling the broad-band, radio-to-X-ray emission of kiloparsec-scale quasar jets, in which the non-thermal continuum is produced by synchrotron radiation and inverse-Compton scattering of the CMB by a single population of relativistic electrons. Rather than approximating the emitting region as a homogeneous one-zone sphere, we described the outflow as a current-carrying, axially symmetric jet with a purely toroidal magnetic field in magneto-hydrostatic equilibrium. The equilibrium condition links the radial profiles of pressure, magnetic field, and bulk velocity, so that the radial stratification of the source is captured without introducing additional free parameters: for any individual source the model retains only three free parameters: the jet inclination $\vartheta$, the total jet power $L_{\rm j}$, and the on-axis bulk Lorentz factor $\Gamma_0$, supplemented by source-specific input quantities ($z$, $R_{\rm j}$, $\ell$, $p$) taken directly from the data. We considered two prescriptions for the radial distribution of the radiating electrons (normalization to the gas pressure, family B, or to the rest-frame magnetic energy density, family A) and two toroidal-field profiles (models 1 and 2), and computed all spectra in \textrm{JAX}. The model was applied to a sample of ten quasar jets at $z \geq 2.5$ resolved in X-rays by \textit{Chandra}, with Bayesian parameter inference and model comparison performed via nested sampling using \texttt{UltraNest}.

The principal results of the analysis are as follows. First, the Bayesian evidence systematically favors electron distributions that follow the gas pressure (family B) over those tracing the magnetic energy density (family A); the preference ranges from weak for the least-constrained jet to decisive for the best-constrained sources (1745$+$624 and B3 0727$+$409), and is driven predominantly by the poor performance of model 1A. For the four sources with multi-frequency radio coverage we additionally constrained the maximum electron Lorentz factor (rather than assuming its fiducial value), without altering the family-B preference. Second, the data discriminate only weakly between the two toroidal field profiles: the family-B variants 1B and 2B are statistically indistinguishable across the entire sample, so that the present observations constrain the normalization of the radiating electrons far better than the detailed shape of the magnetic-field profile. Third, the inferred total jet powers, reaching $\log(L_{\rm j}/{\rm erg\,s^{-1}}) \approx 49$, are systematically larger than those obtained from homogeneous one-zone models, since in the stratified outflow only particular regions contribute efficiently to the observed IC/CMB emission; under simple Eddington-rate scaling these powers imply central black-hole masses that approach $10^{10}\,M_\odot$, although this estimate is highly sensitive to the adopted electron-normalization efficiency $\eta_{\rm e}$. Fourth, the inferred magnetizations are low throughout ($\sigma \ll 1$), a natural consequence of the large electron power required to reproduce the X-ray flux through IC/CMB while the same population accounts for the radio synchrotron emission. Finally, none of the derived quantities ($\Gamma_0$, $L_{\rm j}$, $B(R_{\rm j})$, $\sigma$) shows a significant monotonic trend with redshift, indicating that the inferred properties are not artificially driven by the $(1+z)^4$ growth of the CMB energy density.

While the present data already permit a basic model comparison, a more detailed analysis, and in particular a more precise determination of the model parameters, will require improved datasets. On the radio side, multi-frequency flux measurements spanning broader segments of the synchrotron continuum are needed to constrain the field profile $b(x)$ and the maximum electron energy $\gamma_{\rm
max}$, both of which are currently poorly determined for most sources. In the X-rays, deeper \textit{Chandra} exposures would tighten the photon-index measurements. As discussed in Section~\ref{sec:results}, the jet radius $R_{\rm j}$ is at present often set by the instrumental PSF ($\sim 0\farcs5$) and should be regarded as an upper limit, which directly limits the precision of the inferred energetics. We note that $\gamma$-ray upper limits, while highly constraining for low-redshift sources, lose much of their diagnostic power for the present high-$z$ sample, since the predicted GeV-TeV
fluxes are strongly suppressed by absorption on the EBL over the large path lengths involved; optical/UV upper limits are likewise of limited value here, as the model predicts a deep flux minimum between the synchrotron and IC/CMB components in that band.

Several extensions of the present framework are envisaged for future work:
\begin{itemize}
  \item incorporating EBL absorption, so that the predicted $\gamma$-ray emission can be confronted with \textit{Fermi}-LAT and ground-based Cherenkov-telescope constraints, most informatively for the low-redshift jet population;
  \item allowing a radial stratification of the radiating-electron energy distribution $\zeta(\gamma_e)$, relaxing the assumption of a single, spatially uniform spectral shape;
  \item including a non-zero poloidal magnetic-field component $B_z(r)$, relevant closer to the jet launching region and for a more complete treatment of the magnetization;
  \item modeling the polarization of the synchrotron component, which offers an independent and potentially decisive discriminant between the competing emission scenarios and between the model variants considered here.
\end{itemize}

\begin{acknowledgements}
   P.L. acknowledges that the study was funded under the ``Research support module'' as a part of the ``Excellence Initiative -- Research University'' program at the Jagiellonian University. C.C.C. was supported by NASA DPR S-15633-Y. G.M. acknowledges financial support from INAF GO grant 1.05.24.02.03 Bando Ricerca Fondamentale INAF 2024. A.S. acknowledges support from NASA Contract NAS8-03060 to the Chandra X-ray Center.
\end{acknowledgements}

\bibliography{bibliography}{}
\bibliographystyle{aasjournalv7}

\appendix

\section{Jet Radial Profiles} \label{sec:AppA}

\begin{figure}[htbp]
    \centering
    \labeledimage{(a)}{0.40}{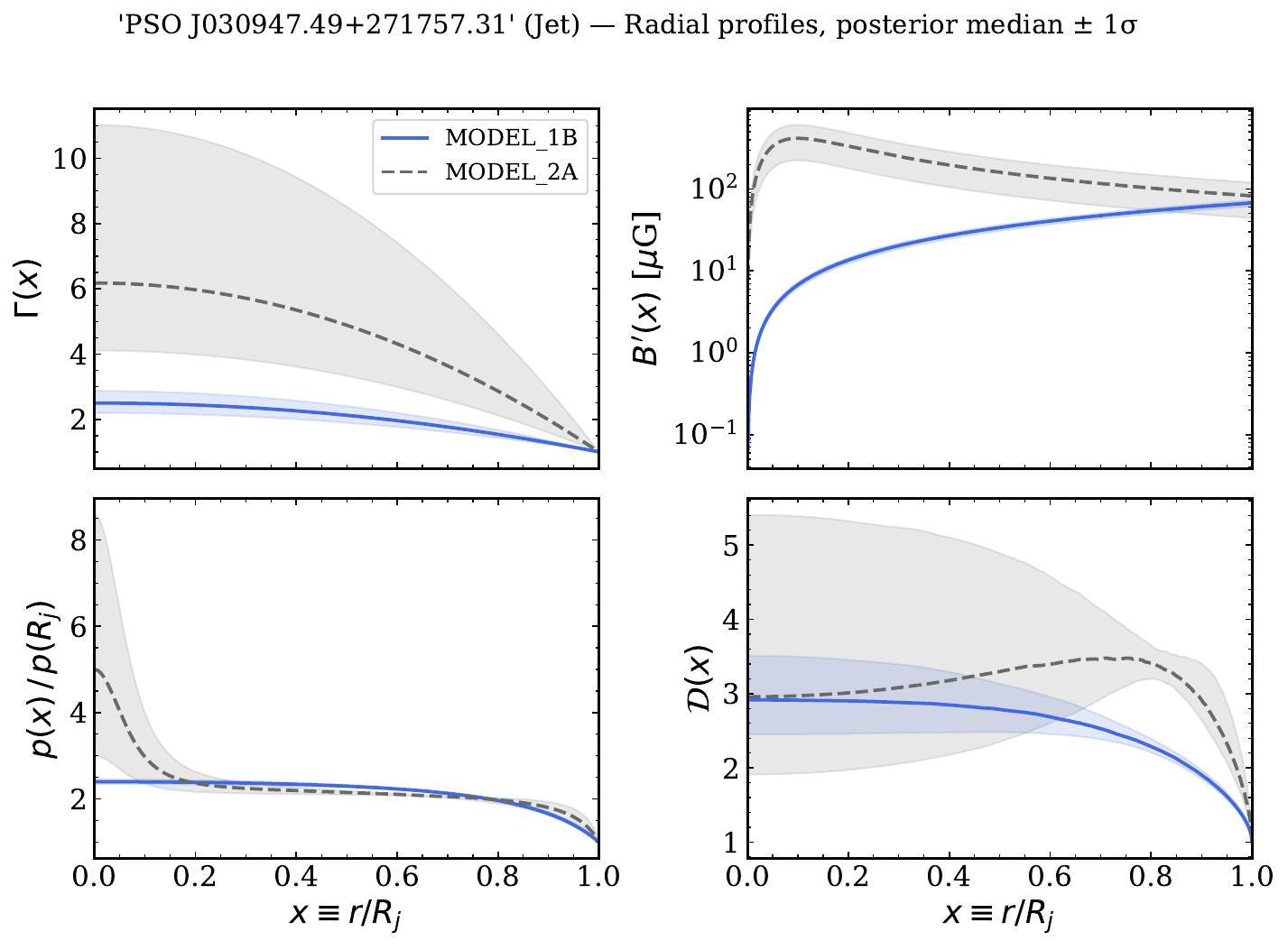} \hspace{1em}
    \labeledimage{(b)}{0.40}{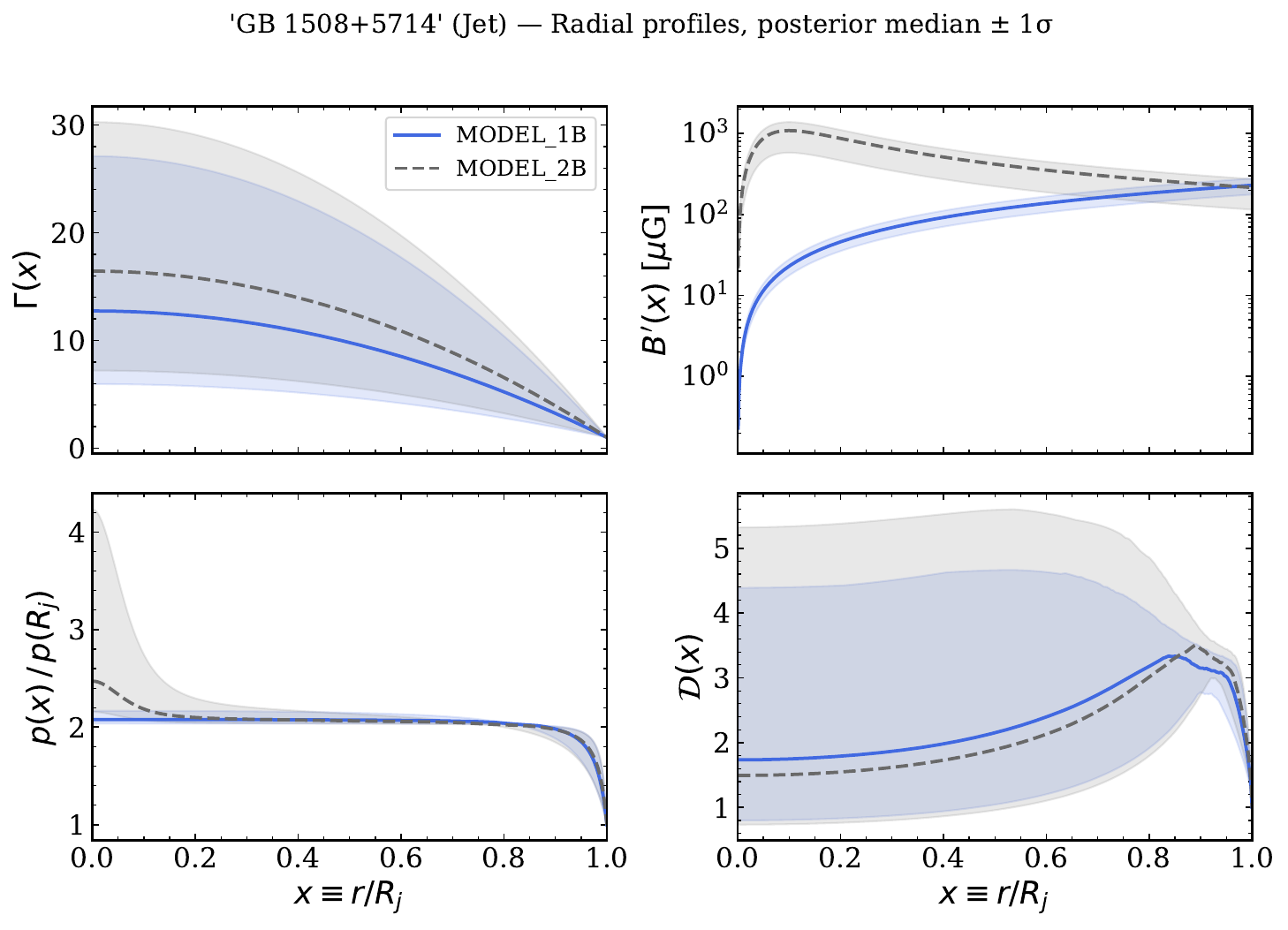}
    
    \vspace{1em}

    \labeledimage{(c)}{0.40}{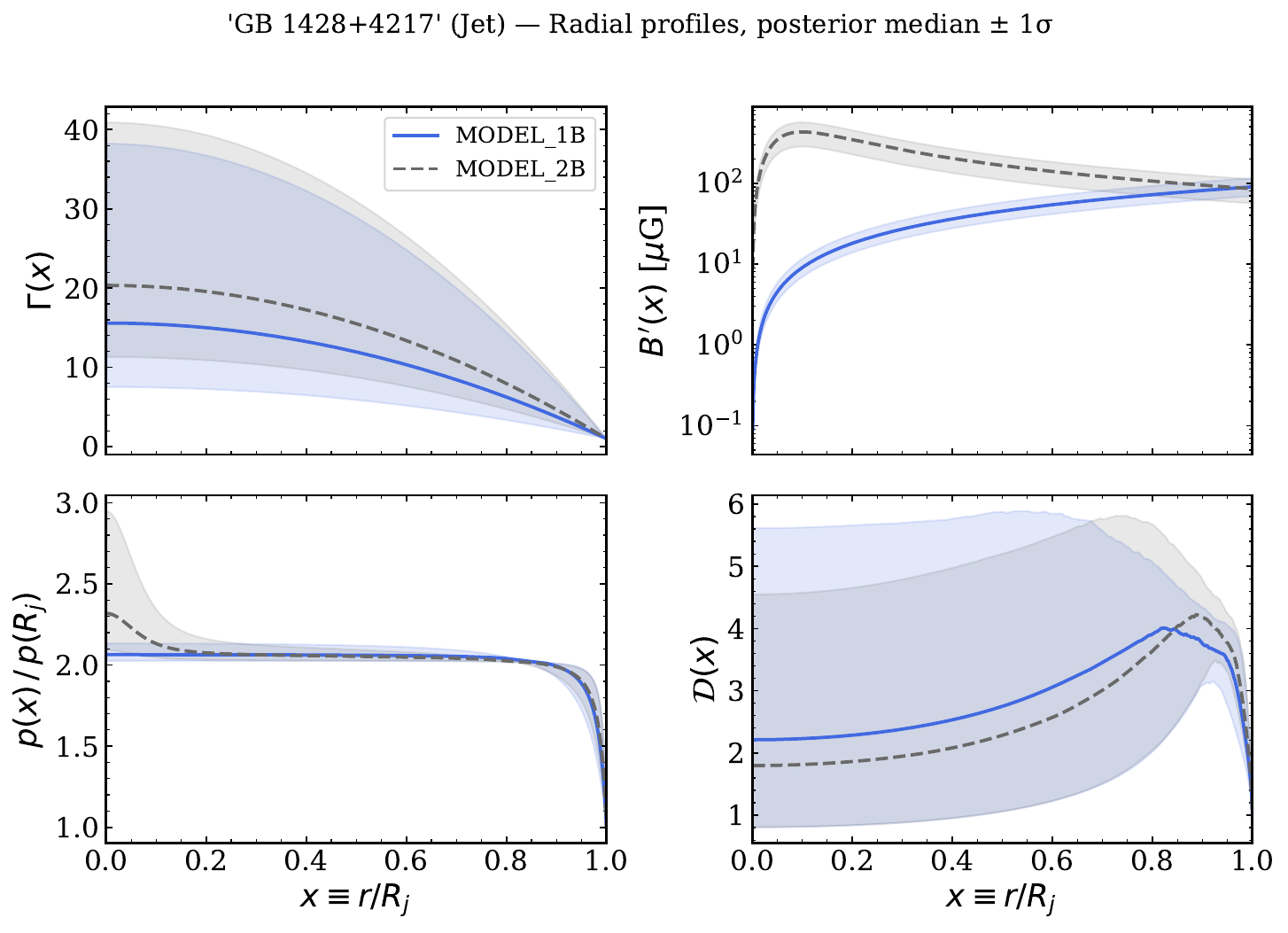} \hspace{1em}
    \labeledimage{(d)}{0.40}{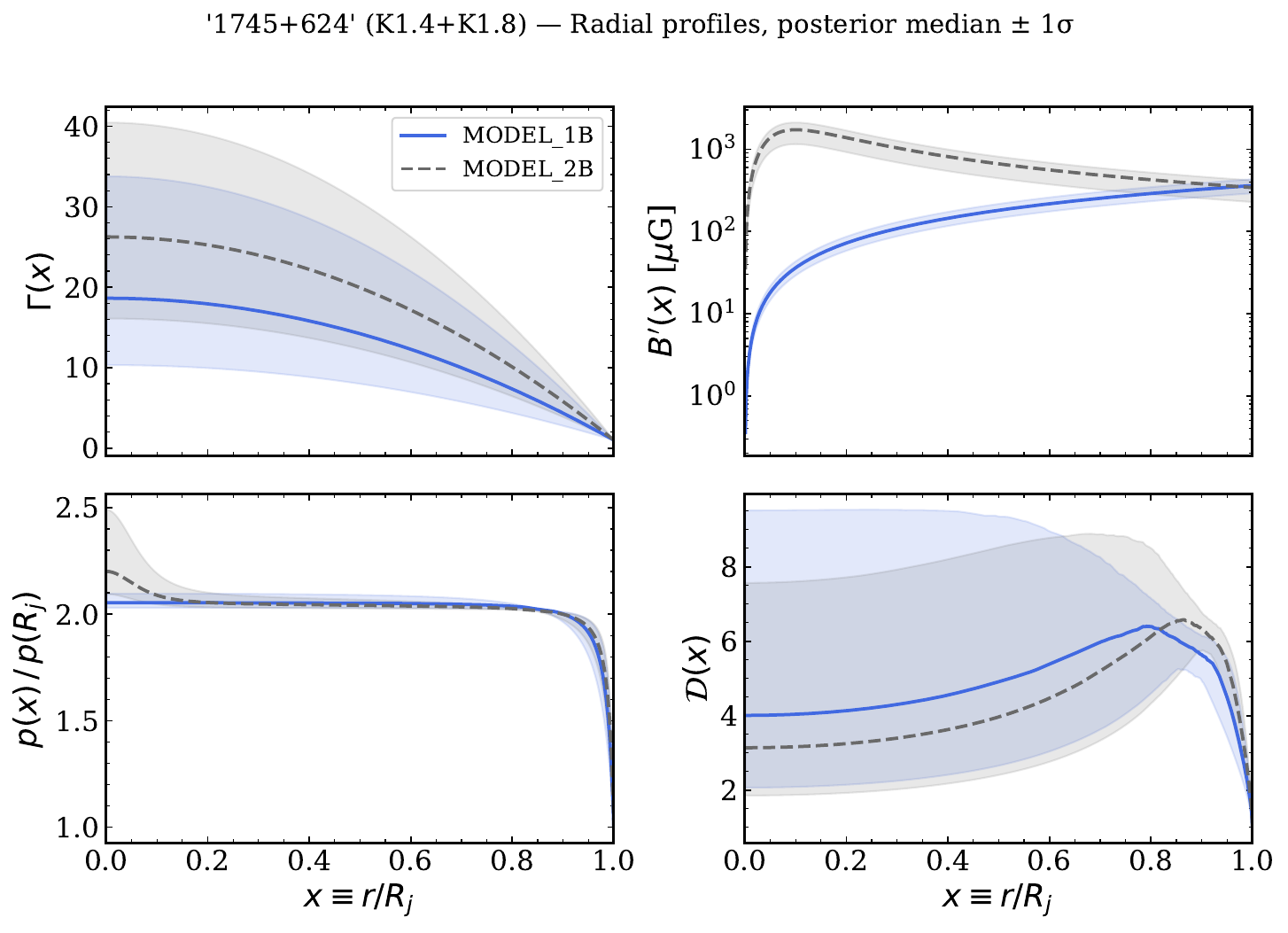}

    \vspace{1em}

    \labeledimage{(e)}{0.40}{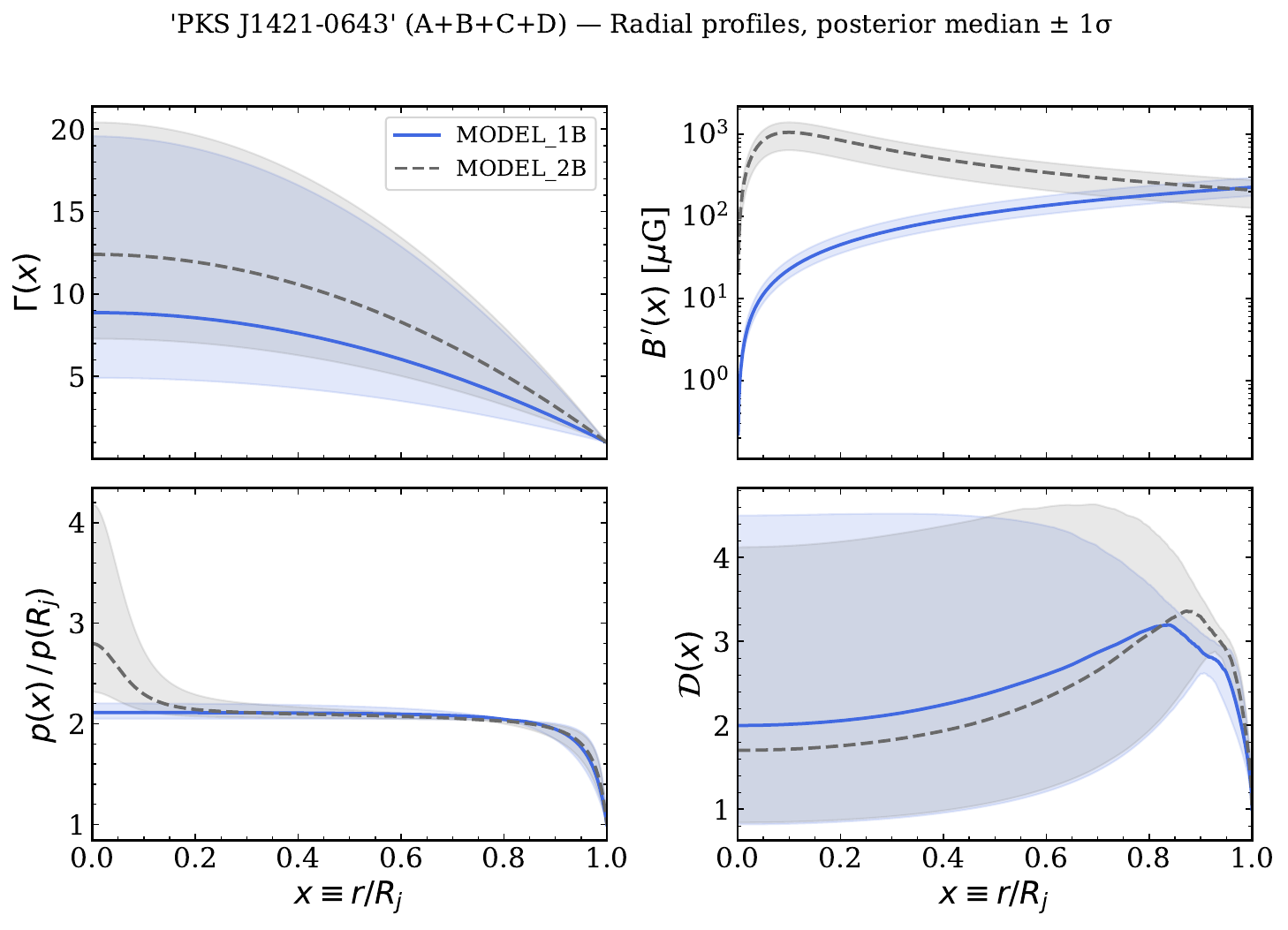} \hspace{1em}
    \labeledimage{(f)}{0.40}{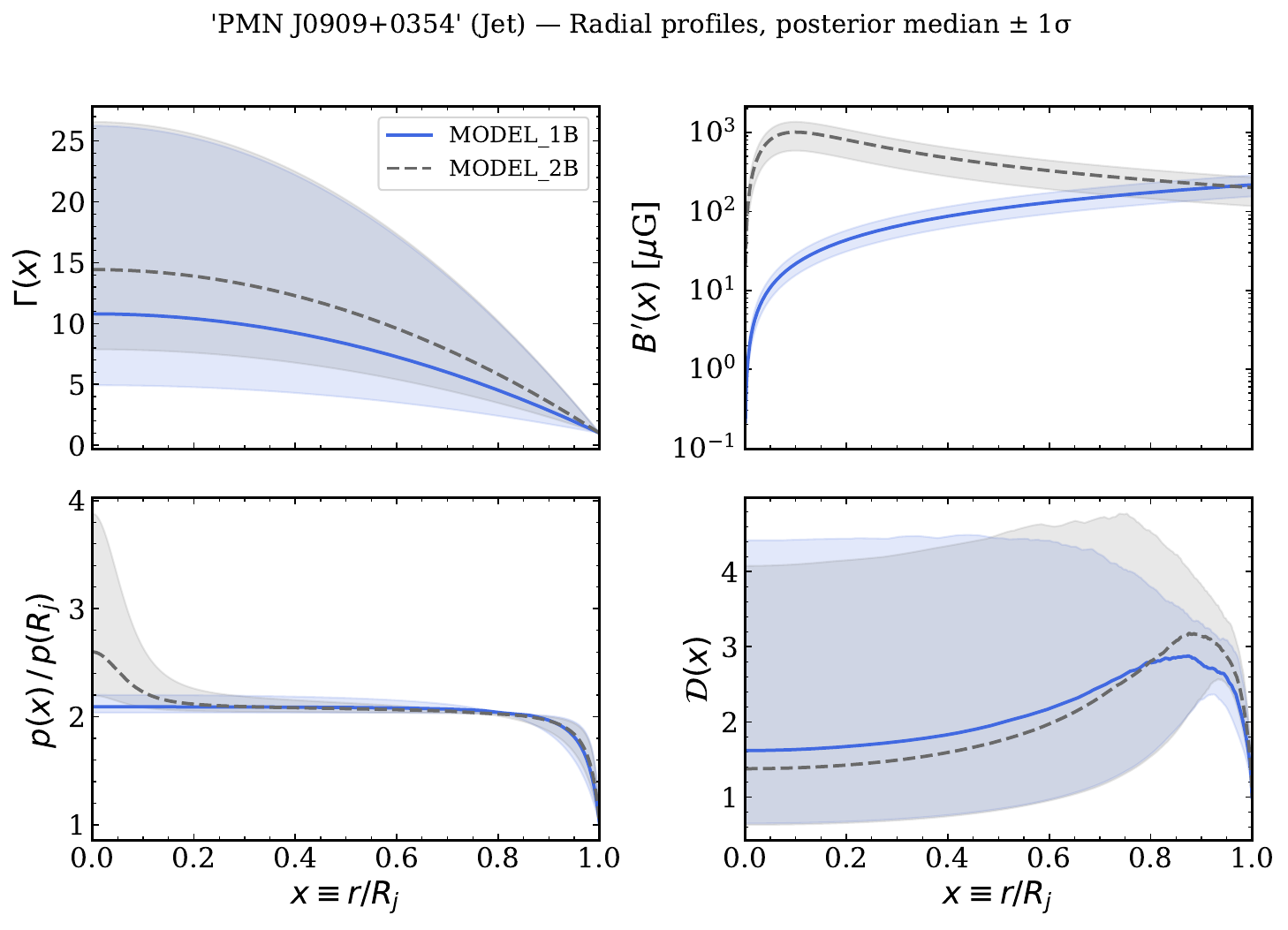}

    \caption{Cross-jet parameter profile plots for the modeled sources, computed for models 1B and 2B: (a) PSO J030947.49+271757.31, (b) GB 1508+5714, (c) GB 1428+4217, (d) 1745+624, (e) PKS J1421-0643, (f) PMN J0909+0354. (Figure continues on the next page.)}
    \label{fig:profiles_part1}
\end{figure}

\begin{figure}[htbp]
    \centering
    \labeledimage{(g)}{0.40}{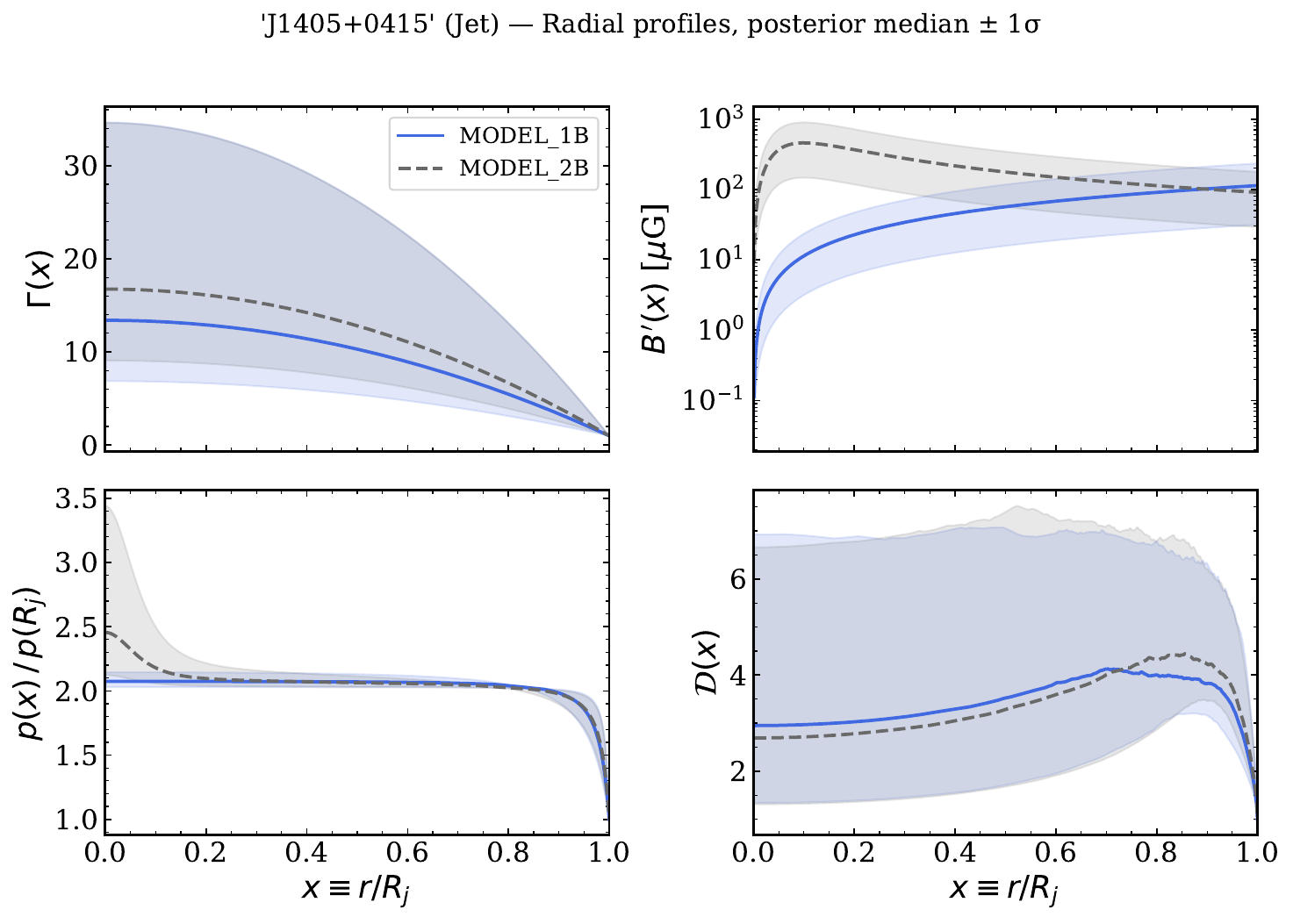} \hspace{1em}
    \labeledimage{(h)}{0.40}{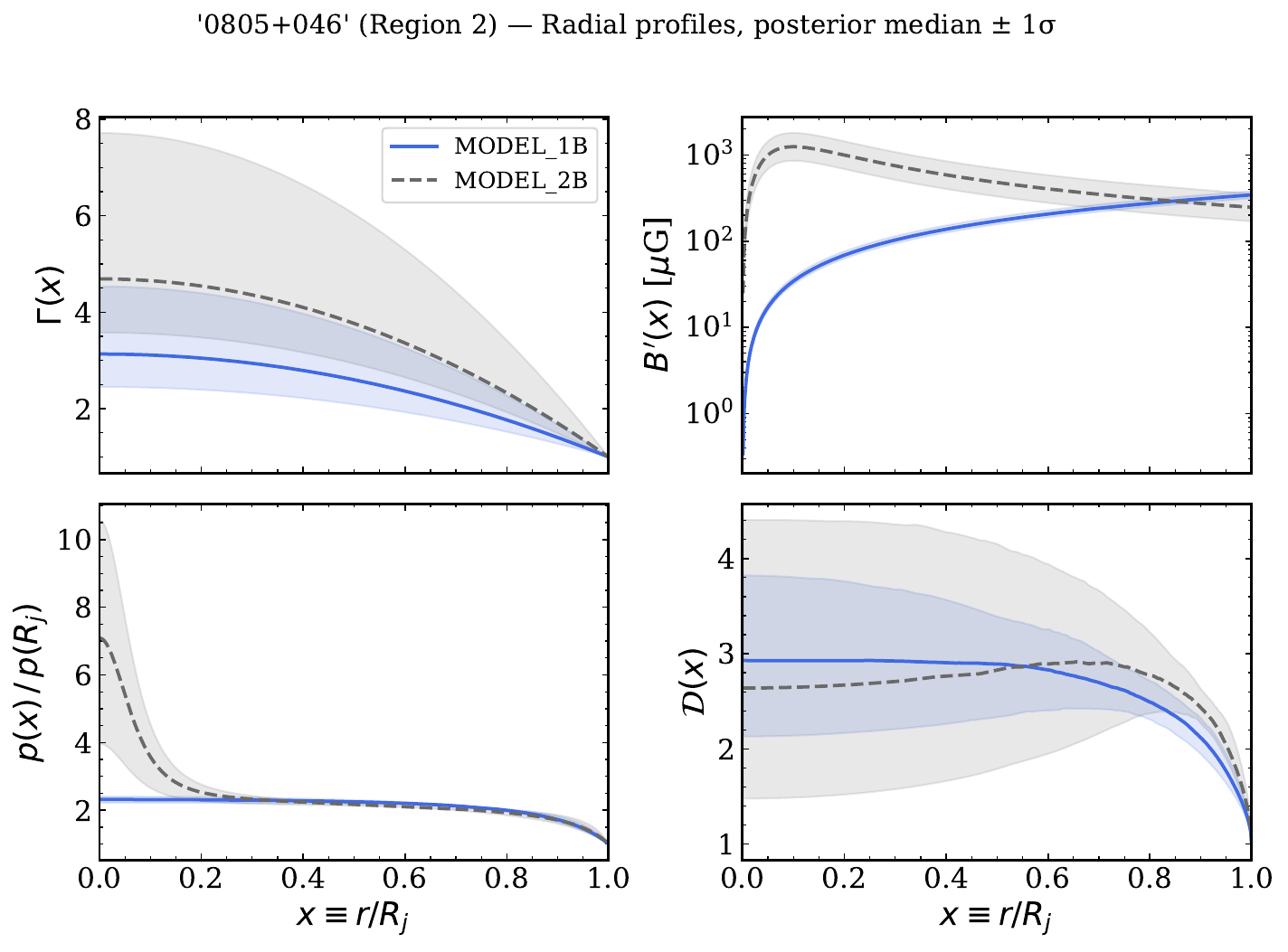}

    \vspace{1em}

    \labeledimage{(i)}{0.40}{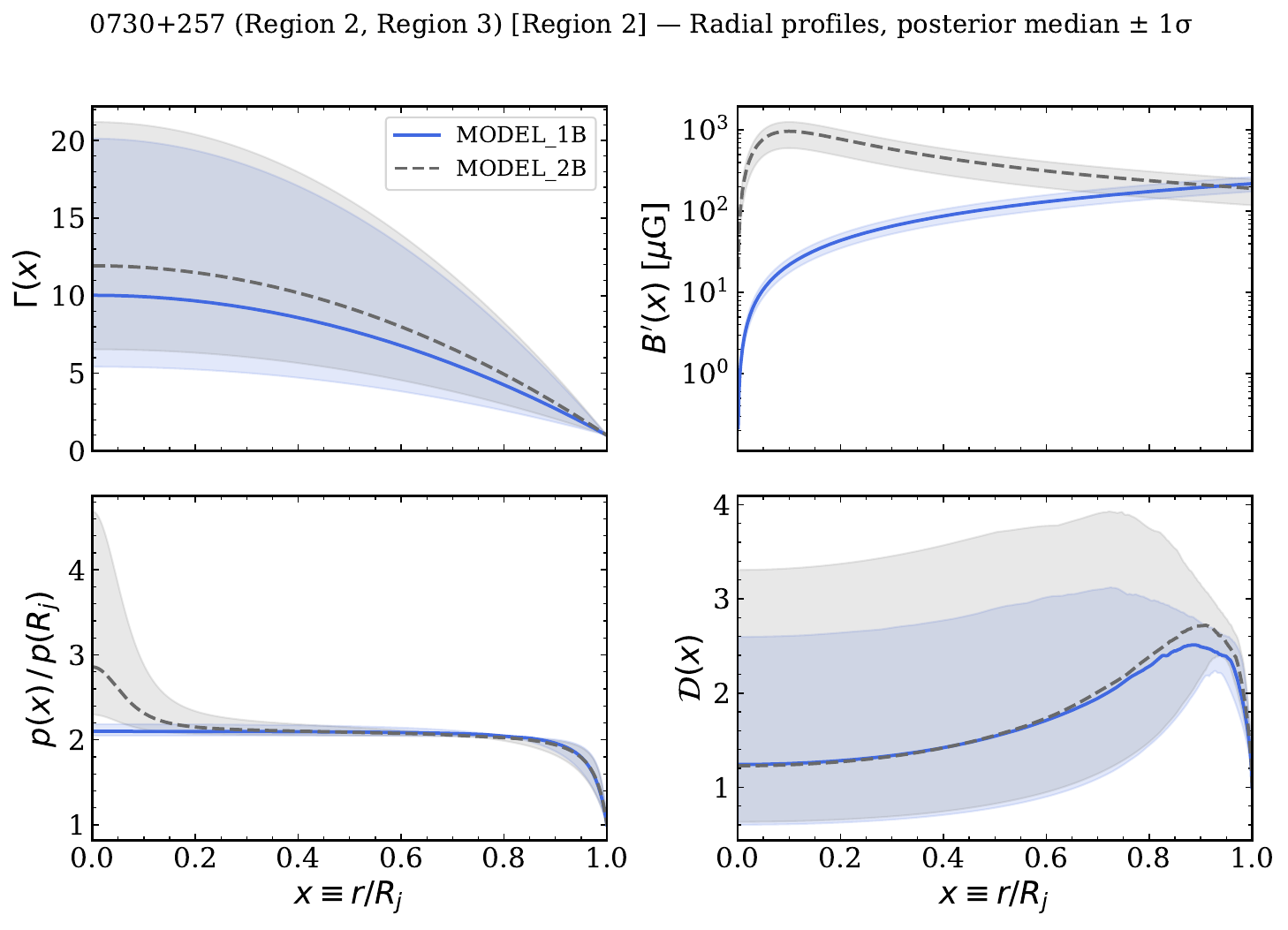} \hspace{1em}
    \labeledimage{(j)}{0.40}{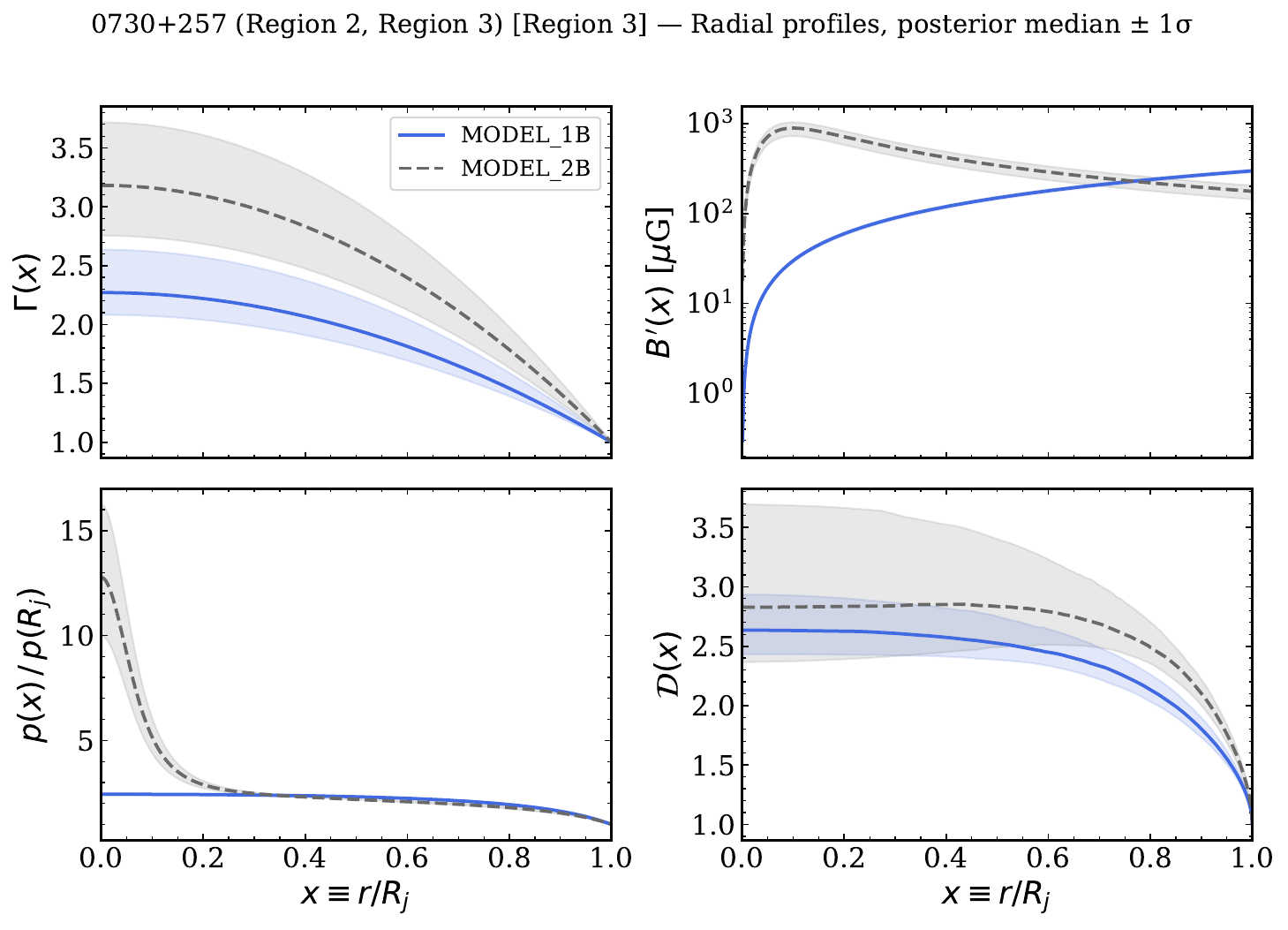}

    \vspace{1em}

    \labeledimage{(k)}{0.40}{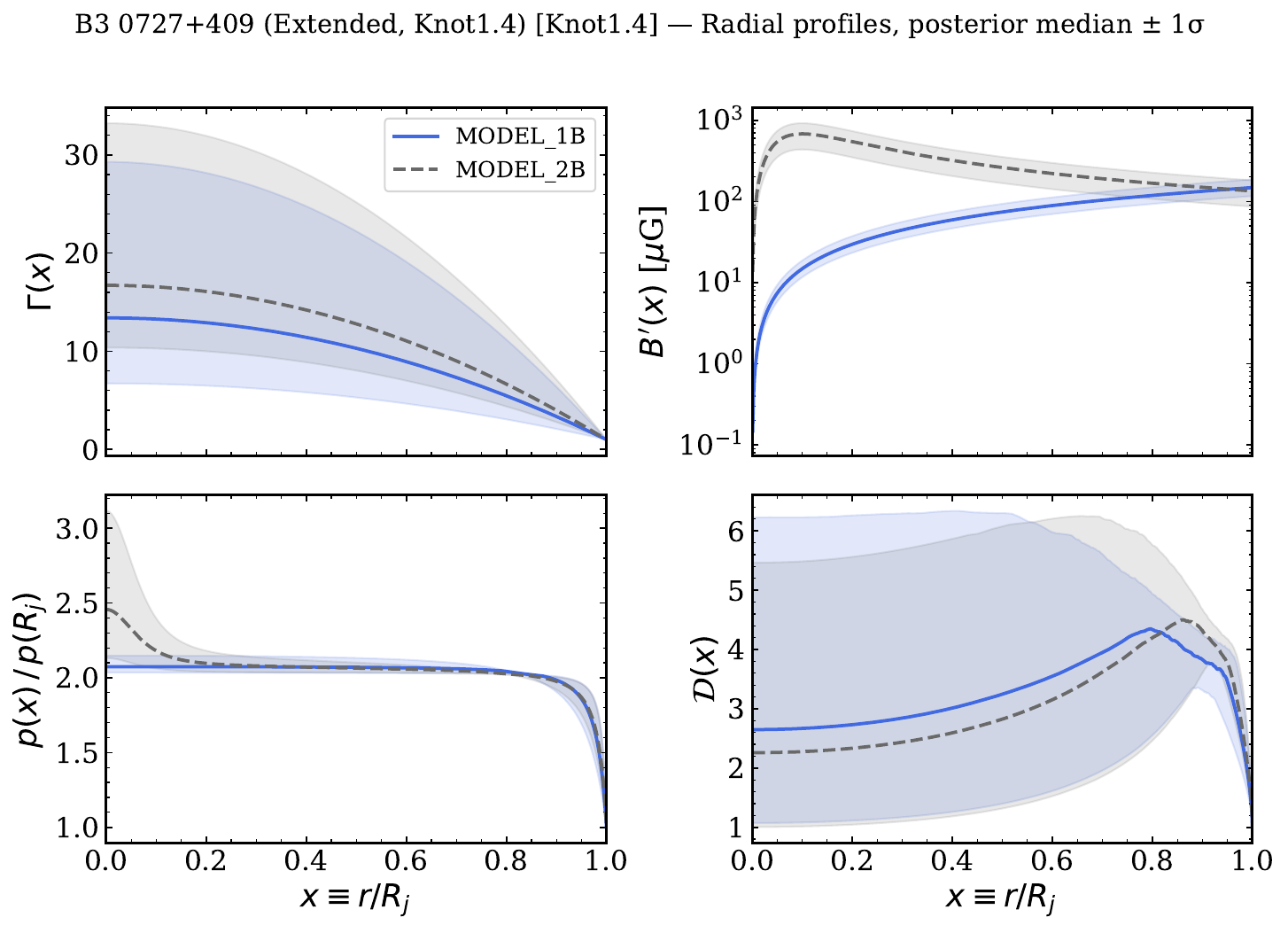} \hspace{1em}
    \labeledimage{(l)}{0.40}{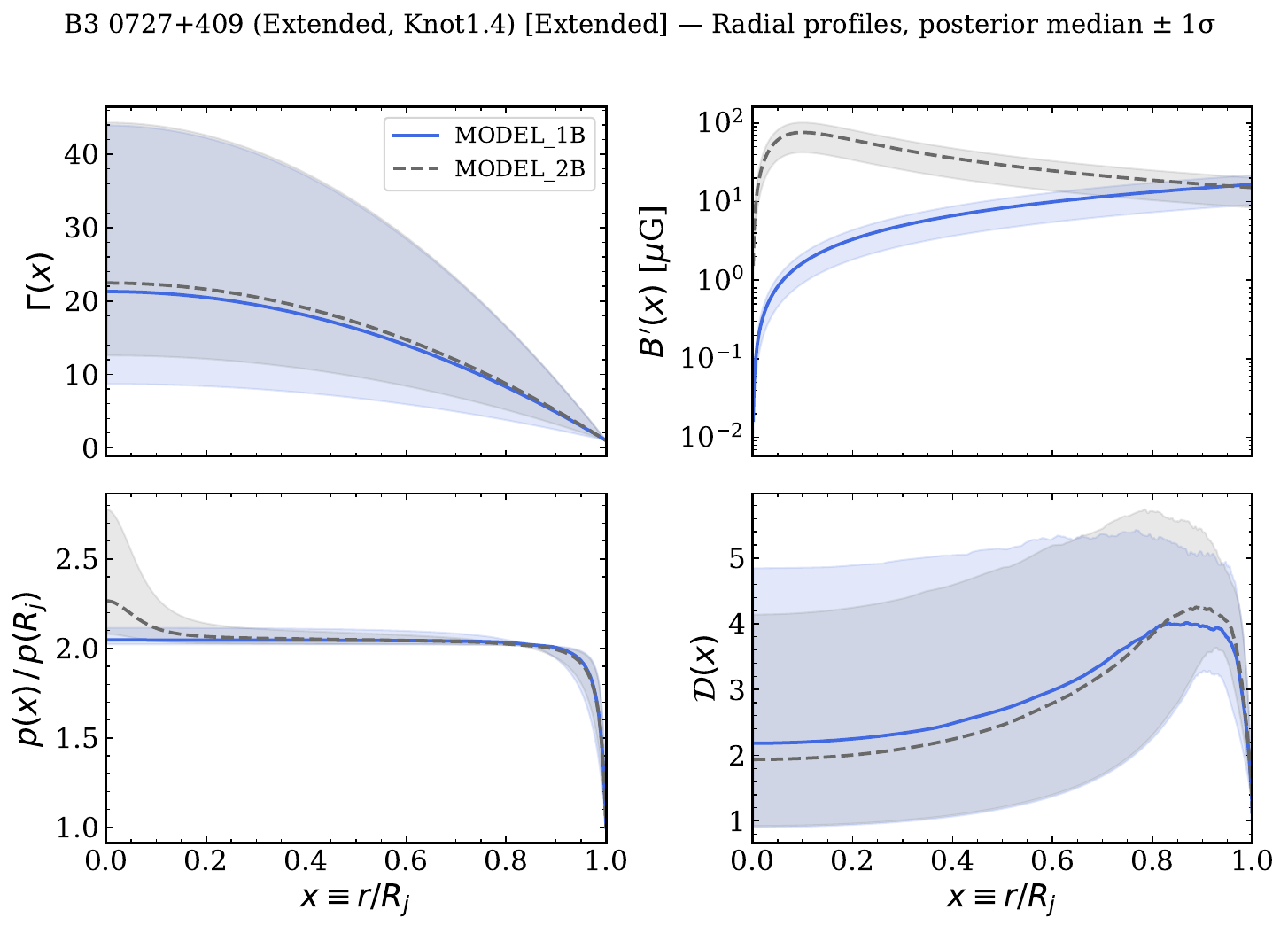}

    \caption{Parameter profile plots (continued). (g) J1405+0415, (h) 0805+046, (i) 0730+257 (Region 2), (j) 0730+257 (Region 3), (k) B3 0727+409 (Knot 1.4), (l) B3 0727+409 (Extended jet).}
    \label{fig:profiles_part2}
\end{figure}

In Figures~\ref{fig:profiles_part1} and \ref{fig:profiles_part2} we present the best-fit transverse (cross-jet) radial profiles of the bulk Lorentz factor $\Gamma(x)$, the comoving toroidal magnetic field $B'(x) = B_\phi(x)/\Gamma(x)$, the normalized gas pressure $P(x)/P(R_{\rm j})$, and the local Doppler factor $\mathcal{D}(x)$, as functions of the normalized jet radius $x \equiv r/R_{\rm j}$, for each source in the sample. The profiles are shown as posterior medians with $1\sigma$ credible bands propagated from 1{,}000 posterior draws. For every source we overplot the two statistically preferred variants identified in Section~\ref{sec:results}: for most sources these are the gas-pressure-normalized models 1B and 2B. For the six sources without multi-frequency radio coverage these are computed at the fiducial value of $\gamma_{\rm max} = 10^5$; for the four sources with the best-sampled radio continua (GB\,1508$+$5714, 1745$+$624, PKS\,J1421$-$0643, and PMN\,J0909$+$0354) they are computed with $\gamma_{\rm max}$ left free in the fit. Together these panels illustrate how the magneto-hydrostatic equilibrium ties the dynamical and radiative structure of the jet together: the gas pressure decreases monotonically outward from the axis, the rest-frame field either rises toward the boundary (model 1) or peaks near $x \simeq 0.1$ before declining (model 2), and the Doppler factor varies across the cross-section through the velocity shear $\Gamma(x)$, producing the radially dependent boosting responsible for the differing effective Doppler factors of the synchrotron and IC/CMB components discussed in Section~\ref{sec:discussion}. 

\section{Corner Plots} \label{sec:AppB}

Figures~\ref{fig:corner_part1} and \ref{fig:corner_part2} show the marginalized one- and two-dimensional posterior distributions of the three free parameters: $\vartheta$, $\log L_{\rm j}$, and $\log(\Gamma_0 - 1)$ for all four model variants and every source, allowing a direct visual comparison of the parameter constraints and degeneracies across models. A pronounced positive $\Gamma_0$--$L_{\rm j}$ correlation is apparent in all cases, reflecting the trade-off between Doppler boosting and intrinsic power in reproducing a fixed observed IC/CMB flux. For the four sources with multi-frequency radio coverage (GB\,1508$+$5714, 1745$+$624, PKS\,J1421$-$0643, and PMN\,J0909$+$0354) the posteriors include $\gamma_{\rm max}$ as an additional free parameter; for the remaining sources $\gamma_{\rm max}$ is fixed at the fiducial value of $10^5$.

\begin{figure}[htbp]
    \centering
    \labeledimage{(a)}{0.35}{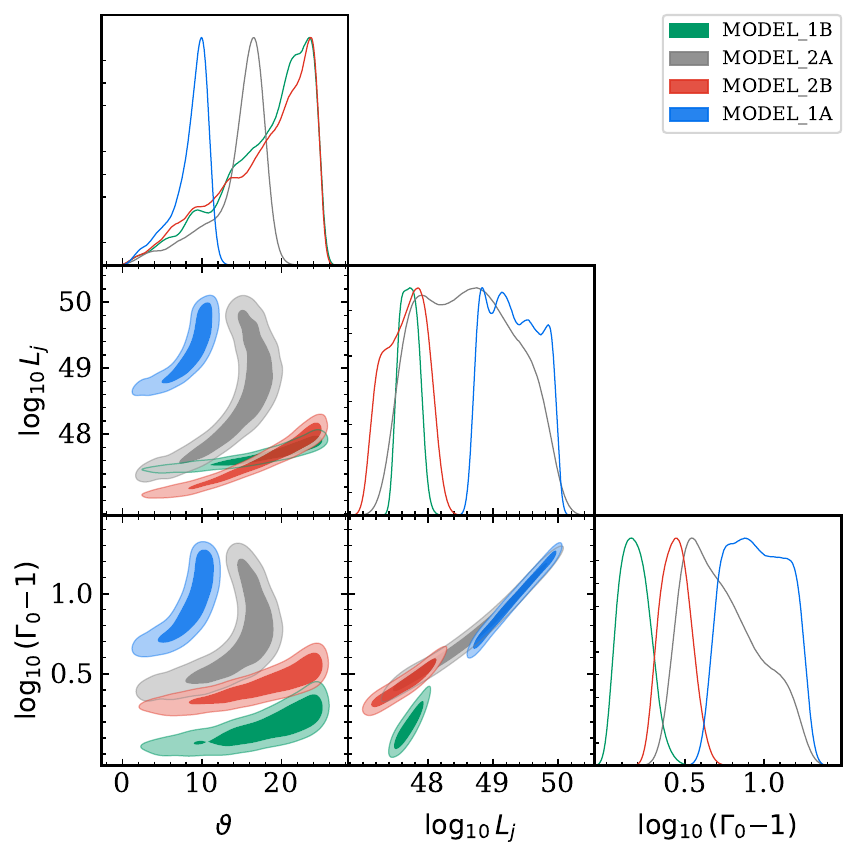} \hspace{1em}
    \labeledimage{(b)}{0.35}{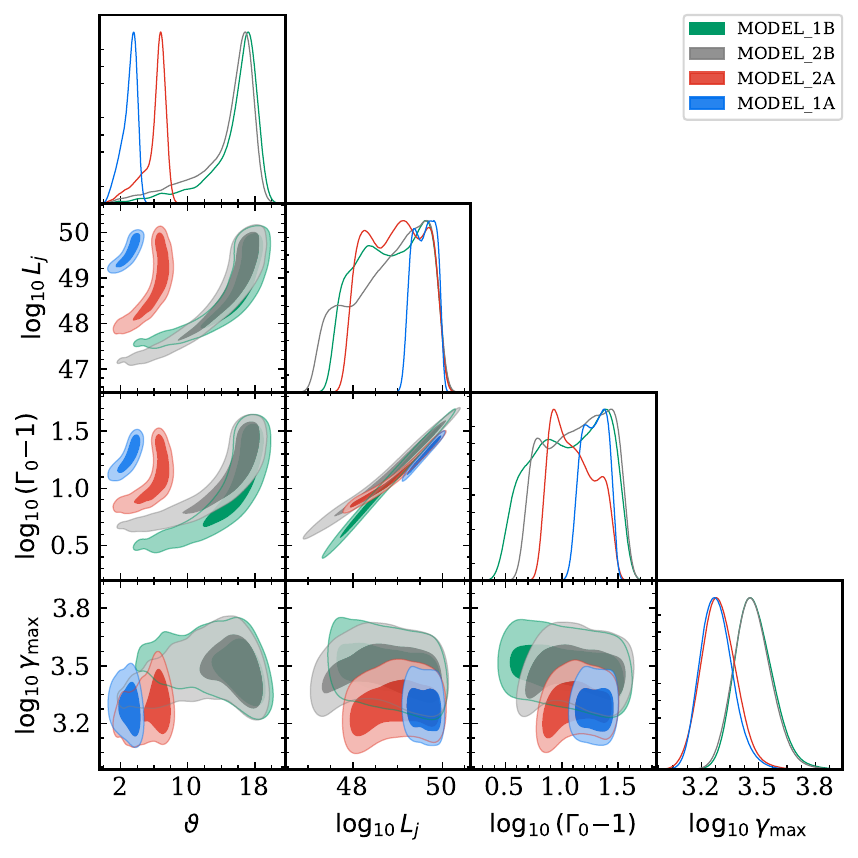}

    \vspace{1em}

    \labeledimage{(c)}{0.35}{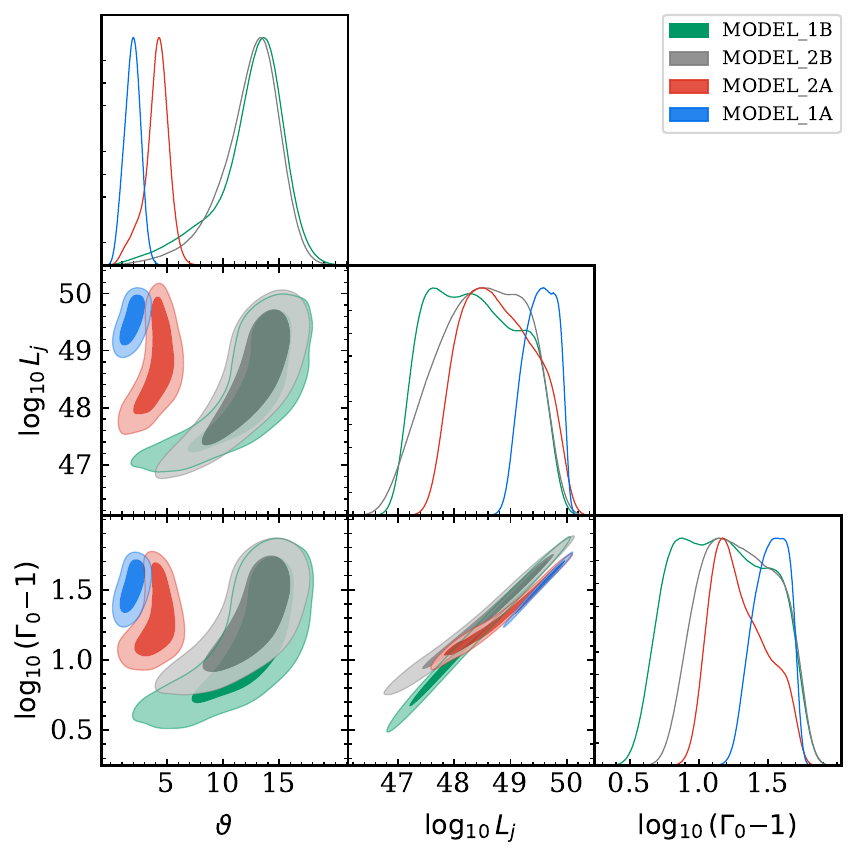} \hspace{1em}
    \labeledimage{(d)}{0.35}{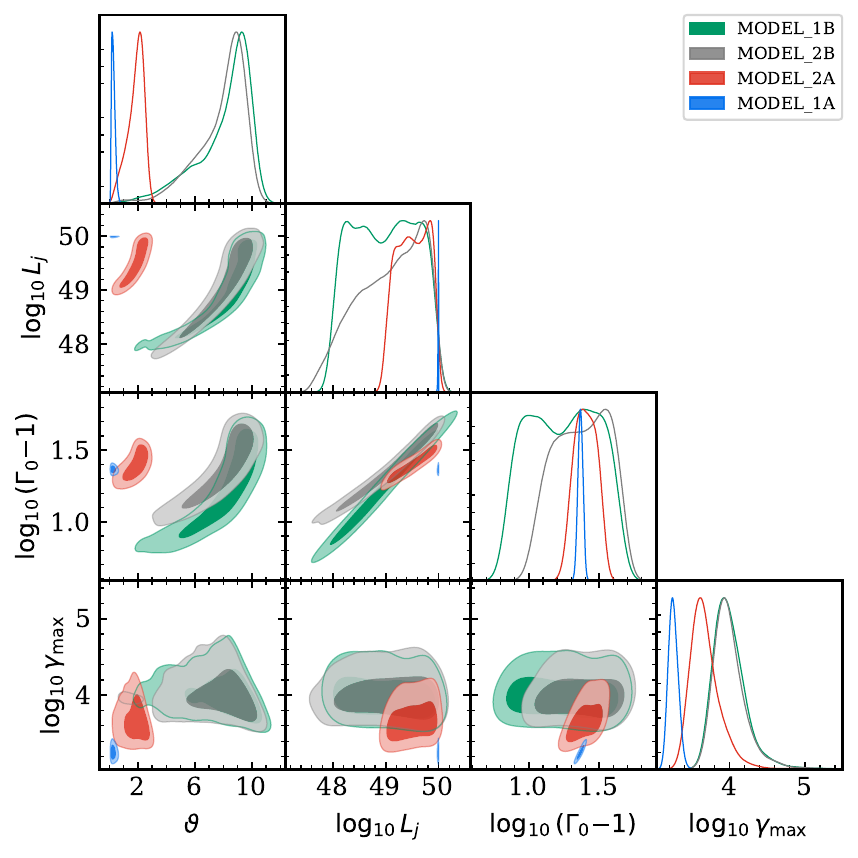}

    \vspace{1em}

    \labeledimage{(e)}{0.35}{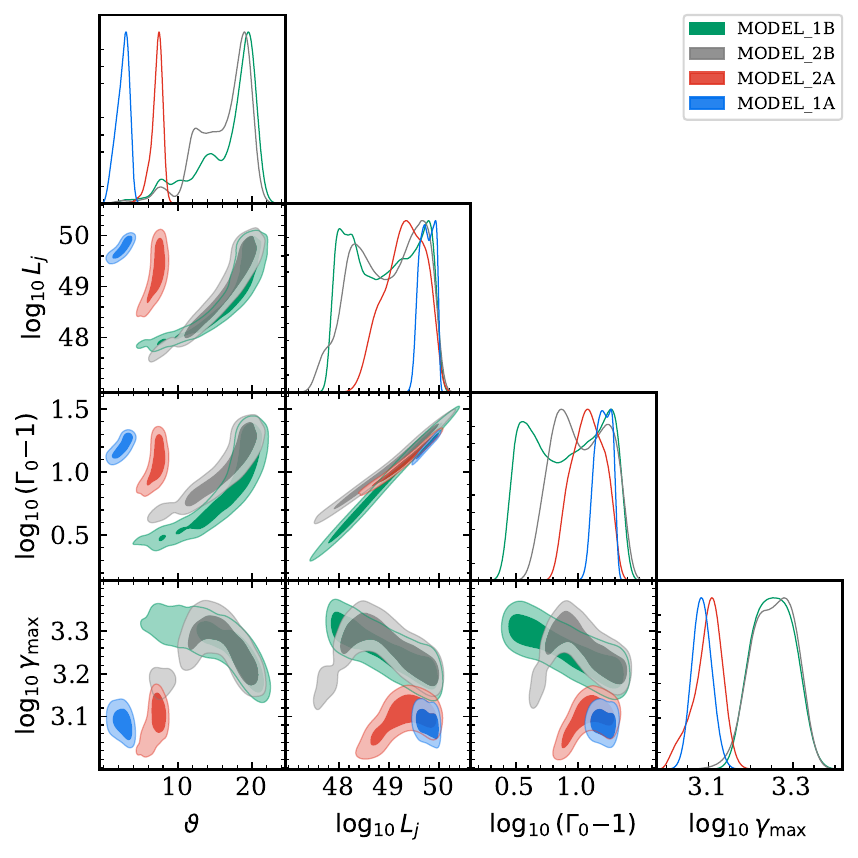}

    \caption{Corner plots for the modeled sources. (a) PSO J030947.49+271757.31, (b) GB 1508+5714, (c) GB 1428+4217, (d) 1745+624, (e) PKS J1421-0643. (Figure continues on the next page.)}
    \label{fig:corner_part1}
\end{figure}

\begin{figure}[htbp]
    \ContinuedFloat 
    \centering
    \labeledimage{(f)}{0.35}{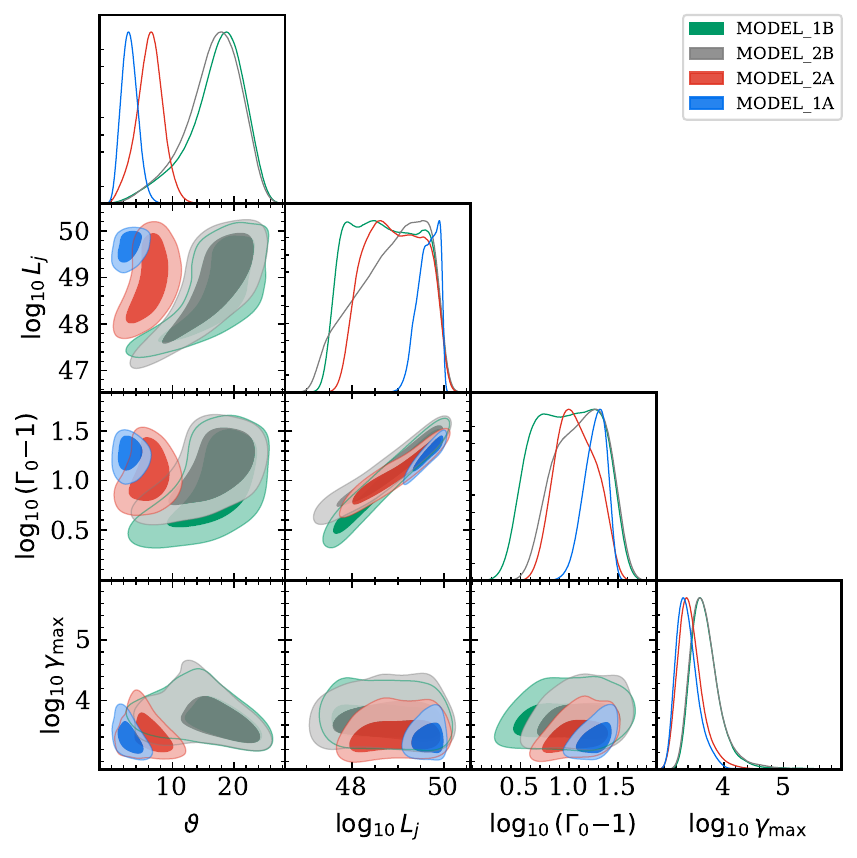} \hspace{1em}
    \labeledimage{(g)}{0.35}{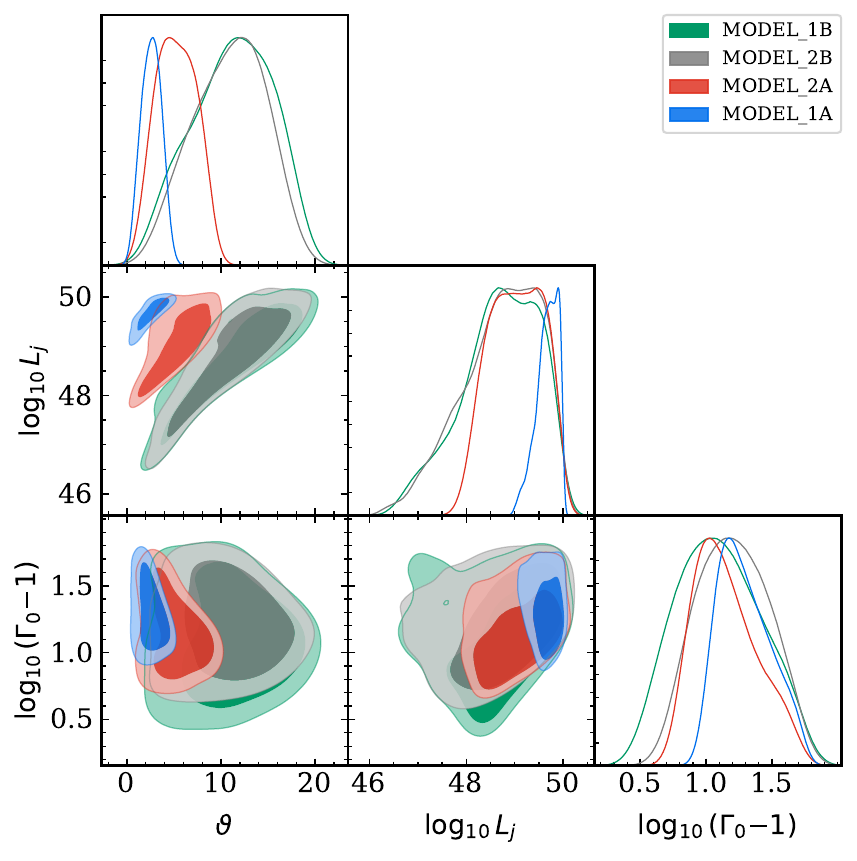}

    \vspace{1em}

    \labeledimage{(h)}{0.35}{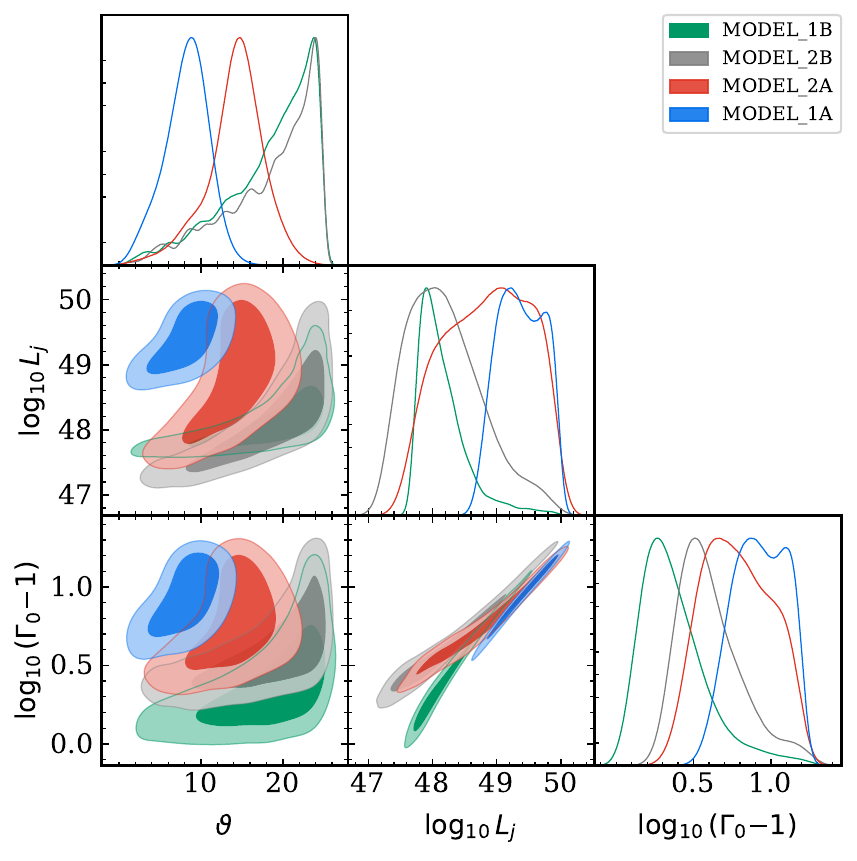} \hspace{1em}
    \labeledimage{(i)}{0.35}{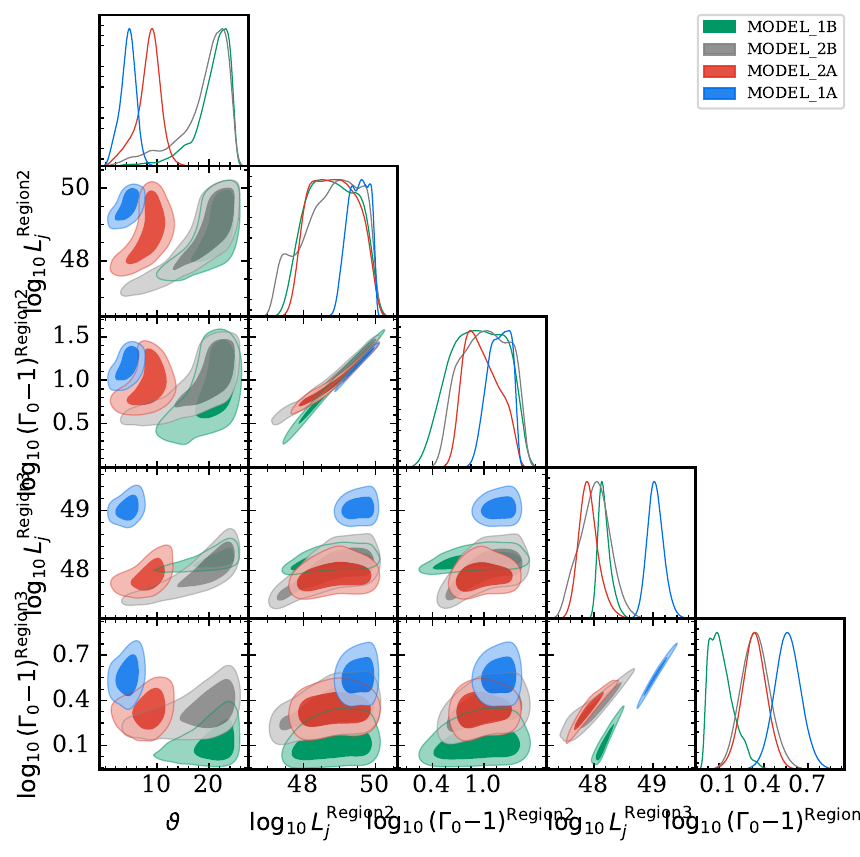}

    \vspace{1em}

    \labeledimage{(j)}{0.35}{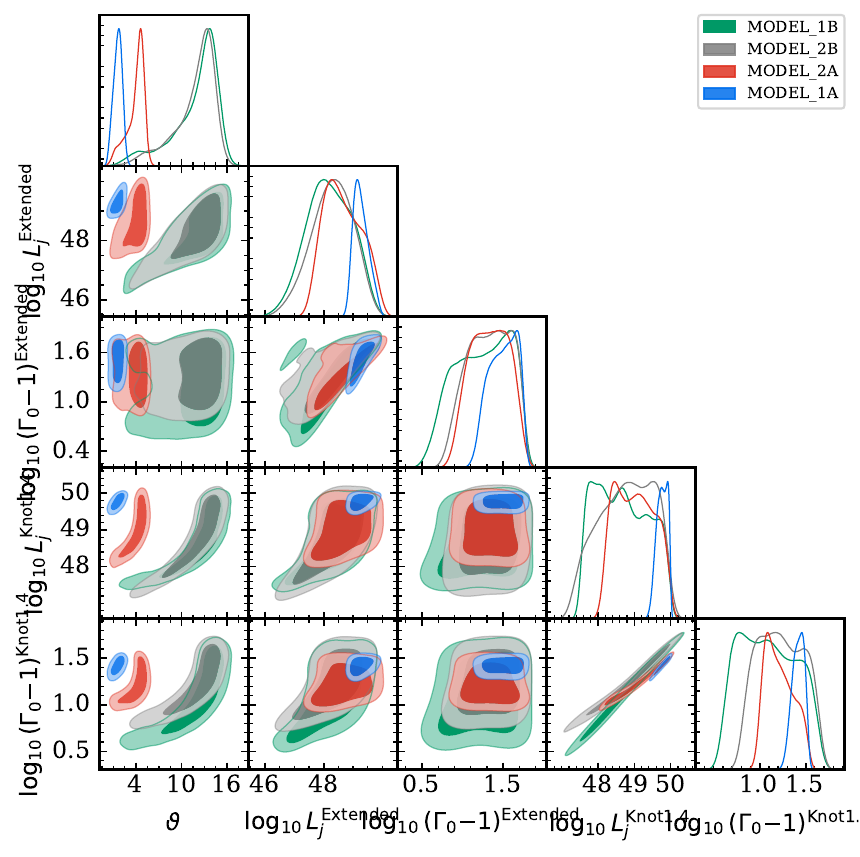}

    \caption[]{Corner plots (continued). (f) PMN J0909+0354, (g) J1405+0415, (h) 0805+046, (i) 0730+257, (j) B3 0727+409.}
    \label{fig:corner_part2}
\end{figure}

\end{document}